\documentclass[a4paper,fleqn,usenatbib,useAMS]{mnras}

\usepackage[T1]{fontenc}
\usepackage{ae,aecompl}

\usepackage{graphicx}	
\usepackage{amsmath}	
\usepackage{amssymb}	
\usepackage{multirow}
\usepackage{longtable}

\usepackage{graphicx}	
\usepackage{multicol}        
\usepackage{bm}		
\usepackage{pdflscape}	

\title[Searching for shocks in SCCs]{Searching for shocks in high-mass starless clump candidates}

\author[F.-Y. Zhu et al.]{
Feng-Yao Zhu,$^{1}$\thanks{E-mail: zhufy@shao.ac.cn}
Jun-Zhi Wang,$^{1}$
Tie Liu$^{1,2}$
Kee-Tae Kim$^{2,3}$
Qing-Feng Zhu$^{4}$
and Fei Li$^{1}$
\\
$^{1}$Shanghai Astronomical Observatory, Chinese Academy of Sciences, Shanghai, 200030, China\\
$^{2}$Korea Astronomy and Space Science Institute, Daejeon, 34055, Korea\\
$^{3}$University of Science and Technology, Korea (UST), 217 Gajeong-ro, Yuseong-gu, Daejeon, 34113, Republic of Korea\\
$^{4}$CAS Key Laboratory for Research in Galaxies and Cosmology, Department of Astronomy, University of Science and Technology of China,\\
Hefei, 230026, China \\
}

\date{Accepted XXX. Received YYY; in original form ZZZ}

\pubyear{2020}

\begin{document}
\label{firstpage}
\pagerange{\pageref{firstpage}--\pageref{lastpage}}
\maketitle

\begin{abstract}
In order to search for shocks in the very early stage of star formation, we performed single-point surveys of SiO J=1-0, 2-1 and 3-2 lines and the H$_2$CO $2_{12}-1_{11}$ line toward a sample of 100 high-mass starless clump candidates (SCCs) by using the Korean VLBI Network (KVN) 21-m radio telescopes. The detection rates of the SiO J=1-0, 2-1, 3-2 lines and the H$_2$CO line are $31.0\%$, $31.0\%$, $19.5\%$ and $93.0\%$, respectively. Shocks seem to be common in this stage of massive star formation. The widths of the observed SiO lines (full width at zero power (FWZP)) range from 3.4 to 55.1 km s$^{-1}$. A significant fraction ($\sim29\%$) of the detected SiO spectra have broad line widths (FWZP $>20~km~s^{-1}$), which are very likely associated with fast shocks driven by protostellar outflows. This result suggests that about one third of the SiO-detected SCCs are not really starless but protostellar. On the other hand, about 40$\%$ of the detected SiO spectra show narrow line widths (FWZP<10 $km~s^{-1}$) probably associated with low-velocity shocks which are not necessarily protostellar in origin. The estimated SiO column densities are mostly $0.31-4.32\times10^{12}~cm^{-2}$. Comparing the SiO column densities derived from SiO J=1-0 and 2-1 lines, we suggest that the SiO molecules in the SCCs may be in the non-LTE condition. The SiO abundances to H$_2$ are usually $0.20-10.92\times10^{-10}$.
\end{abstract}




\section{Introduction} \label{sec:intro}

Massive starless clumps represent the environments that are in the very early stage of star formation, and are suggested as the first phase of high-mass star formation \citep{mot18}. They have not been significantly disrupted by feedback from protostars \citep{mat17,cal18}, and are ideal locations to improve our understanding of massive star and cluster formation \citep{svo19}. Blind surveys of dust continuum emission at (sub)millimeter wavelengths can be used to detect star-forming regions in different evolutionary stages \citep{tra15,cse16,svo16,urq18}. And a sample of massive clumps have been identified as Starless Clump Candidates (SCCs) in the 1.1 mm continuum Bolocam Galactic Plane Survey \citep{svo16}. Although there could be undetected low-mass protostars in these regions since only protostars with bolometric luminosities higher than the $30-140L_\odot$ sensitivity of far-infrared Galactic plane surveys are able to be detected, high-mass protostars are unlikely to exist \citep{svo16,cal18}. So SCCs can still be representatives of the early stage of star formation since weak feedbacks from low-mass protostars can not seriously disturb the early star-forming environment. The study of these SCCs can give us a better understanding of the early evolution of star formation. Abundant in star-forming regions, shocks can significantly influence the environments and provide some information of ongoing star formation activities. Surveys of molecular shock tracers toward these SCCs can help us to determine whether the shocks are common in these environments, and to study the properties of the shocked materials in the early evolutionary stage of star formation.

SiO lines are often used as a probe of shocks in star-forming regions with a wide range of evolutionary stages \citep{jim10,dua14,cse16,cos18}. SiO is believed to form via sputtering and vaporisation of Si atoms from grains due to shock activities \citep{gus08}, and its abundance can be enhanced by fast shocks up to six orders of magnitude higher than those of quiescent regions. This makes SiO emission unlikely being contaminated by the line emission from ambient interstellar medium \citep{li19}. SiO rotational lines have been observed in various conditions such as high-velocity shocks ($v_s\geq20~km~s^{-1}$) caused by powerful outflows from massive protostars and low-velocity shocks ($v_s<10~km~s^{-1}$) \citep{ngu13,dua14,cse16}. Some previous studies reported the observed SiO emission contains both broad and narrow Gaussian components at approximately the same central velocity \citep{ngu11,ngu13,san13}. The broad Gaussian components are commonly attributed to the high-velocity shocks produced by protostellar outflows, while the narrow ones are linked to the low-velocity shocks due to either cloud-cloud collisions or less powerful outflows from low-mass protostars.

The SiO survey toward massive clumps by \citet{cse16} includes the observations of starless clumps selected from the APEX Telescope Large Area Survey of the Galaxy (ATLASGAL), and a high detection rate of $\sim~61\%$ for SiO (2-1) toward starless clumps is resulted. \citet{li19} also observed the SiO (5-4) line toward 201 massive clumps, and find a $57\%$ detection rate for the massive clumps in IRDCs. The high detection rates in \citet{cse16} and \citet{li19} seem to suggest that shocks are ubiquitous in the earliest stage of star formation. Moreover, In these samples, more than $25\%$ of starless clumps identified by infrared colour criteria are found associated with broad SiO line profiles due to outflows of deeply embedded high-mass protostars, and other starless clumps with narrow SiO components could also contain protostellar objects. This suggests that a large fraction of these starless clump candidates are not actually starless, and the SiO lines are very useful to distinguish protostar-embedded clumps and starless clumps. Recently, \citet{svo19} observed 12 of the most massive SCCs in CO 2-1 and SiO 5-4 lines. Most of these SCCs show bipolar molecular outflows indicating star formation. This also suggests that previous infrared surveys are incomplete to detect embedded protostars with low luminosities. More works are still needed to identify reliable samples of starless sources, and to investigate the ubiquity and characteristics of shocks in the previously identified starless clumps.

Besides the information of shocks, the properties of the molecular gas in starless massive clumps can also allow us to understand the early environment of star formation. Formaldehyde, H$_2$CO, is one of the molecules detected in massive star-forming regions and is widely used to estimate the gas temperature and density \citep{tak00,tan17}. It is also used as a diagnostic tool to study the star formation in nearby galaxies \citep{nis19}. The abundance of H$_2$CO is about $10^{-9}$ to $10^{-10}$ in the massive star-forming regions \citep{vic16}.

In the current work, we report surveys of the SiO 1-0, 2-1, 3-2 lines and the H$_2$CO $2_{12}-1_{11}$ line toward 100 SCCs with Korean VLBI Network (KVN) 21m telescopes. By using these molecular lines, the ubiquity of shocks and the levels of star formation activities in these SCCs are investigated. The presence of high-velocity and/or low-velocity shocks is also revealed, and then the targets with fast shocks due to outflows from protostars can be distinguished from the real starless sources. The organization of the present paper is as follows. In Section 2 we describe the observations and the selected sample. The detailed results of the observations are presented in Section 3. The discussions and the analysis about these results are showed in Section 4. In Section 5, the summary is presented.

\section{Observations} \label{sec:method}

\subsection{Sample}

\begin{figure}
  \centering
  \includegraphics[scale=0.5]{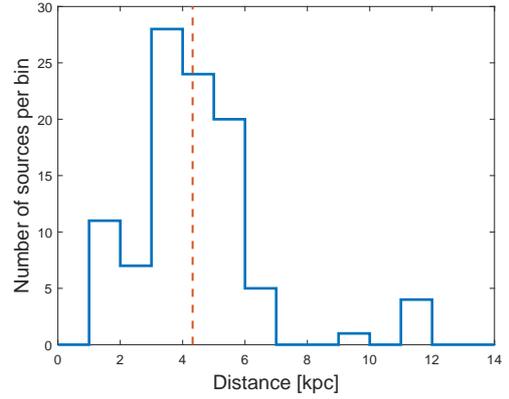}
  \caption{ The distance distribution of the sample. The vertical dashed line represents the average distance.}\label{fig:distance}
\end{figure}

The sample in this study consists of 100 SCCs, which have been observed by \citet{cal18} to search for infall signatures. These targets were bindly selected from the SCC catalog of \citet{svo16} with NH$_3$(1,1) detections. The distances of these sources range from 1.18 to 11.8 kpc, and the average and median distances are 4.32 kpc and 4.19 kpc, respectively \citep{ell15}. The distance distribution of the sample is presented in Figure \ref{fig:distance}. About 70$\%$ of the sources have distances between 3 and 6 kpc. According to the comparison of the mass and mass surface density between the selected sample and the complete sample of SCCs described in \citet{cal18}, the selected sample is representative of the mass range of the complete sample. But the mass surface densities in the selected sample are on average higher by a factor of $\sim$ 2. The $v_{LSR}$ measured from NH$_3$ observations provided in \citet{svo16} is regarded as the systemic velocity of each source, and used for the current observations.

\subsection{Observations and Data reduction}

We conducted single-point molecular line observations using all three (Yonsei, Ulsan and Tamma) KVN 21-m telescopes from March to June 2019 \citep{kim11}. All of the 100 sources were observed in the SiO 1-0, 2-1 and H$_2$CO $2_{12}-1_{11}$ transitions at 43.42376 GHz, 86.84696 GHz and 140.83950 GHz, respectively. And among the sample, 41 targets were also observed in the SiO 3-2 transition at 130.26861 GHz. The backend was a digital spectrometer that provided 64 MHz bandwidth with 4096 channels and 15.63 kHz frequency channel spacing. The position switching observational mode was used with the reference position of 0.2 degree offset from the source in azimuth. The total (ON+OFF) integration times are typically about 1 hour for the SiO 1-0 and 2-1 lines. However, we increased the integration time for some marginally detected source to confirm the detection. For the SiO 3-2 line and the H$_2$CO line, the observation times were commonly about 30 minutes. The system temperatures range from 80 to 400 K at different frequencies. The FWHM beam sizes of the telescopes are about 63$''$, 34$''$, 23$''$ and 21$''$ for SiO 1-0, 2-1, 3-2 lines and H$_2$CO $2_{12}-1_{11}$ line, respectively. The main beam sizes at the average distance of 4.32 kpc for the four lines are 1.33 pc, 0.72 pc, 0.48 pc and 0.44 pc, and the main beam efficiencies are about $47\%$, $43\%$, $35\%$ and $30\%$, respectively.

Table \ref{table_rms1} presents the details of the observations toward individual sources including the rms noise levels and used telescopes. Each source was observed individually with only one telescope except for BGPS 2762, 2931 and 4402. They were observed simultaneously with two telescopes. According to the approximately consistent instrumental parameters of the three KVN 21-m telescopes, the spectra obtained from them are comparable. This was also confirmed by our observations toward W51 E1E2 in the SiO and H$_2$CO lines. Since the properties of the shocked gas in the SCCs are our main research goals, the SiO lines are more important in the current work. In addition, the SiO 2-1 line should provide more representative information than the other SiO lines because it has been observed in more sources than the SiO 3-2 line and its filling factor may be higher than that of the SiO 1-0 line. 

The observations were reduced with CLASS reduction package \citep{gui88}. A linear baseline was removed from each spectrum. The typical rms noise levels of the main-beam brightness temperature scale are about 20 mK for the SiO lines and 40 mK for the H$_2$CO line at the velocity resolution of about 1 km s$^{-1}$. However, the levels are significantly lower or higher than the typical ones for some sources because of different integration times or weather conditions. The effect of the variable observing conditions on the detection rates of the SiO lines is discussed in Appendix A. The histograms of the noise levels and the SiO spectra for individual sources are presented in Figure \ref{fig:noise} and \ref{fig:siospectrum}, respectively. By comparing the noise levels between the sources with and without the SiO detections, we find that the SiO detection rates are meaningful.

\section{results} \label{sec:result}

The properties of global dense gas and shocked gas in the SCCs are investigated from the observations. The H$_2$CO line is used to determine the properties of the dense molecular gas in the SCCs, and the SiO lines can help us to study the physical conditions of the shocked gas.

\subsection{Detection rates}


The H$_2$CO $2_{12}-1_{11}$ line was detected in 93 targets ($93\%$) with a velocity-integrated intensity ($\int T_{\textrm{mb}}dv$) higher than 3 $\sigma_{\textrm{area}}$, where $\sigma_{\textrm{area}}=\sqrt{N}\Delta v\sigma$ represents the rms noise level of the velocity-integrated intensity. $N$ is the number of channels in the velocity range of the line, $\Delta v$ means velocity resolution, and $\sigma$ is rms noise level of a channel. The derived properties of the H$_2$CO line and the J2000 equatorial of the SCCs are listed in Table \ref{table_h2co}. In most targets, the line profiles are Gaussian. The line profiles of 10 sources show two velocity components. Four of them have two emission components probably due to the line-of-sight effect since the two emission components are resolved. The other 6 sources have an absorption component as well as an emission component. These spectra should be due to H$_2$CO emission at the reference positions. Since these spectra contain false absorption components, they are not included in the calculations of the average spectra presented in Section \ref{sec:width}. In addition, only the emission components are used to estimate the luminosities, the column densities and the abundances.

The SiO line profiles are mainly non-Gaussian so that the full width at zero power (FWZP) is used. The FWZP is visually determined as the velocity range where the emission is higher than rms noise level 1 $\sigma$. The numbers of the detections in the SiO lines are shown in Table \ref{table_detection}. The SiO 1-0, 2-1 and 3-2 lines were clearly detected in 31, 31 and 8 sources, respectively. In addition, the numbers of the overlapping detections between the SiO lines are also written in Table \ref{table_detection}. There were 28 sources in which both the SiO 1-0 and 2-1 lines were detected. And in 8 of them, all the three SiO lines were found. BGPS 2945, 3656 and 4230 were detected only in SiO 1-0 emission. This could be caused by different beam sizes for the SiO 1-0 and 2-1 lines. The region where the SiO lines are emitted may be not very close to the center of beam and covered only by the observation at 43.4 GHz. On the other hand, BGPS 3247, 4297 and 4375 were detected only in SiO 2-1 emission. This could be because the SiO emission region is compact. For compact emission, the different beam sizes for the two SiO lines would favour the detection of the SiO 2-1 line because the greater beam dilution effect at 43.4 GHz would potentially reduce the observed SiO 1-0 emission below the instrumental sensitivity limit. The detection rates of the SiO 1-0 and 2-1 lines are both $31\%$. This indicates that shocks are common in SCCs. The star formation activities seem to be abundant even at the very early evolutionary age if the shocks are attributed to outflows from protostellar structures. The large distance range from 1.18-11.8 kpc may lead to a bias in the detection rates. This was checked by using Spearman test relation between detection rates of the SiO 1-0 and 2-1 lines and the distances of the sources. The sources are divided into several groups according to their distances, and the correlations between detection rate and distance are tested. But the p-values for the SiO 1-0 and 2-1 lines are $38\%$ and $73\%$, respectively. This result suggests that the correlations are not significant.


\begin{table*}\tiny 
\centering
\caption{The properties of the H$_2$CO $2_{12}-1_{11}$ line. V$_{\textrm{lsr}}$(NH$_3$) and V$_{\textrm{lsr}}$(H$_2$CO) are line central velocities. The  $\int T_{\textrm{mb}}dv$ is velocity-integrated intensity. $T_{\textrm{mb}}$ is the line peak at main beam temperature. FWHM is full width at half maximum. L$_{\textrm{H}_2\textrm{CO}}$ is luminosity. N(H$_2$CO) and X(H$_2$CO) are column density and abundance, respectively.}\label{table_h2co}
\begin{tabular}{|c|ccccccccc|}
\hline
\multirow{2}*{Name} & \multirow{2}*{RA} & \multirow{2}*{Dec} & V$_{\textrm{lsr}}$(NH$_3$)/V$_{\textrm{lsr}}$(H$_2$CO) & $\int T_{\textrm{mb}}dv$ & $T_{\textrm{mb}}$ & FWHM & L$_{\textrm{H}_2\textrm{CO}}$ & N(H$_2$CO) & X(H$_2$CO) \\
 & & & [km s$^{-1}$] & [K km s$^{-1}$] & [K] & [km s$^{-1}$] & [$10^{-7}$ L$_\odot$] & [cm$^{-2}$] &  \\
\hline
BGPS 2427 & 18:09:33.88 & -20:47:00.76 & $30.66\pm0.65$/... & ... & ... & ... & ... & ... & ... \\
BGPS 2430 & 18:08:49.41 & -20:40:23.82 & $21.27\pm1.19$/... & ... & ... & ... & ... & ... & ... \\
BGPS 2432 & 18:09:44.59 & -20:47:10.21 & $31.08\pm0.71$/... & ... & ... & ... & ... & ... & ... \\
BGPS 2437 & 18:10:19.41 & -20:50:27.45 & $-1.87\pm0.56$/$27.61\pm0.46$ & $1.066\pm0.195$ & $0.23$ & $3.47\pm0.82$ & 145.48 & $5.01\times10^{12}$ & $8.76\times10^{-10}$ \\
BGPS 2533 & 18:10:30.29 & -20:14:44.20 & $31.84\pm0.41$/... & ... & ... & ... & ... & ... & ... \\
BGPS 2564 & 18:10:06.08 & -18:46:05.64 & $29.57\pm0.27$/$29.21\pm0.05$ & $0.480\pm0.045$ & $0.35$ & $0.98\pm0.12$ & 30.23 & $1.94\times10^{12}$ & $6.67\times10^{-10}$ \\
BGPS 2693 & 18:11:13.56 & -17:44:54.85 & $19.13\pm0.21$/$19.08\pm0.06$ & $1.379\pm0.133$ & $0.86$ & $1.17\pm0.15$ & 42.40 & $5.65\times10^{12}$ & $1.45\times10^{-9}$ \\
\multirow{2}*{BGPS 2710} & \multirow{2}*{18:13:49.04} & \multirow{2}*{-17:59:33.25} & \multirow{2}*{$34.59\pm0.42$/}$34.40\pm0.11$ & $3.219\pm0.608$ & $1.13$ & $2.74\pm0.39$ & \multirow{2}*{32.10} & \multirow{2}*{$1.38\times10^{13}$} & \multirow{2}*{$1.74\times10^{-9}$} \\
          &             &              & $~~~~~~~~~~~~~~~~~$$35.21\pm0.36$ & $-3.853\pm0.240$ & $-0.80$ & $5.25\pm0.60$ &  &  &  \\
BGPS 2724 & 18:14:13.61 & -17:59:52.02 & $36.25\pm1.43$/$35.68\pm0.14$ & $4.404\pm0.303$ & $0.80$ & $4.03\pm0.31$ & 42.83 & $2.24\times10^{13}$ & $8.09\times10^{-9}$ \\
BGPS 2732 & 18:14:26.85 & -17:58:50.93 & $37.47\pm0.85$/$36.23\pm0.14$ & $1.560\pm0.100$ & $0.29$ & $4.14\pm0.39$ & 15.33 & $9.70\times10^{12}$ & $8.51\times10^{-9}$\\
BGPS 2742 & 18:14:29.10 & -17:57:21.83 & $36.01\pm1.16$/$40.08\pm1.01$ & $3.833\pm0.543$ & $0.20$ & $13.92\pm2.00$ & 37.16 & $2.08\times10^{13}$ & $1.27\times10^{-8}$ \\
BGPS 2762 & 18:11:39.52 & -17:32:09.40 & $17.85\pm0.69$/$18.37\pm0.05$ & $3.494\pm0.188$ & $1.31$ & $2.10\pm0.15$ & 264.11 & $1.86\times10^{13}$ & $3.79\times10^{-9}$ \\
BGPS 2931 & 18:17:27.51 & -17:06:08.42 & $22.77\pm0.26$/$22.97\pm0.08$ & $3.005\pm0.231$ & $0.86$ & $2.44\pm0.22$ & 224.55 & $1.21\times10^{13}$ & $3.33\times10^{-9}$ \\
BGPS 2940 & 18:17:17.15 & -17:01:07.47 & $20.04\pm0.92$/$19.85\pm0.02$ & $10.088\pm0.100$ & $2.21$ & $3.71\pm0.04$ & 791.58 & $4.81\times10^{13}$ & $5.83\times10^{-9}$ \\
BGPS 2945 & 18:17:27.35 & -17:00:23.66 & $22.68\pm0.25$/$22.68\pm0.04$ & $1.939\pm0.096$ & $0.90$ & $1.68\pm0.13$ & 18.64 & $7.81\times10^{12}$ & $1.30\times10^{-9}$ \\
BGPS 2949 & 18:17:33.74 & -16:59:34.94 & $22.52\pm0.33$/$21.64\pm0.06$ & $2.504\pm0.108$ & $0.61$ & $3.21\pm0.12$ & 28.36 & $1.01\times10^{13}$ & $2.21\times10^{-9}$ \\
BGPS 2970 & 18:17:05.08 & -16:43:28.66 & $40.01\pm0.62$/$40.23\pm0.07$ & $1.820\pm0.115$ & $0.58$ & $2.43\pm0.20$ & 160.41 & $7.69\times10^{12}$ & $7.69\times10^{-10}$ \\
BGPS 2971 & 18:16:48.12 & -16:41:08.91 & $36.55\pm0.47$/$39.22\pm0.31$ & $3.333\pm0.294$ & $0.39$ & $6.05\pm0.65$ & 76.05 & $1.49\times10^{13}$ & $5.48\times10^{-9}$ \\
BGPS 2976 & 18:17:07.84 & -16:41:14.59 & $39.63\pm0.37$/$39.63\pm0.07$ & $1.623\pm0.091$ & $0.45$ & $2.45\pm0.17$ & 37.64 & $6.89\times10^{12}$ & $1.59\times10^{-9}$\\
BGPS 2984 & 18:18:18.23 & -16:44:52.26 & $18.42\pm0.22$/$19.28\pm0.13$ & $0.256\pm0.050$ & $0.19$ & $0.99\pm0.26$ & 6.11 & $1.03\times10^{12}$ & $2.10\times10^{-10}$ \\
BGPS 2986 & 18:18:29.68 & -16:44:50.69 & $19.97\pm0.31$/$20.28\pm0.05$ & $3.386\pm0.170$ & $1.38$ & $1.75\pm0.15$ & 95.16 & $1.36\times10^{13}$ & $9.38\times10^{-10}$ \\
BGPS 3018 & 18:19:13.88 & -16:35:16.47 & $18.96\pm0.47$/$18.93\pm0.04$ & $1.904\pm0.075$ & $0.78$ & $1.80\pm0.10$ & 560.64 & $8.01\times10^{12}$ & $1.21\times10^{-9}$ \\
BGPS 3030 & 18:19:19.68 & -16:31:39.82 & $19.01\pm0.28$/$19.10\pm0.02$ & $2.330\pm0.079$ & $0.96$ & $1.73\pm0.07$ & 51.35 & $1.09\times10^{13}$ & $1.62\times10^{-9}$ \\
BGPS 3110 & 18:20:16.27 & -16:08:51.13 & $17.69\pm0.91$/$17.32\pm0.03$ & $19.574\pm0.331$ & $3.85$ & $3.70\pm0.07$ & 545.00 & $1.16\times10^{14}$ & $6.74\times10^{-9}$ \\
BGPS 3114 & 18:20:31.50 & -16:08:37.80 & .../$22.87\pm0.02$ & $7.185\pm0.079$ & $1.58$ & $3.36\pm0.04$ & 169.96 & $2.99\times10^{13}$ & $5.08\times10^{-10}$ \\
BGPS 3117 & 18:20:06.68 & -16:04:45.75 & $18.57\pm0.74$/$18.67\pm0.03$ & $3.928\pm0.108$ & $1.25$ & $2.29\pm0.07$ & 108.93 & $2.91\times10^{13}$ & $5.18\times10^{-9}$ \\
BGPS 3118 & 18:20:16.17 & -16:05:50.72 & $17.21\pm0.76$/$16.88\pm0.02$ & $11.025\pm0.160$ & $3.20$ & $2.71\pm0.05$ & 305.75 & $6.26\times10^{13}$ & $6.69\times10^{-9}$ \\
BGPS 3125 & 18:20:06.11 & -16:01:58.02 & $21.53\pm0.34$/$21.39\pm0.03$ & $4.508\pm0.123$ & $1.51$ & $2.36\pm0.10$ & 375.06 & $2.39\times10^{13}$ & $4.59\times10^{-9}$ \\
BGPS 3128 & 18:20:35.27 & -16:04:53.81 & $19.75\pm0.76$/$18.36\pm0.06$ & $9.840\pm0.429$ & $2.54$ & $3.08\pm0.15$ & 1185.1 & $5.20\times10^{13}$ & $5.92\times10^{-9}$ \\
BGPS 3129 & 18:20:12.99 & -16:00:24.13 & $19.72\pm0.49$/$19.41\pm0.04$ & $3.084\pm0.149$ & $1.13$ & $1.99\pm0.13$ & 392.96 & $1.48\times10^{13}$ & $4.81\times10^{-9}$ \\
BGPS 3134 & 18:19:52.72 & -15:56:01.56 & $20.47\pm0.59$/$20.95\pm0.01$ & $3.970\pm0.050$ & $1.70$ & $1.70\pm0.02$ & 458.40 & $1.79\times10^{13}$ & $1.11\times10^{-9}$ \\
BGPS 3139 & 18:20.34.24 & -15:58.14.00 & $21.80\pm0.37$/$20.12\pm0.01$ & $6.884\pm0.041$ & $2.31$ & $2.17\pm0.02$ & 1394.5 & $3.23\times10^{13}$ & $4.17\times10^{-9}$ \\
\multirow{2}*{BGPS 3151} & \multirow{2}*{18:20:23.19} & \multirow{2}*{-15:39:31.96} & \multirow{2}*{$39.40\pm0.37$/}$15.50\pm0.04$ & $0.646\pm0.038$ & $0.35$ & $1.38\pm0.09$ & \multirow{2}*{102.43} & \multirow{2}*{$7.94\times10^{12}$} & \multirow{2}*{$7.63\times10^{-10}$} \\
          &             &              & $~~~~~~~~~~~~~~~~~$$39.82\pm0.05$ & $1.295\pm0.050$ & $0.39$ & $2.49\pm0.12$ & & &  \\
BGPS 3220 & 18:24:57.03 & -13:20:32.39 & $46.13\pm1.76$/$46.79\pm0.05$ & $6.001\pm0.184$ & $1.09$ & $4.33\pm0.14$ & 623.76 & $2.93\times10^{13}$ & $2.99\times10^{-9}$ \\
\multirow{2}*{BGPS 3243} & \multirow{2}*{18:25:32.74} & \multirow{2}*{-13:01:31.05} & \multirow{2}*{$68.47\pm0.31$/}$50.61\pm0.13$ & $-1.196\pm0.124$ & $-0.35$ & $2.52\pm0.29$ & \multirow{2}*{194.50} & \multirow{2}*{$5.36\times10^{12}$} & \multirow{2}*{$5.84\times10^{-10}$} \\
          &             &              & $~~~~~~~~~~~~~~~~~$$68.23\pm0.20$ & $1.329\pm0.199$ & $0.33$ & $3.09\pm0.77$ &  & & \\
BGPS 3247 & 18:25:14.45 & -12:54:16.74 & $45.17\pm0.35$/$45.00\pm0.06$ & $1.759\pm0.120$ & $0.68$ & $1.88\pm0.16$ & 240.85 & $7.31\times10^{12}$ & $1.75\times10^{-9}$ \\
BGPS 3276 & 18:26:24.92 & -12:49:30.07 & $67.38\pm0.71$/$67.00\pm0.15$ & $1.134\pm0.124$ & $0.26$ & $2.77\pm0.32$ & 89.60 & $5.31\times10^{12}$ & $1.57\times10^{-9}$ \\
BGPS 3300 & 18:26:28.42 & -12:37:03.98 & $64.15\pm1.04$/$64.17\pm0.19$ & $1.701\pm0.170$ & $0.33$ & $3.71\pm0.38$ & 1604.0 & $7.56\times10^{12}$ & $3.44\times10^{-9}$ \\
BGPS 3302 & 18:27:15.23 & -12:42:56.45 & $66.37\pm0.75$/$66.57\pm0.09$ & $4.470\pm0.158$ & $0.64$ & $5.10\pm0.20$ & 4300.4 & $2.19\times10^{13}$ & $2.73\times10^{-9}$ \\
BGPS 3306 & 18:23:34.02 & -12:13:52.79 & $57.11\pm0.36$/$57.16\pm0.07$ & $0.965\pm0.091$ & $0.48$ & $1.46\pm0.18$ & 152.64 & $3.93\times10^{12}$ & $1.13\times10^{-9}$ \\
BGPS 3312 & 18:25:44.52 & -12:28:34.11 & $47.30\pm0.30$/$47.10\pm0.08$ & $0.883\pm0.084$ & $0.39$ & $1.90\pm0.22$ & 169.91 & $4.23\times10^{12}$ & $2.15\times10^{-9}$ \\
BGPS 3315 & 18:25:33.24 & -12:26:50.63 & $44.31\pm0.68$/$47.29\pm0.10$ & $2.761\pm0.178$ & $0.64$ & $3.16\pm0.25$ & 439.44 & $1.38\times10^{13}$ & $1.70\times10^{-9}$ \\
\multirow{2}*{BGPS 3344} & \multirow{2}*{18:26:40.00} & \multirow{2}*{-12:25:15.81} & \multirow{2}*{$65.69\pm0.36$/}$47.09\pm0.17$ & $-1.243\pm0.211$ & $-0.39$ & $2.35\pm0.53$ & \multirow{2}*{180.24} & \multirow{2}*{$6.35\times10^{12}$} & \multirow{2}*{$1.28\times10^{-9}$} \\
          &             &              & $~~~~~~~~~~~~~~~~~$$65.53\pm0.34$ & $1.399\pm0.249$ & $0.26$ & $3.86\pm0.89$ &  & & \\
BGPS 3442 & 18:28:13.51 & -11:40:44.94 & $65.75\pm0.46$/$66.88\pm0.32$ & $1.319\pm0.153$ & $0.19$ & $5.06\pm0.77$ & 108.16 & $5.43\times10^{12}$ & $1.05\times10^{-9}$ \\
BGPS 3444 & 18:28:27.26 & -11:41:33.99 & $69.69\pm0.44$/... & ... & ... & ... & ... & ... & ... \\
BGPS 3475 & 18:28:28.28 & -11:06:44.16 & $75.86\pm1.18$/$77.19\pm0.30$ & $0.828\pm0.145$ & $0.18$ & $3.41\pm0.63$ & 67.25 & $3.93\times10^{12}$ & $1.37\times10^{-10}$ \\
BGPS 3484 & 18:29:15.74 & -10:58:28.73 & $56.35\pm0.89$/$56.03\pm0.33$ & $1.408\pm0.236$ & $0.23$ & $4.44\pm0.86$ & 118.30 & $6.85\times10^{12}$ & $3.89\times10^{-9}$ \\
BGPS 3487 & 18:29:22.77 & -10:58:01.69 & $54.60\pm1.22$/$55.70\pm0.13$ & $2.214\pm0.141$ & $0.39$ & $4.24\pm0.30$ & 186.71 & $1.48\times10^{13}$ & $7.67\times10^{-9}$ \\
BGPS 3534 & 18:30:33.45 & -10:24:19.00 & $65.06\pm0.99$/... & ... & ... & ... & ... & ... & ... \\
BGPS 3604 & 18:30:43.92 & -9:34:42.15 & $51.52\pm0.44$/$51.32\pm0.06$ & $2.396\pm0.134$ & $0.78$ & $2.45\pm0.17$ & 2011.5 & $9.69\times10^{12}$ & $1.73\times10^{-9}$ \\
BGPS 3606 & 18:29:41.95 & -9:24:49.10 & $49.56\pm0.42$/$49.49\pm0.16$ & $0.443\pm0.083$ & $0.20$ & $1.60\pm0.30$ & 54.50 & $1.74\times10^{12}$ & $5.16\times10^{-10}$ \\
BGPS 3608 & 18:31:54.82 & -9:39:05.03 & $63.77\pm1.36$/$63.96\pm0.27$ & $1.995\pm0.220$ & $0.26$ & $5.45\pm0.62$ & 230.13 & $8.45\times10^{12}$ & $1.85\times10^{-9}$ \\
BGPS 3627 & 18:31:42.32 & -9:24:29.17 & $81.30\pm0.51$/$82.25\pm0.54$ & $0.596\pm0.149$ & $0.11$ & $3.99\pm1.05$ & 71.83 & $2.45\times10^{12}$ & $7.18\times10^{-10}$ \\
BGPS 3656 & 18:32:49.54 & -9:21:29.26 & $77.25\pm0.79$/$27.91\pm0.29$ & $0.749\pm0.141$ & $0.19$ & $2.89\pm0.74$ & 79.14 & $3.04\times10^{12}$ & $8.49\times10^{-10}$ \\
BGPS 3686 & 18:34:14.58 & -9:18:35.84 & $77.28\pm0.50$/$78.14\pm0.29$ & $1.979\pm0.199$ & $0.24$ & $5.92\pm0.69$ & 118.35 & $8.61\times10^{12}$ & $1.43\times10^{-9}$ \\
BGPS 3705 & 18:34:32.69 & -9:14:09.40 & $61.58\pm0.90$/$74.42\pm0.57$ & $5.559\pm0.303$ & $0.19$ & $21.00\pm1.23$ & 373.78 & $2.31\times10^{13}$ & $4.90\times10^{-9}$ \\
BGPS 3710 & 18:34:20.55 & -9:10:01.94 & $74.68\pm1.04$/$75.11\pm0.13$ & $3.014\pm0.195$ & $0.45$ & $4.85\pm0.47$ & 130.96 & $1.30\times10^{13}$ & $2.09\times10^{-9}$ \\
\hline
\end{tabular}
\end{table*}

\begin{table*}\tiny 
\centering
\caption{continued}\label{table_h2co2}
\begin{tabular}{|c|ccccccccc|}
\hline
\multirow{2}*{Name} & \multirow{2}*{RA} & \multirow{2}*{Dec} & V$_{\textrm{lsr}}$(NH$_3$)/V$_{\textrm{lsr}}$(H$_2$CO) & $\int T_{\textrm{mb}}dv$ & $T_{\textrm{mb}}$ & FWHM & L$_{\textrm{H}_2\textrm{CO}}$ & N(H$_2$CO) & X(H$_2$CO) \\
 & & & [km s$^{-1}$] & [K km s$^{-1}$] & [K] & [km s$^{-1}$] & [$10^{-7}$ L$_\odot$] & [cm$^{-2}$] &  \\
\hline
BGPS 3716 & 18:34:24.15 & -9:08:03.60 & $75.87\pm1.57$/$76.81\pm0.12$ & $3.738\pm0.149$ & $0.48$ & $5.83\pm0.26$ & 256.25 & $1.73\times10^{13}$ & $5.67\times10^{-9}$ \\
BGPS 3736 & 18:33:28.22 & -8:55:04.36 & $65.39\pm0.71$/$65.74\pm0.06$ & $1.544\pm0.095$ & $6.44$ & $1.70\pm0.10$ & 270.75 & $6.43\times10^{12}$ & $1.44\times10^{-9}$ \\
BGPS 3822 & 18:33:32.06 & -8:32:26.27 & $54.55\pm0.64$/$54.86\pm0.10$ & $1.909\pm0.158$ & $0.51$ & $2.71\pm0.26$ & 150.15 & $7.70\times10^{12}$ & $8.56\times10^{-10}$ \\
\multirow{2}*{BGPS 3833} & \multirow{2}*{18:33:36.50} & \multirow{2}*{-8:30:50.70} & \multirow{2}*{$55.59\pm0.55$/}$55.23\pm0.30$ & $0.903\pm0.156$ & $0.19$ & $3.59\pm0.85$ & \multirow{2}*{126.16} & \multirow{2}*{$7.53\times10^{12}$} & \multirow{2}*{$1.29\times10^{-9}$} \\
          &             &             & $~~~~~~~~~~~~~~~~~$$81.71\pm0.33$ & $0.953\pm0.156$ & $0.19$ & $4.17\pm0.84$ & & & \\
BGPS 3892 & 18:35:59.74 & -8:38:56.48 & $64.46\pm0.95$/$63.33\pm0.13$ & $0.531\pm0.080$ & $0.23$ & $1.63\pm0.28$ & 103.48 & $2.13\times10^{12}$ & $1.76\times10^{-10}$ \\
BGPS 3922 & 18:33:40.98 & -8:14:55.30 & $89.22\pm0.48$/$89.28\pm0.12$ & $0.578\pm0.080$ & $0.26$ & $1.75\pm0.29$ & 3897.1 & $2.36\times10^{12}$ & $6.15\times10^{-10}$ \\
BGPS 3924 & 18:34:51.17 & -8:23:40.02 & $81.29\pm0.87$/$81.35\pm0.14$ & $0.440\pm0.073$ & $0.23$ & $1.59\pm0.29$ & 101.84 & $1.78\times10^{12}$ & $5.55\times10^{-10}$ \\
BGPS 3982 & 18:34:30.79 & -8:02:07.36 & $53.94\pm0.32$/$54.28\pm0.15$ & $1.640\pm0.145$ & $0.39$ & $3.47\pm0.35$ & 1523.8 & $6.81\times10^{12}$ & $2.01\times10^{-9}$ \\
BGPS 4029 & 18:35:54.40 & -7:59:44.60 & $81.52\pm0.77$/$81.43\pm0.30$ & $2.179\pm0.409$ & $0.55$ & $3.05\pm0.53$ & 188.95 & $8.81\times10^{12}$ & $9.13\times10^{-10}$ \\
BGPS 4082 & 18:35:10.07 & -7:39:43.55 & $99.52\pm0.51$/$99.83\pm0.40$ & $4.683\pm0.631$ & $0.61$ & $6.04\pm1.09$ & 838.26 & $1.99\times10^{13}$ & $3.90\times10^{-9}$ \\
BGPS 4095 & 18:35:04.00 & -7:36:06.46 & $112.99\pm0.44$/$118.07\pm0.24$ & $1.120\pm0.344$ & $0.48$ & $1.84\pm0.94$ & 222.38 & $4.74\times10^{12}$ & $6.90\times10^{-10}$ \\
BGPS 4119 & 18:36:29.65 & -7:42:06.09 & $55.33\pm1.20$/$55.36\pm0.17$ & $2.223\pm0.136$ & $0.29$ & $5.42\pm0.40$ & 478.76 & $9.86\times10^{12}$ & $4.65\times10^{-9}$ \\
\multirow{2}*{BGPS 4135} & \multirow{2}*{18:37:44.06} & \multirow{2}*{-7:48:15.35} & \multirow{2}*{$59.63\pm1.71$/}$58.39\pm0.07$ & $0.708\pm0.066$ & $0.29$ & $1.70\pm0.20$ & \multirow{2}*{109.59} & \multirow{2}*{$5.39\times10^{12}$} & \multirow{2}*{$1.20\times10^{-9}$} \\
          &             &             & $~~~~~~~~~~~~~~~~~$$61.72\pm0.11$ & $0.534\pm0.070$ & $0.19$ & $1.93\pm0.35$ & & & \\
BGPS 4140 & 18:36:49.66 & -7:40:36.83 & $96.00\pm0.40$/$96.03\pm0.14$ & $0.969\pm0.178$ & $0.35$ & $1.73\pm0.43$ & 87.79 & $4.29\times10^{12}$ & $1.39\times10^{-9}$ \\
\multirow{2}*{BGPS 4145} & \multirow{2}*{18:36:52.95} & \multirow{2}*{-7:39:49.20} & \multirow{2}*{$96.53\pm0.52$/}$96.44\pm0.12$ & $0.514\pm0.111$ & $0.35$ & $1.02\pm0.25$ & \multirow{2}*{88.43} & \multirow{2}*{$2.21\times10^{12}$} & \multirow{2}*{$5.74\times10^{-10}$} \\
          &             &             & $~~~~~~~~~~~~~~~~~$$101.44\pm0.14$ & $-2.661\pm0.208$ & $-0.55$ & $3.56\pm0.33$ & & &  \\
BGPS 4191 & 18:37:04.58 & -7:33:12.26 & $97.51\pm0.56$/$97.14\pm0.14$ & $1.639\pm0.174$ & $0.45$ & $2.61\pm0.30$ & 298.01 & $6.85\times10^{12}$ & $1.17\times10^{-9}$ \\
BGPS 4230 & 18:35:50.85 & -7:12:23.58 & $107.44\pm0.50$/$107.42\pm0.08$ & $1.341\pm0.091$ & $0.45$ & $2.18\pm0.08$ & 235.36 & $6.18\times10^{12}$ & $8.06\times10^{-10}$ \\
BGPS 4294 & 18:38:51.58 & -6:55:36.52 & $56.17\pm0.56$/$56.39\pm0.11$ & $1.780\pm0.100$ & $0.33$ & $4.17\pm0.26$ & 398.71 & $7.88\times10^{12}$ & $1.79\times10^{-9}$ \\
\multirow{2}*{BGPS 4297} & \multirow{2}*{18:38:56.37} & \multirow{2}*{-6:55:08.44} & \multirow{2}*{$58.65\pm0.56$/}$58.60\pm0.21$ & $1.224\pm0.318$ & $0.35$ & $2.67\pm0.93$ & \multirow{2}*{209.50} & \multirow{2}*{$6.23\times10^{12}$} & \multirow{2}*{$2.10\times10^{-9}$} \\
          &             &             & $~~~~~~~~~~~~~~~~~$$63.72\pm0.60$ & $-1.178\pm0.275$ & $-0.19$ & $4.48\pm1.19$ & & & \\
BGPS 4346 & 18:38:49.58 & -6:31:27.06 & $92.59\pm1.71$/$93.35\pm0.09$ & $4.450\pm0.133$ & $0.51$ & $6.24\pm0.19$ & 1045.3 & $2.80\times10^{13}$ & $1.96\times10^{-8}$ \\
BGPS 4347 & 18:38:42.93 & -6:30:27.83 & $93.47\pm0.82$/$93.66\pm0.06$ & $3.523\pm0.100$ & $0.61$ & $4.13\pm0.14$ & 684.06 & $1.58\times10^{13}$ & $5.72\times10^{-9}$ \\
BGPS 4354 & 18:38:51.42 & -6:29:15.38 & $93.97\pm0.86$/$92.97\pm0.97$ & $2.795\pm0.608$ & $0.26$ & $9.19\pm2.46$ & 660.38 & $1.16\times10^{13}$ & $3.20\times10^{-9}$ \\
BGPS 4356 & 18:37:29.48 & -6:18:12.13 & $109.94\pm0.72$/$110.83\pm0.29$ & $2.669\pm0.371$ & $0.51$ & $4.20\pm0.64$ & 366.49 & $1.20\times10^{13}$ & $1.18\times10^{-9}$ \\
BGPS 4375 & 18:39:10.19 & -6:21:15.90 & $93.07\pm0.42$/$93.21\pm0.12$ & $0.965\pm0.199$ & $0.45$ & $1.56\pm0.54$ & 96.23 & $3.90\times10^{12}$ & $8.59\times10^{-10}$ \\
\multirow{2}*{BGPS 4396} & \multirow{2}*{18:38:34.74} & \multirow{2}*{-5:56:43.97} & \multirow{2}*{$112.73\pm0.47$/}$97.02\pm0.29$ & $0.894\pm0.166$ & $0.23$ & $2.88\pm0.59$ & \multirow{2}*{165.93} & \multirow{2}*{$9.04\times10^{12}$} & \multirow{2}*{$8.14\times10^{-10}$} \\
          &             &             & $~~~~~~~~~~~~~~~~~$$113.30\pm0.26$ & $1.316\pm0.240$ & $0.29$ & $3.50\pm1.02$ & & & \\
\multirow{2}*{BGPS 4402} & \multirow{2}*{18:39:28.64} & \multirow{2}*{-5:57:58.57} & \multirow{2}*{$99.21\pm0.59$/}$99.17\pm0.21$ & $1.391\pm0.145$ & $0.23$ & $4.46\pm0.63$ & \multirow{2}*{177.00} & \multirow{2}*{$5.80\times10^{12}$} & \multirow{2}*{$9.31\times10^{-10}$} \\
          &             &             & $~~~~~~~~~~~~~~~~~$$113.93\pm0.29$ & $-0.845\pm0.119$ & $-0.16$ & $4.65\pm0.64$ & & & \\
BGPS 4422 & 18:38:47.88 & -5:36:16.38 & $110.71\pm0.46$/$110.83\pm0.16$ & $0.835\pm0.100$ & $0.23$ & $3.05\pm0.49$ & 91.38 & $3.59\times10^{12}$ & $9.60\times10^{-10}$ \\
BGPS 4472 & 18:41:17.32 & -5:09:56.83 & $46.88\pm0.40$/$47.71\pm0.08$ & $3.419\pm0.100$ & $0.41$ & $6.10\pm0.23$ & 244.90 & $1.48\times10^{13}$ & $2.57\times10^{-9}$ \\
BGPS 4732 & 18:44:23.40 & -4:02:01.21 & $88.32\pm0.54$/$88.20\pm0.14$ & $1.458\pm0.133$ & $0.33$ & $3.17\pm0.33$ & 144.33 & $5.96\times10^{12}$ & $5.62\times10^{-10}$ \\
BGPS 4827 & 18:44:42.45 & -3:44:21.63 & $86.10\pm0.50$/$86.30\pm0.07$ & $2.198\pm0.124$ & $0.64$ & $2.48\pm0.18$ & 369.51 & $9.91\times10^{12}$ & $9.92\times10^{-10}$ \\
BGPS 4841 & 18:42:15.65 & -3:22:26.19 & $83.98\pm0.66$/$83.40\pm0.10$ & $1.188\pm0.104$ & $0.39$ & $2.29\pm0.23$ & 149.66 & $4.83\times10^{12}$ & $7.88\times10^{-10}$ \\
BGPS 4902 & 18:46:11.36 & -3:42:55.73 & $84.14\pm1.05$/$84.12\pm0.07$ & $4.078\pm0.145$ & $0.74$ & $3.98\pm0.17$ & 612.16 & $1.73\times10^{13}$ & $3.93\times10^{-9}$ \\
BGPS 4953 & 18:45:51.82 & -3:26:24.16 & $90.78\pm0.77$/$90.89\pm0.11$ & $1.759\pm0.116$ & $0.41$ & $3.14\pm0.26$ & 368.68 & $7.25\times10^{12}$ & $2.03\times10^{-9}$ \\
BGPS 4962 & 18:45:59.61 & -3:25:14.53 & $88.31\pm0.39$/$88.87\pm0.25$ & $1.354\pm0.215$ & $0.33$ & $3.10\pm0.56$ & 347.76 & $5.56\times10^{12}$ & $1.57\times10^{-9}$ \\
BGPS 4967 & 18:43:27.80 & -3:05:14.94 & $80.31\pm0.54$/$80.80\pm0.09$ & $0.891\pm0.165$ & $0.61$ & $1.06\pm5.33$ & 83.70 & $3.63\times10^{12}$ & $7.30\times10^{-10}$ \\
BGPS 5021 & 18:44:37.07 & -2:55:04.40 & $80.05\pm0.65$/$79.86\pm0.13$ & $2.653\pm0.169$ & $0.48$ & $4.29\pm0.33$ & 192.28 & $1.08\times10^{13}$ & $1.27\times10^{-9}$ \\
BGPS 5064 & 18:45:48.44 & -2:44:31.65 & $100.72\pm0.59$/$100.80\pm0.10$ & $2.873\pm0.199$ & $0.58$ & $3.64\pm0.37$ & 539.90 & $1.29\times10^{13}$ & $2.70\times10^{-9}$ \\
BGPS 5089 & 18:48:49.88 & -2:59:47.86 & $85.20\pm0.64$/$85.54\pm0.07$ & $1.724\pm0.100$ & $0.55$ & $2.53\pm0.18$ & 509.91 & $7.59\times10^{12}$ & $1.28\times10^{-9}$ \\
BGPS 5090 & 18:46:35.81 & -2:42:30.19 & $96.30\pm0.64$/$96.83\pm0.11$ & $0.803\pm0.111$ & $0.41$ & $1.54\pm0.25$ & 149.05 & $3.21\times10^{12}$ & $7.97\times10^{-10}$ \\
BGPS 5114 & 18:50:23.54 & -3:01:31.58 & $65.82\pm0.76$/$64.51\pm0.35$ & $2.275\pm0.283$ & $0.35$ & $5.33\pm0.69$ & 213.46 & $9.75\times10^{12}$ & $1.39\times10^{-9}$ \\
BGPS 5166 & 18:47:54.26 & -2:26:07.11 & $102.73\pm0.53$/$102.40\pm0.06$ & $2.175\pm0.145$ & $0.93$ & $1.86\pm0.14$ & 559.08 & $9.68\times10^{12}$ & $2.42\times10^{-9}$ \\
BGPS 5183 & 18:47:00.29 & -2:16:38.63 & $113.78\pm0.44$/... & ... & ... & ... & ... & ... & ... \\
BGPS 5243 & 18:47:54.70 & -2:11:10.72 & $95.89\pm0.44$/$96.06\pm0.13$ & $1.170\pm0.153$ & $0.45$ & $2.12\pm0.35$ & 219.85 & $4.83\times10^{12}$ & $1.02\times10^{-9}$ \\
\hline
\end{tabular}
\end{table*}

\begin{table}\footnotesize
\centering
\caption{The numbers of overlapping detections between the SiO and H$_2$CO lines. The numbers in the diagonal represent the total numbers of the detections in the corresponding lines.}\label{table_detection}
\begin{tabular}{|c||c|c|c|c|}
\hline
   & SiO 1-0 & SiO 2-1 & SiO 3-2 & H$_2$CO \\
\hline
SiO 1-0 & 31 & 28 & 8 & 31 \\
SiO 2-1 & 28 & 31 & 8 & 31 \\
SiO 3-2 & 8 & 8 & 8 & 8 \\
H$_2$CO & 31 & 31 & 8 & 93 \\
\hline
\end{tabular}
\end{table}

\subsection{Line Width}\label{sec:width}

\begin{figure}
  \centering
  \includegraphics[scale=0.5]{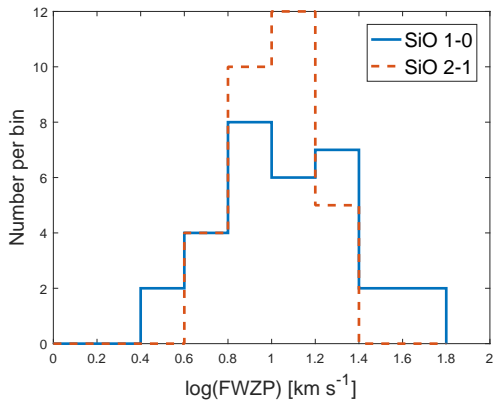}
  \caption{ The histogram of the line widths (FWZP) of the SiO 1-0 and 2-1 lines. }\label{fig:siofwzphist}
\end{figure}

\begin{figure}
  \centering
  \includegraphics[scale=0.5]{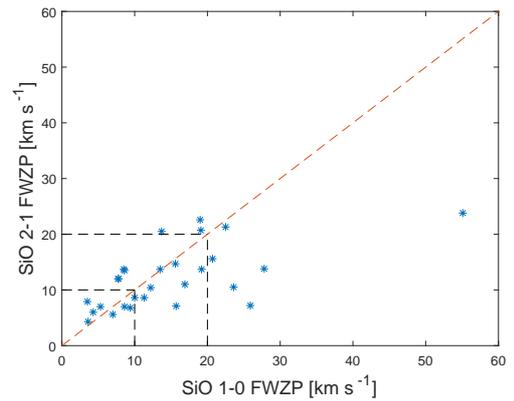}
  \caption{ The comparison of the FWZPs of the SiO 1-0 and 2-1 lines of corresponding sources. The dashed line indicates the position where the FWZPs of the SiO lines are equal. }\label{fig:fwzpcom}
\end{figure}

The FWZPs of the detected SiO 1-0 and 2-1 lines are listed in Table \ref{table_sio1}. Following \citet{beu07}, the SiO spectra are divided into 3 groups as low velocity (FWZP $<$10 km s$^{-1}$), intermediate velocity (10 km s$^{-1}\leq$ FWZP $<$20 km s$^{-1}$) and high velocity (FWZP $\geq$20 km s$^{-1}$). For the SiO 1-0 line, there are 13 sources in the low-velocity group, 11 sources in the intermediate-velocity group and 7 sources in the high-velocity group. For the SiO 2-1 line, the corresponding numbers are 14, 12 and 5, respectively. The mean FWZPs are 14.88 $km~s^{-1}$ and 11.59 $km~s^{-1}$ for the SiO 1-0 and 2-1 lines, and the median values are 11.75 $km~s^{-1}$ and 10.50 $km~s^{-1}$, respectively. The histograms of FWZPs are displayed in Figure \ref{fig:siofwzphist}. More than half of the sources have FWZPs $>10~km~s^{-1}$ probably attributed to outflows from undetected embedded protostars while the number of the sources associated with narrow velocity ranges (FWZP $<10~km~s^{-1}$) is still considerable. In the current work, the proportions of sources in the low-velocity group are $\sim40\%$ for both SiO lines. The comparison of the FWZP values for the SiO 1-0 and 2-1 lines is presented in Figure \ref{fig:fwzpcom}. The relation between the FWZPs of the two SiO lines is obvious.  The mean and median FWZPs of the two SiO lines are roughly equal although the covered regions in the observations of the SiO 1-0 and 2-1 lines are different due to the different beam sizes. Other detailed properties of the SiO lines are also written in Table \ref{table_sio1}.

\begin{figure}
  \centering
  \includegraphics[scale=0.5]{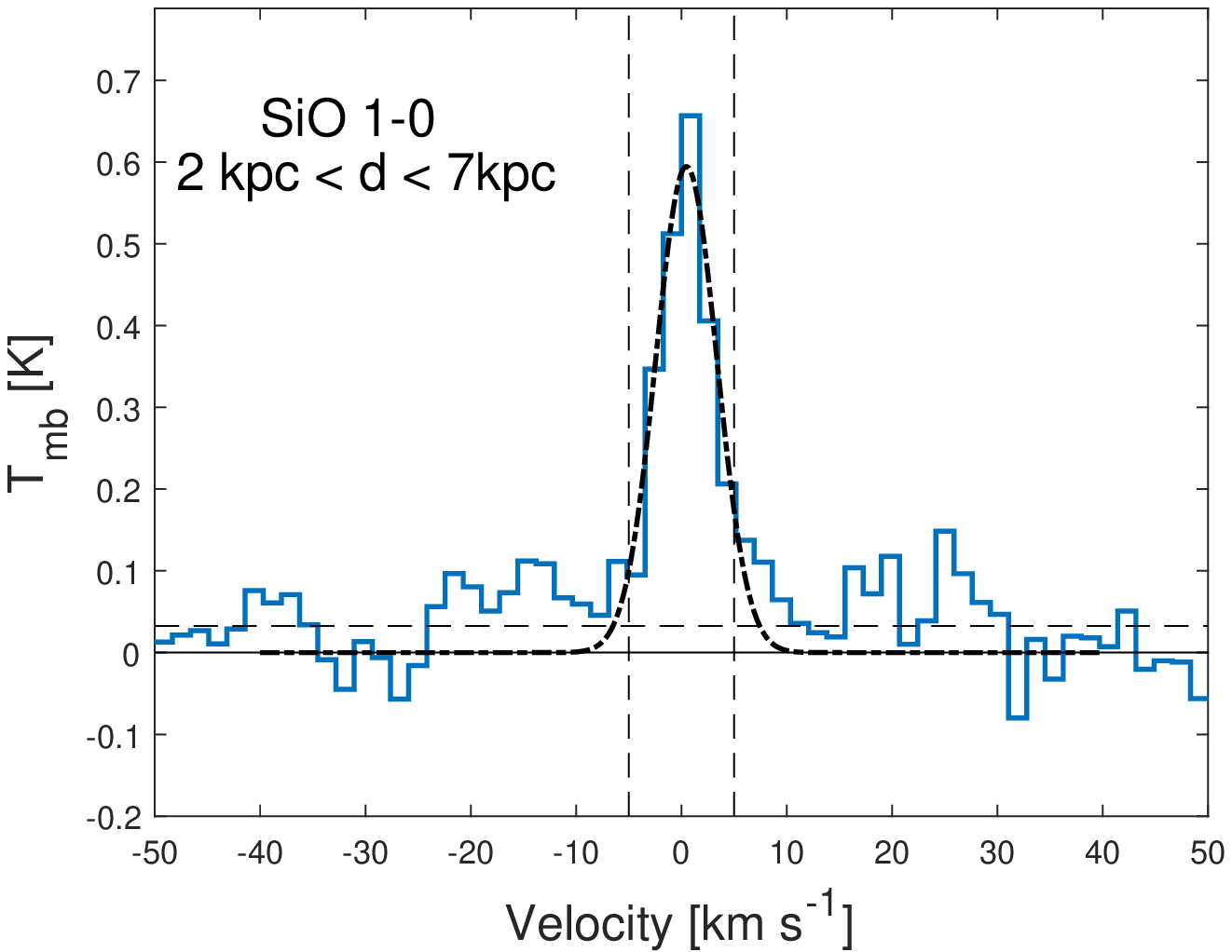}
  \includegraphics[scale=0.5]{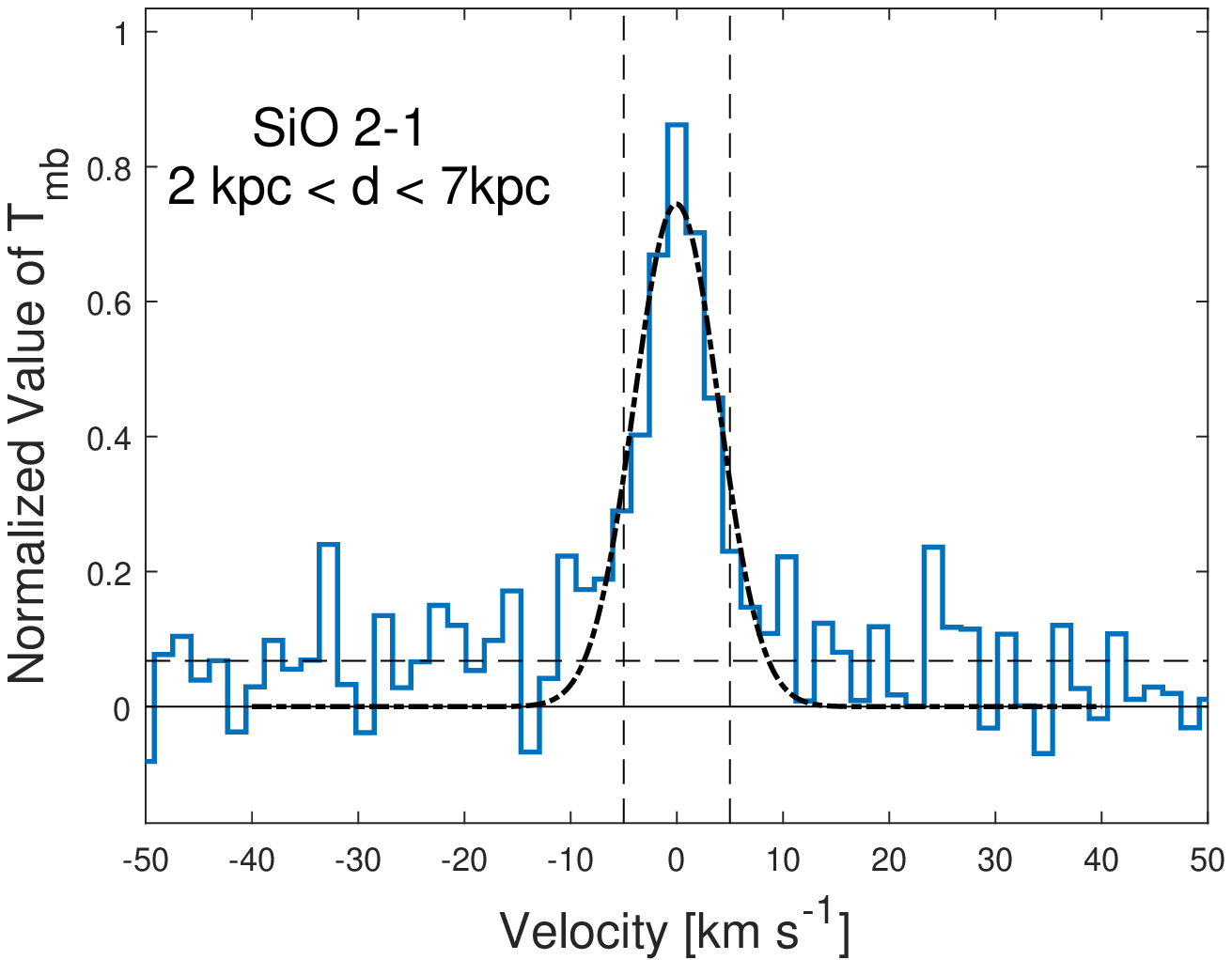}
  \caption{ The distance-weighted average SiO line profiles of the detected sources located between 2-7 kpc. The vertical and horizontal dashed lines indicate the velocity range from -5 to 5 $km~s^{-1}$ and the rms noise levels. The Gaussin fittings to the profiles are plotted as dash-dot lines. }\label{fig:linewidths}
\end{figure}

\begin{figure}
  \centering
  \includegraphics[scale=0.5]{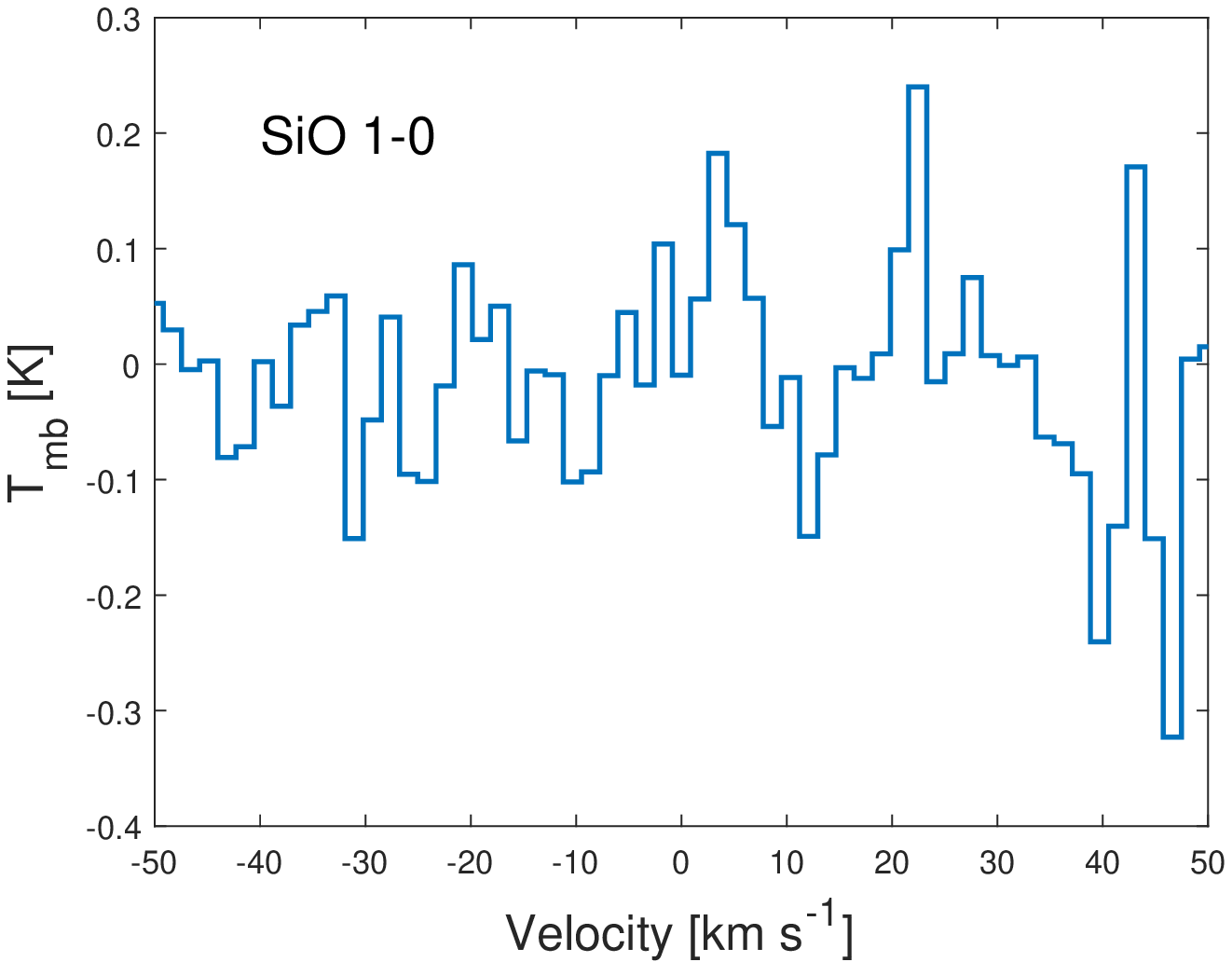}
  \includegraphics[scale=0.5]{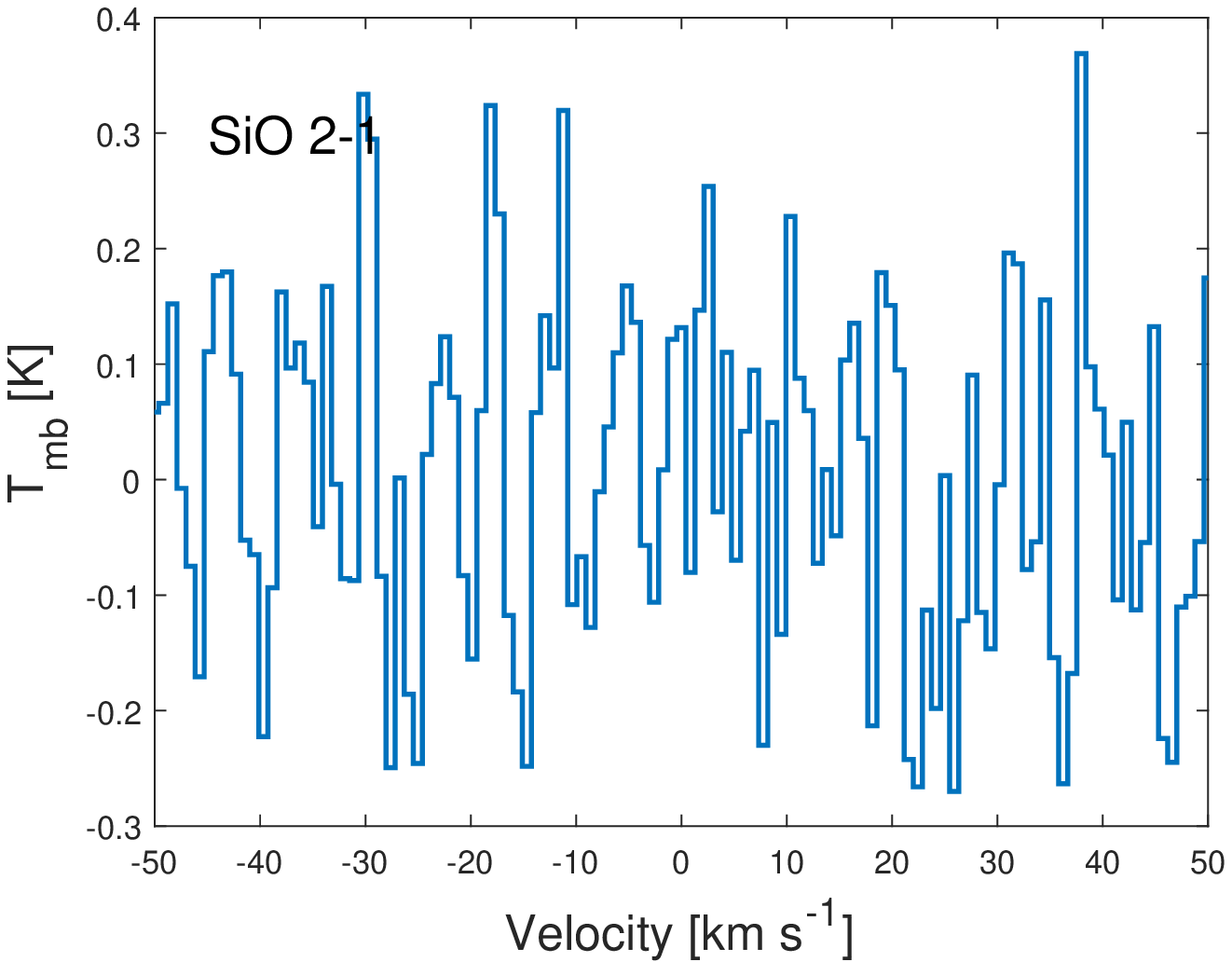}
  \caption{ The distance-weighted average profiles of the SiO 1-0 and 2-1 lines for the targets without clear detections. }\label{fig:noSiO}
\end{figure}

The SiO line profiles of most sources do not have high signal-to-noise ratios so that the high-velocity wings can not be clearly distinguished from noise for individual sources. The average line profiles of the SiO 1-0 and 2-1 emission from detected sources located between 2-7 kpc are presented in Figure \ref{fig:linewidths}. The value of main beam temperature is adjusted as if the distances of these sources are uniformly 1 kpc. The detected sources with a large distance $>7$ kpc are excluded lest the spectra of the most distant sources dominate the average spectra. Single Gaussians are fitted to the average spectra, and the fitted curves are also presented. The fitted curve to the SiO 1-0 spectrum has a 6.8 km s$^{-1}$ FWHM, and the high-velocity wings are also shown. The FWHM of the fitted curve to the SiO 2-1 spectrum is 9.4 km s$^{-1}$. In addition, the distance-weighted average SiO line profiles from the sources without clear detections are plotted in Figure \ref{fig:noSiO}. The profiles suggest that the SiO emission is undetected in these sources mainly due to intrinsic property and not because of the instrumental sensitivity limit.

The widths of the SiO 3-2 transition are provided in Table \ref{table_sio3}. The mean and median values of the FWZPs for the SiO 3-2 line are $8.46$ and 8.0 km s$^{-1}$, respectively. There is no source with FWZP>20 km s$^{-1}$, and almost all of the sources associated with broad SiO 1-0 and 2-1 line profiles are not detected in the SiO 3-2 line. It could be caused by the low sensitivity limitation in the SiO 3-2 observations since the noise of the SiO 3-2 spectrum is higher than those of the SiO 1-0 and 2-1 spectra in most of these sources. But the different beam sizes can also be a factor. The conclusion can not be made before mapping observations are performed.

\begin{figure}
  \centering
  \includegraphics[scale=0.5]{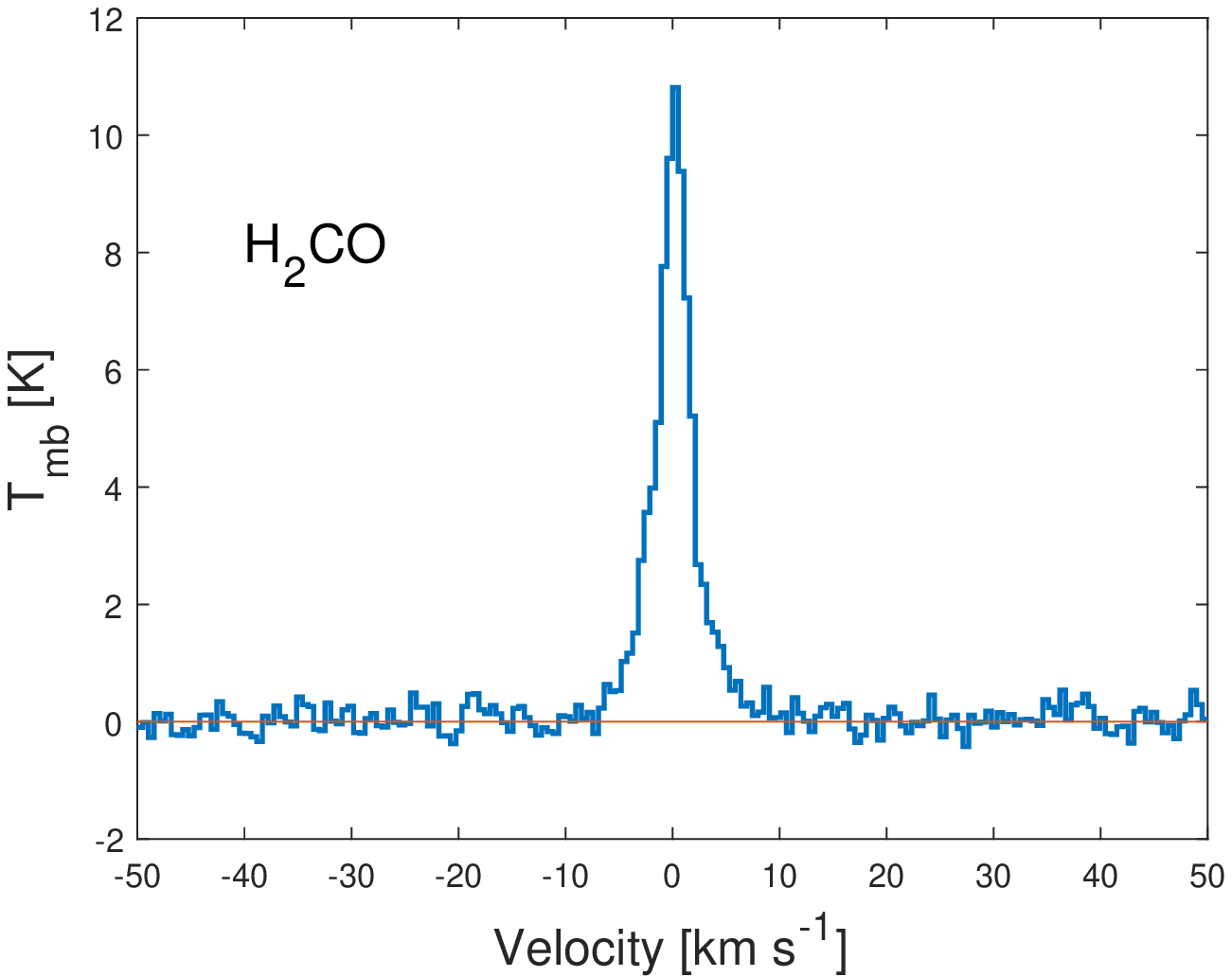}
  \includegraphics[scale=0.5]{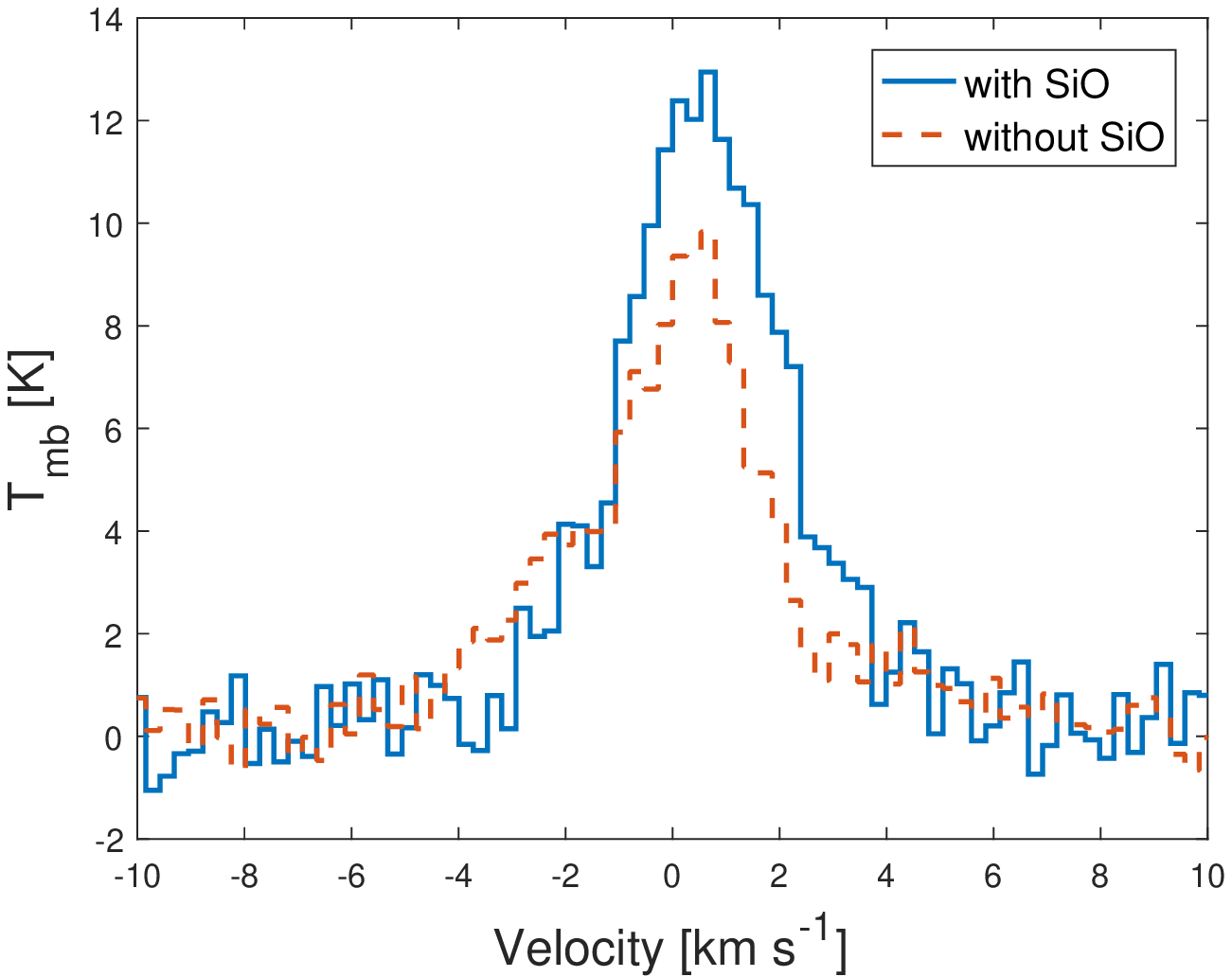}
  \caption{ The distance-weighted average line profile of H$_2$CO emissions is presented in the top panel. The comparison between the distance-weighted average H$_2$CO line profiles from the sources with and without SiO emissions is displayed in the lower panel. }\label{fig:h2cocom}
\end{figure}

The FWHMs of the  H$_2$CO lines are included in Table \ref{table_h2co}. The average value of the FWHMs of the H$_2$CO lines is 3.47 km s$^{-1}$, and the median value is 3.05 $km~s^{-1}$. The average H$_2$CO line profile is also calculated and displayed in the top panel of Figure \ref{fig:h2cocom}. The comparison between the distance-weighted average H$_2$CO line profiles from the sources with and without SiO emissions is displayed in the lower panel. The FWHMs of the average H$_2$CO line profiles from the sources associated with SiO emissions and without SiO emissions are 3.46 and 2.79 $km~s^{-1}$, respectively. But the difference of the two average FWHMs is not statistically significant. The line widths of the SiO and H$_2$CO molecules can not be explained by thermal broadenings with reasonable temperature of dense gases. So the higher values of the line widths are mostly attributed to turbulence or bulk motions. The significant difference in the velocity fields showed from line widths between SiO and H$_2$CO emissions suggests the different origins and distributions between these two molecules.

\begin{table*}\tiny 
\centering
\caption{The properties of the SiO 1-0 and 2-1 lines emitted from corresponding sources. FWZP is full width at zero power. L$_{\textrm{SiO}1-0}$ and L$_{\textrm{SiO}2-1}$ are the luminosities for the SiO 1-0 and 2-1 lines, respectively.}\label{table_sio1}
\begin{tabular}{|c|ccccc|ccccc|}
\hline
   &   \multicolumn{5}{c|}{SiO 1-0} & \multicolumn{5}{c|}{SiO 2-1} \\
\hline
\multirow{2}*{Name} & V$_{\textrm{lsr}}$ & $\int T_{\textrm{mb}}dv$ & $T_{\textrm{mb}}$ & FWZP & L$_{\textrm{SiO}1-0}$ & V$_{lsr}$ & $\int T_{\textrm{mb}}dv$ & $T_{\textrm{mb}}$ & FWZP & L$_{\textrm{SiO}2-1}$ \\
& [km s$^{-1}$] & [K km s$^{-1}$] & [K] & [km s$^{-1}$] & $10^{-7}$ L$_\odot$ & [km s$^{-1}$] & [K km s$^{-1}$] & [K] & [km s$^{-1}$] & $10^{-7}$ L$_\odot$ \\
\hline
BGPS 2724 & $34.15\pm0.40$ & $0.098\pm0.028$ & 0.049 & 3.5 & 0.26 & $34.80\pm0.76$ & $0.243\pm0.055$ & 0.049 & 7.9 & 1.98 \\
BGPS 2945 & $20.85\pm0.28$ & $0.201\pm0.035$ & 0.080 & 6.1 & 0.43 & ... & ... & ... & ... & ... \\
BGPS 2970 & $40.81\pm0.84$ & $0.688\pm0.088$ & 0.061 & 25.9 & 13.04 & $43.03\pm1.84$ & $0.324\pm0.065$ & 0.061 & 7.2 & 26.89 \\
BGPS 2976 & $40.78\pm1.05$ & $0.420\pm0.061$ & 0.041 & 19.0 & 1.61 & $40.03\pm1.84$ & $0.403\pm0.066$ & 0.026 & 22.6 & 6.10 \\
BGPS 3110 & $17.96\pm0.25$ & $0.604\pm0.061$ & 0.131 & 8.5 & 4.18 & $17.48\pm0.18$ & $0.935\pm0.070$ & 0.170 & 13.7 & 12.58 \\
BGPS 3114 & $22.27\pm0.50$ & $0.355\pm0.053$ & 0.068 & 10.0 & 2.29 & $22.69\pm0.27$ & $0.548\pm0.053$ & 0.115 & 8.6 & 7.69 \\
BGPS 3118 & $18.18\pm0.75$ & $0.430\pm0.056$ & 0.045 & 20.7 & 2.39 & $18.06\pm0.59$ & $0.659\pm0.078$ & 0.070 & 15.6 & 13.00 \\
BGPS 3128 & $17.84\pm0.56$ & $0.268\pm0.048$ & 0.061 & 8.6 & 8.71 & $18.22\pm0.47$ & $0.404\pm0.081$ & 0.093 & 7.0 & 29.31 \\
BGPS 3139 & $19.76\pm0.34$ & $0.163\pm0.026$ & 0.064 & 5.3 & 8.33 & $19.84\pm0.85$ & $0.141\pm0.035$ & 0.026 & 7.0 & 16.88 \\
BGPS 3220 & $46.34\pm0.20$ & $0.205\pm0.035$ & 0.086 & 7.0 & 5.15 & $44.77\pm0.45$ & $0.345\pm0.056$ & 0.100 & 5.6 & 23.48 \\
BGPS 3247 & ... & ... & ... & ... & ... & $45.41\pm0.48$ & $0.555\pm0.075$ & 0.115 & 7.7 & 67.19 \\
BGPS 3344 & $62.09\pm2.34$ & $0.258\pm0.056$ & 0.019 & 15.7 & 9.98 & $64.87\pm1.48$ & $0.223\pm0.049$ & 0.035 & 7.1 & 34.39 \\
BGPS 3442 & $73.09\pm2.76$ & $0.180\pm0.035$ & 0.023 & 13.5 & 10.63 & $77.31\pm2.49$ & $0.374\pm0.075$ & 0.026 & 13.7 & 59.83 \\
BGPS 3604 & $48.00\pm0.42$ & $0.704\pm0.061$ & 0.070 & 23.6 & 138.3 & $46.26\pm0.92$ & $0.314\pm0.059$ & 0.041 & 10.5 & 146.34 \\
BGPS 3627 & $79.61\pm3.54$ & $0.608\pm0.148$ & 0.033 & 27.8 & 20.63 & $80.16\pm2.28$ & $0.454\pm0.125$ & 0.041 & 13.8 & 41.80 \\
BGPS 3656 & $36.05\pm2.38$ & $0.219\pm0.066$ & 0.074 & 5.1 & 10.15 & ... & ... & ... & ... \\
BGPS 3686 & $77.99\pm0.61$ & $0.275\pm0.041$ & 0.041 & 13.7 & 4.05 & $80.55\pm1.67$ & $0.454\pm0.079$ & 0.033 & 20.5 & 21.30 \\
BGPS 3710 & $74.47\pm0.46$ & $0.258\pm0.054$ & 0.049 & 19.2 & 1.56 & $73.60\pm1.09$ & $0.238\pm0.053$ & 0.033 & 13.7 & 6.38 \\
BGPS 3822 & $56.49\pm5.06$ & $0.576\pm0.106$ & 0.016 & 55.1 & 12.48 & $50.09\pm2.49$ & $0.526\pm0.084$ & 0.029 & 23.8 & 35.04 \\
BGPS 3982 & $53.85\pm0.31$ & $0.298\pm0.026$ & 0.070 & 7.7 & 80.58 & $54.14\pm0.58$ & $0.556\pm0.056$ & 0.064 & 12.0 & 316.89 \\
BGPS 4029 & $81.45\pm1.01$ & $0.435\pm0.048$ & 0.039 & 22.5 & 10.36 & $82.68\pm0.75$ & $0.936\pm0.081$ & 0.074 & 21.3 & 47.90 \\
BGPS 4082 & $99.33\pm1.05$ & $0.310\pm0.045$ & 0.035 & 16.9 & 18.38 & $98.64\pm0.36$ & $0.640\pm0.059$ & 0.100 & 11.0 & 93.33 \\
BGPS 4230 & $108.83\pm2.92$ & $0.680\pm0.170$ & 0.029 & 41.4 & 30.20 & ... & .... & ... & ... & ... \\
BGPS 4294 & $56.10\pm0.29$ & $0.111\pm0.021$ & 0.058 & 3.6 & 7.54 & $55.88\pm0.27$ & $0.171\pm0.029$ & 0.068 & 4.3 & 24.11 \\
BGPS 4297 & ... & ... & ... & ... & ... & $56.01\pm0.50$ & $0.295\pm0.038$ & 0.058 & 6.9 & 43.88 \\
BGPS 4356 & $110.81\pm0.46$ & $0.338\pm0.038$ & 0.064 & 9.4 & 12.76 & $111.31\pm0.87$ & $0.205\pm0.053$ & 0.039 & 6.8 & 16.86 \\
BGPS 4375 & ... & ... & ... & ... & ... & $91.66\pm0.40$ & $0.185\pm0.033$ & 0.061 & 5.1 & 11.55 \\
BGPS 4396 & $110.04\pm1.30$ & $0.155\pm0.048$ & 0.029 & 8.6 & 6.64 & $107.46\pm1.67$ & $0.414\pm0.079$ & 0.035 & 13.5 & 44.49 \\
BGPS 4402 & $101.44\pm0.73$ & $0.279\pm0.043$ & 0.041 & 12.2 & 8.60 & $99.33\pm1.03$ & $0.404\pm0.071$ & 0.051 & 10.4 & 48.51 \\
BGPS 4472 & $47.04\pm0.88$ & $0.501\pm0.075$ & 0.051 & 19.1 & 9.28 & $49.12\pm0.89$ & $1.071\pm0.108$ & 0.086 & 20.7 & 76.84 \\
BGPS 5021 & $80.08\pm0.30$ & $0.274\pm0.038$ & 0.078 & 7.8 & 13.11 & $79.65\pm0.21$ & $0.731\pm0.063$ & 0.135 & 12.0 & 68.09 \\
BGPS 5064 & $99.81\pm0.42$ & $0.425\pm0.061$ & 0.074 & 15.6 & 17.23 & $100.07\pm0.90$ & $0.541\pm0.049$ & 0.061 & 14.7 & 75.41 \\
BGPS 5114 & $65.63\pm0.26$ & $0.420\pm0.038$ & 0.093 & 11.3 & 9.66 & $64.93\pm0.25$ & $0.598\pm0.053$ & 0.138 & 8.6 & 30.46\\
BGPS 5243 & $96.79\pm0.21$ & $0.146\pm0.024$ & 0.068 & 4.3 & 6.10 & $96.93\pm0.25$ & $0.270\pm0.040$ & 0.070 & 6.0 & 32.04 \\
\hline
\end{tabular}
\end{table*}

\begin{table*}\tiny 
\caption{The properties of the SiO 3-2 line emitted from corresponding sources. L$_{\textrm{SiO}3-2}$ is the luminosity}\label{table_sio3}
\centering
\begin{tabular}{|c|ccccc|}
\hline
\multirow{2}*{Name} & V$_{\textrm{lsr}}$ & $\int T_{\textrm{mb}}dv$ & $T_{\textrm{mb}}$ & FWZP & L$_{\textrm{SiO}3-2}$ \\
& km s$^{-1}$ & K km s$^{-1}$ & K & km s$^{-1}$ & $10^{-7}$ L$_\odot$ \\
\hline
BGPS 3110 & $17.92\pm0.14$ & $0.588\pm0.035$ & 0.148 & 6.8 & 15.85 \\
BGPS 3114 & $22.20\pm0.31$ & $0.400\pm0.081$ & 0.148 & 3.5 & 8.75 \\
BGPS 3118 & $19.77\pm1.66$ & $0.573\pm0.138$ & 0.166 & 9.2 & 19.46 \\
BGPS 3686 & $80.79\pm1.18$ & $0.454\pm0.085$ & 0.041 & 13.8 & 26.89 \\
BGPS 4472 & $50.94\pm0.66$ & $1.300\pm0.118$ & 0.103 & 16.1 & 94.59 \\
BGPS 5064 & $101.30\pm1.24$ & $0.153\pm0.043$ & 0.029 & 5.6 & 43.26 \\
BGPS 5114 & $65.47\pm0.25$ & $0.184\pm0.039$ & 0.084 & 3.4 & 14.49 \\
BGPS 5243 & $97.30\pm0.57$ & $0.261\pm0.059$ & 0.049 & 9.3 & 30.11 \\
\hline
\end{tabular}
\end{table*}

\subsection{Luminosity}

Since the spatial distributions of the gases that emit the molecular lines are not known, we use fluxes per beam calculated from velocity-integrated intensity to estimate the luminosities of the SiO and H$_2$CO lines for simplicity. So the derived values represent the lower limit of the real luminosities. In table \ref{table_sio1}, the luminosities of the SiO 1-0 and 2-1 lines are listed. And those of the SiO 3-2 line and the H$_2$CO line are provided in Table \ref{table_sio3} and \ref{table_h2co}, respectively. The luminosity ranges from $0.26\times10^{-7}$ to $1.38\times10^{-5}$ L$_\odot$ for the SiO 1-0 line, from $1.98\times10^{-7}$ to $3.17\times10^{-5}$ L$_\odot$ for the SiO 2-1 line, from $8.75\times10^{-7}$ to $9.46\times10^{-6}$ L$_\odot$ for the SiO 3-2 line, and $6.11\times10^{-7}$ to $4.30\times10^{-4}$ L$_\odot$ for the H$_2$CO line. The median values of the luminosities are $9.28\times10^{-7}$, $3.20\times10^{-6}$, $2.32\times10^{-6}$ and $1.89\times10^{-5}$ L$_\odot$ for the corresponding lines, respectively.

\subsection{SiO column density and abundance} \label{sec:cln}

Under the assumption of a beam-filling factor of unity and a LTE condition for the SiO 1-0 and 2-1 lines and the optically thin assumption, the molecular column density can be derived by the equation below \citep{man15}:

\begin{equation}
N=\frac{3h}{8\pi^3S\mu^2}\frac{Q_{\textrm{rot}}(T_{\textrm{ex}})}{g_jg_kg_s}\frac{exp(E_u/kT_{\textrm{ex}})}{exp(h\nu/kT_{\textrm{ex}})-1}\frac{\int T_{\textrm{mb}}dv}{J_\nu(T_{\textrm{ex}})-J_\nu(T_{\textrm{bg}})},  \\
\label{eq:cln}
\end{equation}

where h is the Planck constant, S is the line strength, and $\mu$ is the dipole moment. $Q_{\textrm{rot}}(T_{\textrm{ex}})$ is the partition function for the excitation temperature $T_{\textrm{ex}}$. $T_{\textrm{bg}}=2.73$ K is the temperature of the background radiation. $g_J=2J_u+1$ is the rotational degeneracy, and $g_K=1$ and $g_s=1$ are the K degeneracy and spin degeneracy for SiO molecules, respectively. $J_\nu(T)=\frac{h\nu}{k}(exp(h\nu/kT)-1)^{-1}$ is the Rayleigh-Jeans brightness temperature.

The median gas kinetic temperature of these SCCs is 14 K \citep{svo16}, but the temperature of the shocked gas should be different. The excitation temperature of the shocked gas emitting the SiO lines is unclear, but T$_{\textrm{ex}}=10~K$ is considered as a good assumption of excitation temperature for the physical properties of SCCs by previous studies \citep{leu14,cse16}. \citet{cse16} also pointed out that the estimated column density of the SiO molecules does not vary significantly with a appropriate excitation temperature ranging from 5 to 30 K. The estimated SiO column densities are written in Table \ref{table_cln}. The estimated SiO column densities derived from the observed fluxes of the SiO 1-0 line and the SiO 2-1 line range from $5.99\times10^{11}$ to $4.32\times10^{12}~cm^{-2}$ and from $3.07\times10^{11}$ to $2.32\times10^{12}$ cm$^{-2}$, respectively. The median values are $1.83\times10^{12}$ cm$^{-2}$ and $8.75\times10^{11}~cm^{-2}$, and the mean values are $2.16\times10^{12}$ and $9.92\times10^{11}~cm^{-2}$. The range of the estimated SiO column densities is not significantly different from the results for the starless clumps in the previous works \citep{cse16,li19}. In most sources, the SiO column densities estimated from the SiO 1-0 and 2-1 lines can not match even if the $T_{\textrm{ex}}$ is changed in a range of 5-100 K. What causes the difference between the estimated values derived from the SiO 1-0 and 2-1 lines is discussed in Section \ref{sec:excitation}.

\begin{figure}
  \centering
  \includegraphics[scale=0.5]{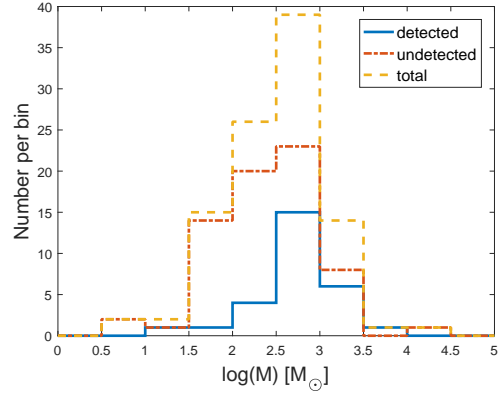}
  \caption{ The mass distributions of the sources. The solid line represents the mass distribution of the sources with SiO emission. The dash-dotted line shows the mass distribution of the sources without SiO emission. And the dashed line display the mass distribution of all the sources. }\label{fig:massdistribution}
\end{figure}

In order to estimate the SiO abundance, the column density of H$_2$ should be also calculated. The 1.1 mm continuum thermal dust emission from BGPS is used to estimate the gas mass and column density of the SCCs \citep{gin13} through the formula \citep{li19} below:

\begin{equation}
M_{\textrm{gas}}=\eta\frac{S_\nu d^2}{B_\nu(T)\kappa_\nu}
\end{equation}

and

\begin{equation}
N(\textrm{H}_2)=\frac{\eta F_\nu}{B_\nu(T)\Omega\kappa_\nu\mu_{H_2}m_\textrm{H}}~~~,
\end{equation}

where $M_{\textrm{gas}}$ is the clump mass. $\eta=100$ is the gas-to-dust mass ratio \citep{sch08}, $S_\nu$ is the integrated continuum flux at the frequency of $\nu$. $d$ is the source distance given in \citet{cal18}, $B_\nu(T)$ is the intensity of a blackbody at the temperature T and frequency $\nu$, and $\kappa_\nu=10(\nu/(1.2\times10^{12}))^{1.5}~cm^2~g^{-1}$ is the dust opacity \citep{hil83}. $N(\textrm{H}_2)$ is the column density of H$_2$, $F_\nu$ is the flux density, and $\Omega$ is the beam solid angle. $\mu_{\textrm{H}_2}=2.8$ is the molecular weight of a hydrogen molecule \citep{kau08}, and $m_H$ is the mass of a hydrogen atom. The gas mass $M_{\textrm{gas}}$ is derived from the total 1.1 mm continuum flux toward a starless clump candidate, but the H$_2$ column density is calculated by using the continuum flux with a beam size of 40$''$. Assuming that the dust temperature is equal to the NH$_3$ derived gas kinetic temperature provided by \citet{svo16}, the estimated gas masses and H$_2$ column densities of the SCCs which exhibit SiO emission are listed in Table \ref{table_cln}. The mass distribution of the starless clump candidates is showed in Figure \ref{fig:massdistribution}. The mass distributions of the sources with and without SiO emission are also showed in Figure \ref{fig:massdistribution}. The mean and median gas masses of the total SCC sample are 697.1 and 372.2 M$_\odot$, respectively. And the mean masses of the SCCs with and without SiO emission are 937.3 and 620.6 M$_\odot$. The median masses of the SCCs with and without SiO emission are 535.5 and 243.4 M$_\odot$. By using Kolmogorov-Smirnov test, it is found that the mass distributions of the two samples are not significantly different.

After the column densities of SiO and H$_2$ molecules are calculated, the SiO abundances are obtained and also presented in Table \ref{table_cln}. In the SCCs, the H$_2$ column density ranges from $1.14\times10^{21}$ to $5.89\times10^{22}~cm^{-2}$, and the mean and median values are $6.18\times10^{21}$ and $4.84\times10^{21}~cm^{-2}$, respectively. The SiO abundance estimated from the SiO 1-0 line ranges from $3.70\times10^{-11}$ to $1.09\times10^{-9}$, and the average and median values are $3.36\times10^{-10}$ and $2.77\times10^{-10}$, respectively. Correspondingly, The SiO abundance derived from the properties of the SiO 2-1 line ranges from $2.02\times10^{-11}$ to $4.03\times10^{-10}$. In addition, the mean and median values are $1.59\times10^{-10}$ and $1.41\times10^{-10}$, respectively.

\begin{table*} \tiny 
\centering
\caption{The column densities and abundances of the SiO molecules estimated from the flux densities of the SiO 1-0, 2-1 and 3-2 lines, and the excitation temperature is assumed to be 10 K.}\label{table_cln}
\begin{tabular}{|c|cc|cc|cc|cc|}
\hline
   &   &   & \multicolumn{2}{c|}{SiO 1-0} & \multicolumn{2}{c|}{SiO 2-1} & \multicolumn{2}{c|}{SiO 3-2} \\
\hline
Name & M [M$_\odot$] & N(H$_2$) [cm$^{-2}$] & N(SiO) [cm$^{-2}$] & X(SiO) & N(SiO) [cm$^{-2}$] & X(SiO) & N(SiO) [cm$^{-2}$] & X(SiO) \\
\hline
BGPS 2724 & 18.5 & $2.77\times10^{21}$ & $5.99\times10^{11}$ & $2.16\times10^{-10}$ & $5.27\times10^{11}$ & $1.90\times10^{-10}$ & ... & ...\\
BGPS 2945 & 52.9 & $6.03\times10^{21}$ & $1.24\times10^{12}$ & $2.05\times10^{-10}$ & ... & ... & ... & ... \\
BGPS 2970 & 597.6 & $1.00\times10^{22}$ & $4.23\times10^{12}$ & $4.23\times10^{-10}$ & $7.01\times10^{11}$ & $7.01\times10^{-11}$ & ... & ... \\
BGPS 2976 & 36.6 & $4.34\times10^{21}$ & $2.58\times10^{12}$ & $5.94\times10^{-10}$ & $8.75\times10^{11}$ & $2.02\times10^{-10}$ & ... & ... \\
BGPS 3110 & 370.8 & $1.72\times10^{22}$ & $3.71\times10^{12}$ & $2.16\times10^{-10}$ & $2.03\times10^{12}$ & $1.18\times10^{-10}$ & $1.01\times10^{12}$ & $5.87\times10^{-11}$ \\
BGPS 3114 & 5371.3 & $5.89\times10^{22}$ & $2.18\times10^{12}$ & $3.71\times10^{-11}$ & $1.19\times10^{12}$ & $2.02\times10^{-11}$ & $6.85\times10^{11}$ & $1.16\times10^{-11}$ \\
BGPS 3118 & 365.4 & $9.36\times10^{21}$ & $2.64\times10^{12}$ & $2.82\times10^{-10}$ & $1.44\times10^{12}$ & $1.54\times10^{-10}$ & $9.83\times10^{11}$ & $1.05\times10^{-10}$ \\
BGPS 3128 & 695.2 & $8.79\times10^{21}$ & $1.65\times10^{12}$ & $1.87\times10^{-10}$ & $8.75\times10^{11}$ & $9.96\times10^{-11}$ & ... & ... \\
BGPS 3139 & 1417.0 & $7.75\times10^{21}$ & $1.00\times10^{12}$ & $1.29\times10^{-10}$ & $3.07\times10^{11}$ & $3.96\times10^{-11}$ & ... & ... \\
BGPS 3220 & 415.4 & $9.80\times10^{21}$ & $1.26\times10^{12}$ & $1.29\times10^{-10}$ & $7.47\times10^{11}$ & $7.65\times10^{-11}$ & ... & ... \\
BGPS 3247 & 380.6 & $4.18\times10^{21}$ & ... & ... & $1.21\times10^{12}$ & $2.89\times10^{-10}$ & ... & ... \\
BGPS 3344 & 1776.1 & $4.96\times10^{21}$ & $1.58\times10^{12}$ & $3.20\times10^{-10}$ & $4.83\times10^{11}$ & $9.73\times10^{-11}$ & ... & ... \\
BGPS 3442 & 254.3 & $5.15\times10^{21}$ & $1.11\times10^{12}$ & $2.15\times10^{-10}$ & $8.11\times10^{11}$ & $1.58\times10^{-10}$ &... & ... \\
BGPS 3604 & 2676.0 & $5.61\times10^{21}$ & $4.32\times10^{12}$ & $7.69\times10^{-10}$ & $6.84\times10^{11}$ & $1.22\times10^{-10}$ &... & ... \\
BGPS 3627 & 766.4 & $3.41\times10^{21}$ & $3.73\times10^{12}$ & $1.10\times10^{-9}$ & $9.85\times10^{11}$ & $2.89\times10^{-10}$ &... & ... \\
BGPS 3656 & 161.4 & $3.58\times10^{21}$ & $1.35\times10^{12}$ & $3.77\times10^{-10}$ & ... & ... &... & ... \\
BGPS 3686 & 404.7 & $6.01\times10^{21}$ & $1.69\times10^{12}$ & $2.82\times10^{-10}$ & $9.85\times10^{11}$ & $1.64\times10^{-10}$ & $7.76\times10^{11}$ & $1.29\times10^{-10}$ \\
BGPS 3710 & 133.4 & $6.22\times10^{21}$ & $1.58\times10^{12}$ & $2.55\times10^{-10}$ & $5.16\times10^{11}$ & $8.28\times10^{-11}$ &... & ... \\
BGPS 3822 & 392.8 & $9.00\times10^{21}$ & $3.54\times10^{12}$ & $3.93\times10^{-10}$ & $1.15\times10^{12}$ & $1.27\times10^{-10}$ &... & ... \\
BGPS 3982 & 2177.2 & $3.39\times10^{21}$ & $1.83\times10^{12}$ & $5.39\times10^{-10}$ & $1.21\times10^{12}$ & $3.57\times10^{-10}$ &... & ... \\
BGPS 4029 & 473.4 & $9.65\times10^{21}$ & $2.67\times10^{12}$ & $2.77\times10^{-10}$ & $2.03\times10^{12}$ & $2.11\times10^{-10}$ &... & ... \\
BGPS 4082 & 336.3 & $5.10\times10^{21}$ & $1.91\times10^{12}$ & $3.74\times10^{-10}$ & $1.39\times10^{12}$ & $2.73\times10^{-10}$ &... & ... \\
BGPS 4230 & 623.2 & $7.67\times10^{21}$ & $4.18\times10^{12}$ & $5.43\times10^{-10}$ & ... & ... &... & ... \\
BGPS 4294 & 613.0 & $4.39\times10^{21}$ & $6.86\times10^{11}$ & $1.56\times10^{-10}$ & $3.73\times10^{11}$ & $8.52\times10^{-11}$ &... & ... \\
BGPS 4297 & 142.8 & $2.96\times10^{21}$ & ... & ... & $6.43\times10^{11}$ & $2.17\times10^{-10}$ &... & ... \\
BGPS 4356 & 1727.4 & $1.02\times10^{22}$ & $2.07\times10^{12}$ & $2.03\times10^{-10}$ & $4.46\times10^{11}$ & $4.37\times10^{-11}$ &... & ... \\
BGPS 4375 & 117.5 & $4.54\times10^{21}$ & ... & ... & $4.02\times10^{11}$ & $8.86\times10^{-11}$ &... & ... \\
BGPS 4396 & 808.3 & $1.11\times10^{22}$ & $9.58\times10^{11}$ & $8.61\times10^{-11}$ & $8.98\times10^{11}$ & $8.11\times10^{-11}$ &... & ... \\
BGPS 4402 & 253.0 & $6.23\times10^{21}$ & $1.72\times10^{12}$ & $2.76\times10^{-10}$ & $8.75\times10^{11}$ & $1.40\times10^{-10}$ &... & ... \\
BGPS 4472 & 130.3 & $5.76\times10^{21}$ & $3.08\times10^{12}$ & $5.34\times10^{-10}$ & $2.32\times10^{12}$ & $4.03\times10^{-10}$ & $2.23\times10^{12}$ & $3.87\times10^{-10}$ \\
BGPS 5021 & 832.0 & $8.51\times10^{21}$ & $1.68\times10^{12}$ & $1.98\times10^{-10}$ & $1.59\times10^{12}$ & $1.87\times10^{-10}$ &... & ... \\
BGPS 5064 & 1958.2 & $4.77\times10^{21}$ & $2.62\times10^{12}$ & $5.48\times10^{-10}$ & $1.18\times10^{12}$ & $2.47\times10^{-10}$ & $2.61\times10^{11}$ & $5.48\times10^{-11}$ \\
BGPS 5114 & 821.9 & $7.02\times10^{21}$ & $2.58\times10^{12}$ & $3.67\times10^{-10}$ & $1.30\times10^{12}$ & $1.85\times10^{-10}$ & $3.14\times10^{11}$ & $4.48\times10^{-11}$ \\
BGPS 5243 & 421.9 & $4.75\times10^{21}$ & $8.98\times10^{11}$ & $1.89\times10^{-10}$ & $5.85\times10^{11}$ & $1.23\times10^{-10}$ & $4.48\times10^{11}$ & $9.41\times10^{-11}$ \\
\hline
\end{tabular}
\end{table*}

\subsection{H$_2$CO column density and abundance}

The column density of the H$_2$CO molecules is also calculated through Equation \ref{eq:cln} under the same assumptions but with corresponding parameters for H$_2$CO molecules. The excitation temperature is assumed to be the NH$_3$ derived gas kinetic temperature. The derived column densities are provided in Table \ref{table_h2co}, and they range from $1.03\times10^{12}$ to $1.16\times10^{14}$ cm$^{-2}$. The average column density is $1.27\times10^{13}$ cm$^{-2}$, and the median value is $7.94\times10^{12}$ cm$^{-2}$. The H$_2$CO abundances estimated by comparing column densities of H$_2$ and H$_2$CO molecules are also written in Table \ref{table_h2co}. The abundances range from $1.77\times10^{-10}$ to $1.96\times10^{-8}$. The average and median values are $2.57\times10^{-9}$ and $1.45\times10^{-9}$, respectively. The estimated values of the H$_2$CO abundances are similar to those provided in previous works \citep{vic16}.

\subsection{The uncertainties of the estimation for gas mass and molecular hydrogen column density}

According to the BGPS v2.1 data provided by \citet{gin13}, the average $1\sigma$ uncertainty of the 1.1 mm continuum flux density with a beam size of 40$''$  is $34.69\%$, and the maximum uncertainty is $81.48\%$. The average uncertainty of the 1.1 mm continuum integrated flux density is $20.14\%$, and the maximum value is $46.04\%$. The NH$_3$ derived gas kinetic temperature is used to estimate the gas masses and column densities. For most sources, the uncertainty of the gas kinetic temperature is lower than $3\%$. The uncertainties of the gas-to-dust ratio $\eta$ and the dust opacity $\kappa_\nu$ are assumed to be $23\%$ and $28\%$ as in \citet{li19}, respectively. Then the uncertainties of gas mass and H$_2$ column density for the SCCs are $\sim44\%$ and $\sim53\%$, respectively. When analyzing some trends about the gas masses and column densities among the SCCs, these uncertainties are not important because of the large ranges of gas masses and H$_2$ column densities among the SCCs. And the estimated molecular abundances calculated from H$_2$ column densities are still meaningful since very accurate values of the abundances are not indispensable.

\section{discussions} \label{sec:discussion}

\subsection{The excitation condition of the shocked gas} \label{sec:excitation}

\begin{figure}
  \centering
  \includegraphics[scale=0.5]{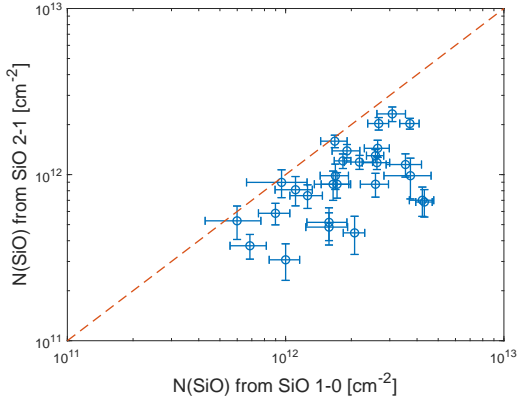}
  \caption{ The comparison between the SiO column densities estimated from the SiO 1-0 and 2-1 lines of the sources both detected in the two SiO lines. The estimates are calculated under the LTE assumption. The dashed line represents the position where the two estimated column densities are equal.}\label{fig:com_siocln}
\end{figure}

From Table \ref{table_cln}, it is showed that the SiO column densities estimated from the SiO 1-0 and 2-1 lines are different. And in Figure \ref{fig:com_siocln}, it is obvious that the estimated column densities from the SiO 1-0 line are higher than those from the SiO 2-1 line in almost all of the sources with two line detections. Generally, the beam-filling factor and the non-LTE condition could make this difference. Since the SiO column density estimated from the SiO 1-0 line is higher, the beam-filling factor of the SiO 1-0 line would be higher if this was the main cause. Considering the region covered by the SiO 2-1 observation is also covered by the SiO 1-0 observation, the outer region only covered by the SiO 1-0 observation must be denser than those in the center region if the beam-filling factor of the SiO 1-0 line is higher. This may be true in several sources, but can not be true in most SCCs with detected SiO 2-1 emission because the pointing position should not always be very far from the shocked gas. So the beam-filling factor and observational coverage should not be the main cause. Then the lower estimated column densities from the SiO 2-1 line are probably attributed to the non-LTE condition due to low density of H$_2$ molecules. Commonly, the SiO column densities estimated by using the line fluxes of low SiO rotational transitions tend to be overestimated if the calculation is treated with the LTE assumption that is not satisfied in fact. Therefore we suggest that the SiO column densities from the SiO 1-0 line could be overestimated because of the non-LTE condition if the beam dilution effect is not important. In order to check this suggestion, we perform the non-LTE analysis by using RADEX \citep{tak07} with a plane parallel slab geometry to estimate the SiO column densities under the assumption of the beam-filling factors of unity. The resulting SiO column densities are lower than those estimated from the SiO 1-0 line under the LTE assumption in almost all the sources. Although these non-LTE estimates are not necessarily more accurate than the LTE values due to the unknown beam-filling factors, the analysis still supports a non-LTE environment of the SiO molecules.

\subsection{The fraction of the shocked gas in SCCs}

The comparison between the distance-weighted average H$_2$CO line profiles from the sources associated with and without SiO emission is plotted in the lower panel of Figure \ref{fig:h2cocom}. There is no essential difference in these two line profiles. The corresponding average line areas are 15.00 and 12.08 $K~km~s^{-1}$ for the sources with and without SiO emission, respectively. The average line area for the sources associated with SiO emission is higher, but the difference between these two average values is not statistically significant at the $5\%$ significance level.

\begin{figure}
  \centering
  \includegraphics[scale=0.5]{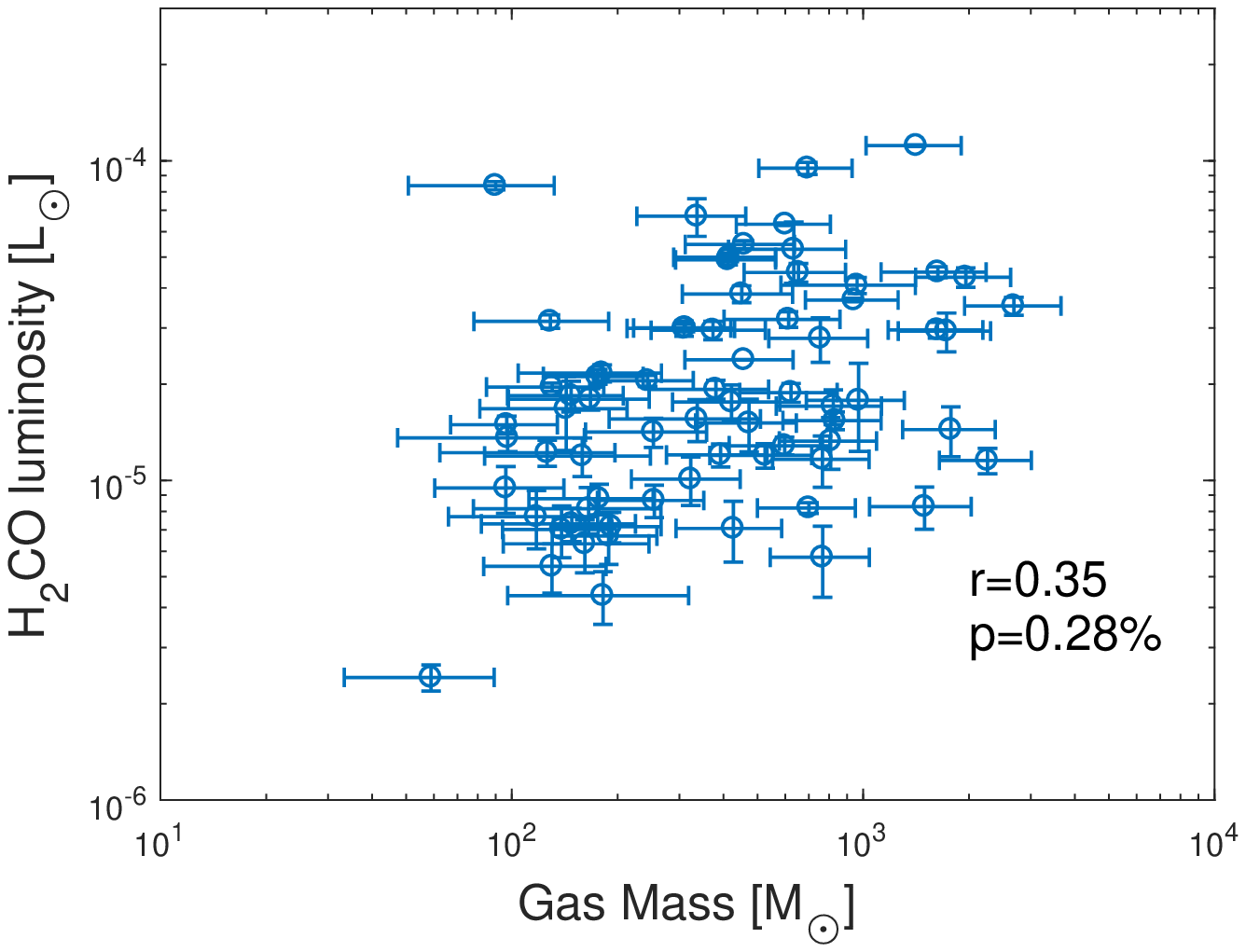}
  \includegraphics[scale=0.5]{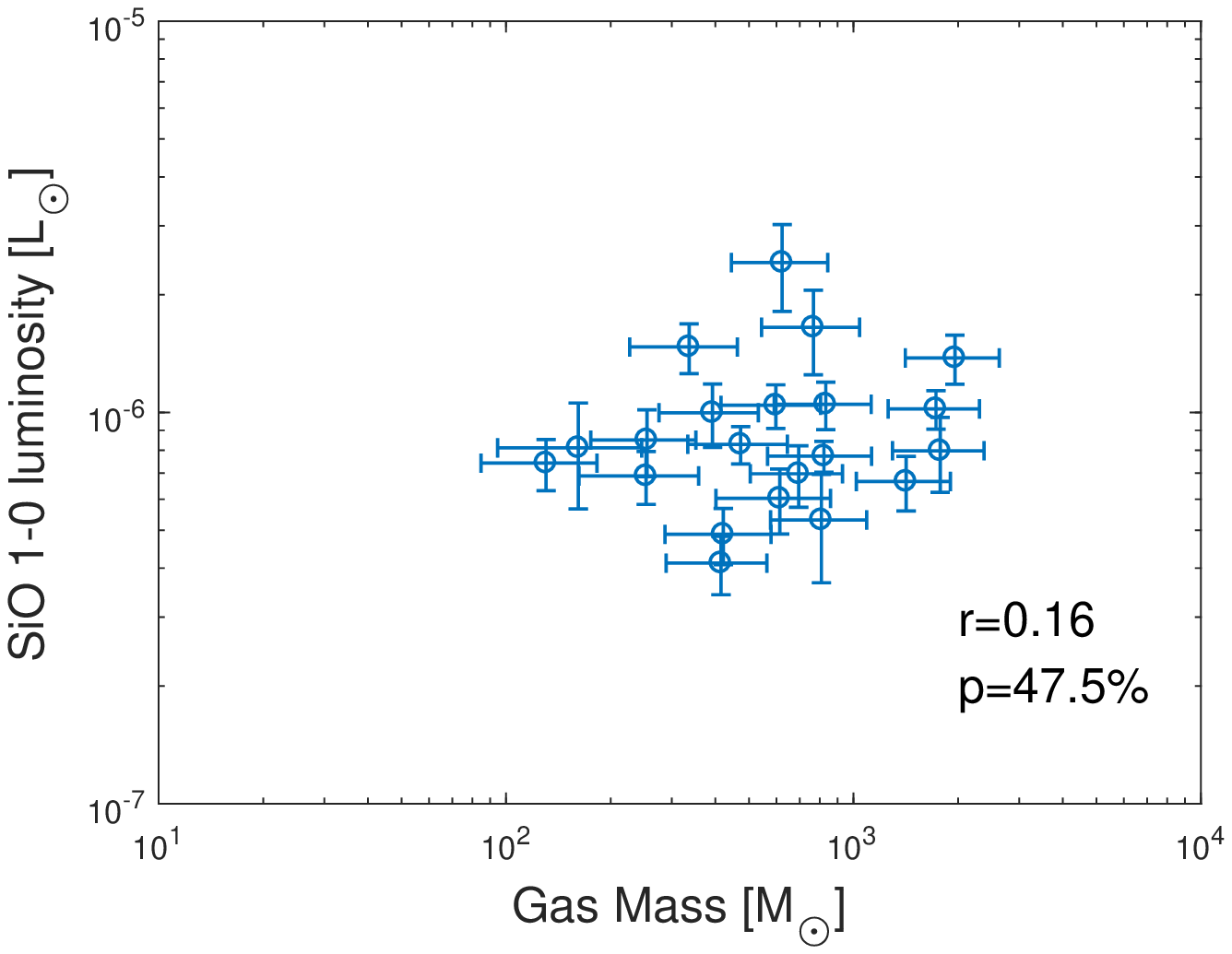}
  \includegraphics[scale=0.5]{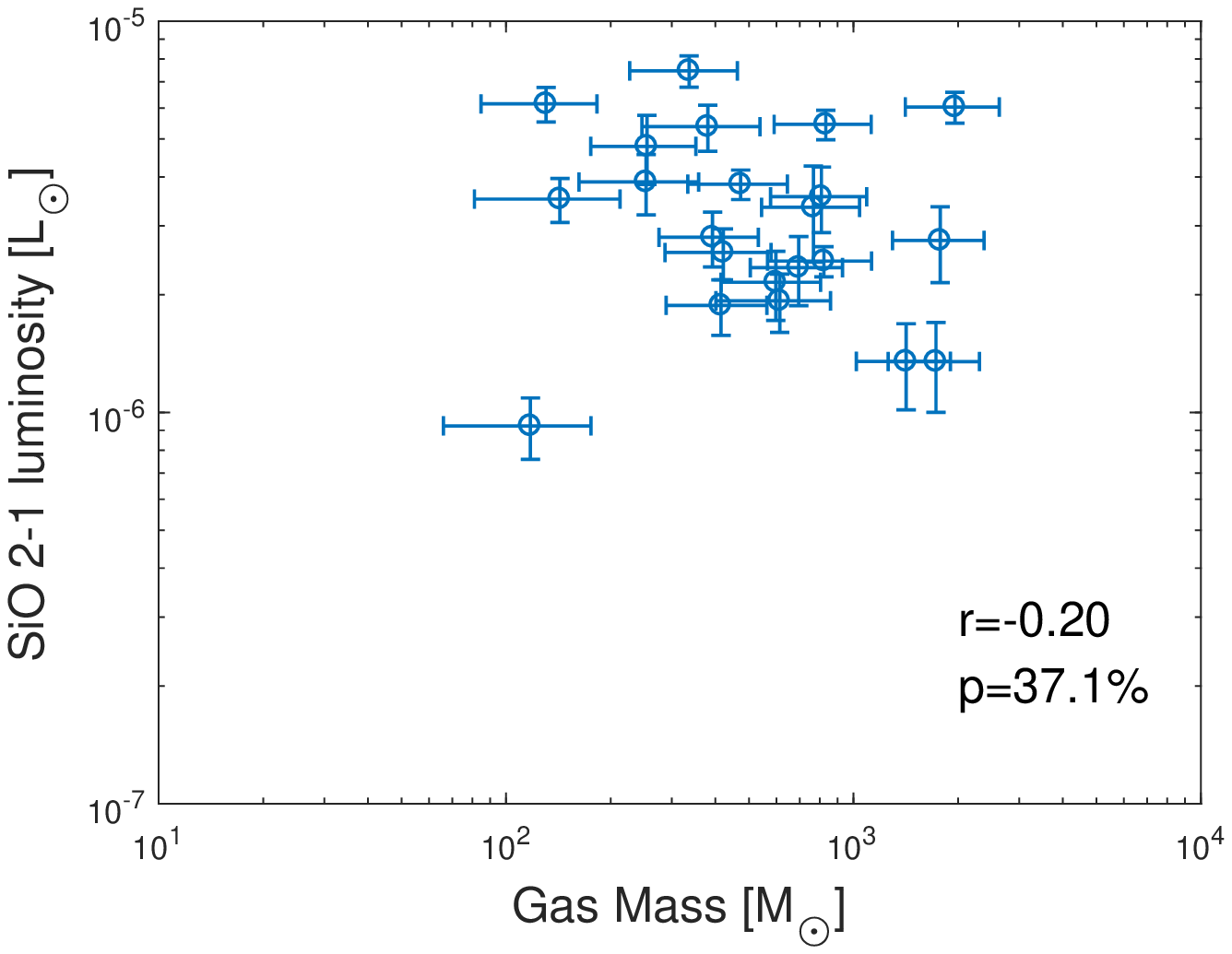}
  \caption{ The luminosity of the H$_2$CO and SiO lines as a function of the gas mass of the SCCs. The distance limited sample includes the detected sources located between 3-7 kpc. }\label{fig:h2co-mass}
\end{figure}

The luminosities of the SiO and H$_2$CO lines as a function of the SCC mass are showed in Figure \ref{fig:h2co-mass}. The sample is distance limited and only includes the sources located between 3-7 kpc. The spearman test is used to evaluate the correlation between the H$_2$CO line luminosity and the gas mass. The Spearman's rank correlation coefficient is 0.35, and the p-value is $0.28\%$. So the positive correlation between the H$_2$CO line luminosity and the clump mass is statistically significant. The correlation can also be be found from the top panel of Figure \ref{fig:h2co-mass}. This result is not strange since H$_2$CO molecules are regarded as a good tracer of dense gas of molecular cloud \citep{guo16}. On the contrary, the luminosity of the SiO lines is not significantly correlated to the clump mass in the spearman test. The coefficients of the correlation between the luminosities of the SiO 1-0 and 2-1 lines and the clump mass are $0.16$ and $-0.20$, and the corresponding p-values are $47.5\%$ and $37.1\%$, respectively. This implies that the relation between the shocks and the clump mass is not obvious.

\subsection{The origin of the shocks in SCCs}

The SCCs can be representative of the very early stage of star formation. In dense clouds, SiO is the best tracer to probe the shocks and is very useful to evaluate the level of star-forming activities. According to our observations of the SiO lines, shocks seem to be quite common even in the early evolutionary age of star formation. The broad SiO line profiles (FWZP$>20~km~s^{-1}$) are obtained in 10 sources. They are probably associated with fast shocks due to protostellar outflows \citep{mar92,jim10}. For the distance-weight average SiO 1-0 and 2-1 spectra presented in Figure \ref{fig:linewidths}, the fitted Gaussian curve to the SiO 1-0 spectrum has relatively narrow line width with a FWHM of 6.8 km s$^{-1}$, but the high-velocity wings are obvious. Moreover, the fitted curve to the SiO 2-1 spectrum with a 9.4 km s$^{-1}$ FWHM can be classified as a broad component in \citet{cse16}. These results suggest some Class 0-like protostars deeply embedded in these SCCs, and SiO line emission seems to be a more powerful tool to detect deeply embedded protostars and their driving shocks than infrared emission \citep{cse16}. In the current work, the signal-to-noise ratios of the detected SiO line profiles are mostly not high ($\int T_{mb}dv<10\sigma_{area}$). Considering the limiting sensitivity of the observations that make the SiO line widths possibly underestimated, there must be more sources hosting protostars. The other sources show relatively narrower line widths (FWZP $<20~km~s^{-1}$). About the formation of the narrow velocity component, several explanations are provided. One is low-velocity shocks due to large scale cloud-cloud collisions \citep{jim10,lou16}, and another is less powerful outflows from intermediate and low-mass protostars \citep{beu07,ino13,san13}. So far, the origin of the shocks in SCCs can not be solved because of the single-dish observations in the current works.

\section{Summary} \label{sec:conclusion}

In this paper, we used the KVN 21m radio telescopes to perform the single-point observations of the SiO 1-0, 2-1 and 3-2 lines and the H$_2$CO $2_{12}-1_{11}$ line toward 100 SCCs. The sample of SCCs is provided by \citet{cal18} and was identified from the BGPS v2.1 catalog \citep{gin13}. The detection rates of the four lines are $31.0\%$, $31.0\%$, $19.5\%$ and $93.0\%$, respectively. Since SiO is a tracer of shocks, the detection rates of the SiO lines suggest that shocks are quite common in the SCCs as very early stage of star formation. The SiO-detected sources provide a sample of targets for further studies of the origin of shocks in SCCs using high angular resolution observations. The details of our results are summarized as follows.

1. According to the FWZP velocity ranges of the detected SiO lines, a significant fraction ($\sim29.4\%$) of the SiO spectra have broad line widths (FWZP $>20~km~s^{-1}$). These broad line widths are regarded as indicators of the fast shocks due to outflows from high-mass protostars. The sources with fast shocks should be not real starless but protostellar clumps. In addition, about $40\%$ of the SiO detections have narrow line widths (FWZP $<10~km~s^{-1}$) possibly associated with low-velocity shocks. This result implies that slow shocks are not rare in the early stage of star formation. The origins of the shocks marked by the SiO emissions in the SCCs need to be studied by further high-resolution mapping observations.

2. The H$_2$CO line widths are $\sim3.0~km~s^{-1}$ and relatively much narrower than those of the SiO lines . This indicate the locations of SiO (shocked gas) and the H$_2$CO molecules are different.

3. The estimated SiO column densities in the SCCs are $\sim10^{12}~cm^{-2}$ under the LTE assumption. By comparing the estimated values derived from the SiO 1-0 and 2-1 line fluxes, the SiO molecules are found to be non-LTE, and the estimated values from the line fluxes of low transitions should be overestimated without the consideration about beam dilution effect. The estimated SiO abundances measured against H$_2$ range from $2.02\times10^{-11}$ to $1.09\times10^{-9}$, and they should be also overestimated. On the other hand, the estimated column densities and abundances could be underestimated due to a beam dilution.

4. The estimated H$_2$CO column densities in the SCCs range from $1.03\times10^{12}$ to $1.16\times10^{14}~cm^{-2}$. The median column density is $7.94\times10^{12}~cm^{-2}$. The estimated H$_2$CO abundances are from $1.77\times10^{-10}$ to $1.96\times10^{-8}$, and the median value is $1.45\times10^{-9}$.

5. The positive correlation of the observed luminosity of the H$_2$CO line and the gas mass of the SCCs is obvious. On the other hand, the SiO line luminosities seem to be not significantly correlated to the gas mass. This result also implies that the clump mass is not a direct cause of the shocks formed in the early stage of star formation.

\section*{Acknowledgements}

We are grateful to all staff members in Korean VLBI Network (KVN) who helped to operate the array and to correlate the data. The KVN is a facility operated by Korea Astronomy and Space Science Institute (KASI). The work is supported by funding from National Key Basic Research and Development Program of China (Grant No. 2017YFA0402604), National Science Foundation of China Grant No. 11590783, U1731237, and China Postdoctoral Science Foundation Grant No. 2020M671267.

\section*{DATA AVAILABILITY STATEMENT}

The data underlying this paper will be shared on reasonable request to the corresponding author.








\appendix
\section{The noise levels and the SiO spectra for individual sources}

\begin{figure}
  \centering
  \includegraphics[scale=0.5]{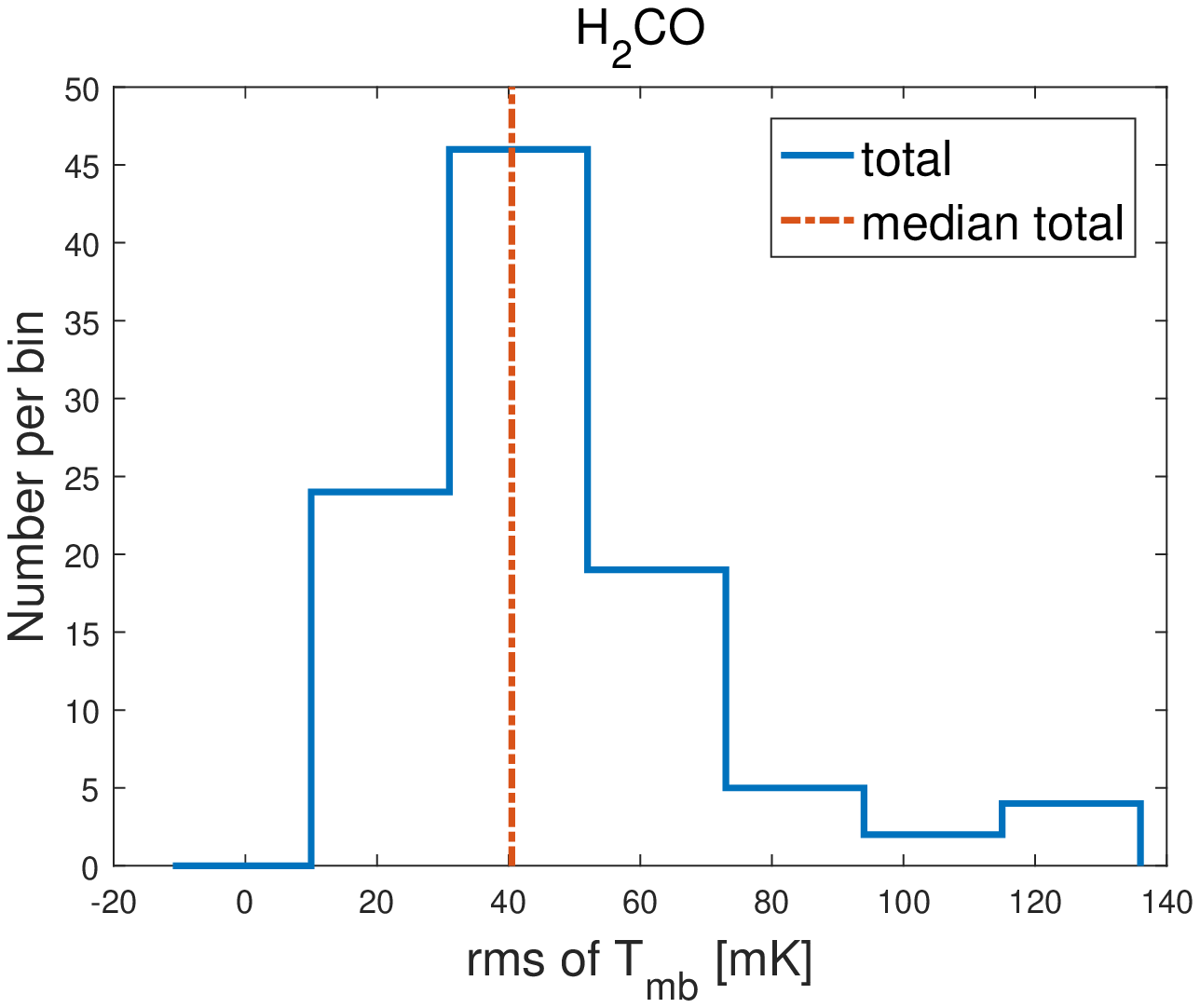}
  \includegraphics[scale=0.5]{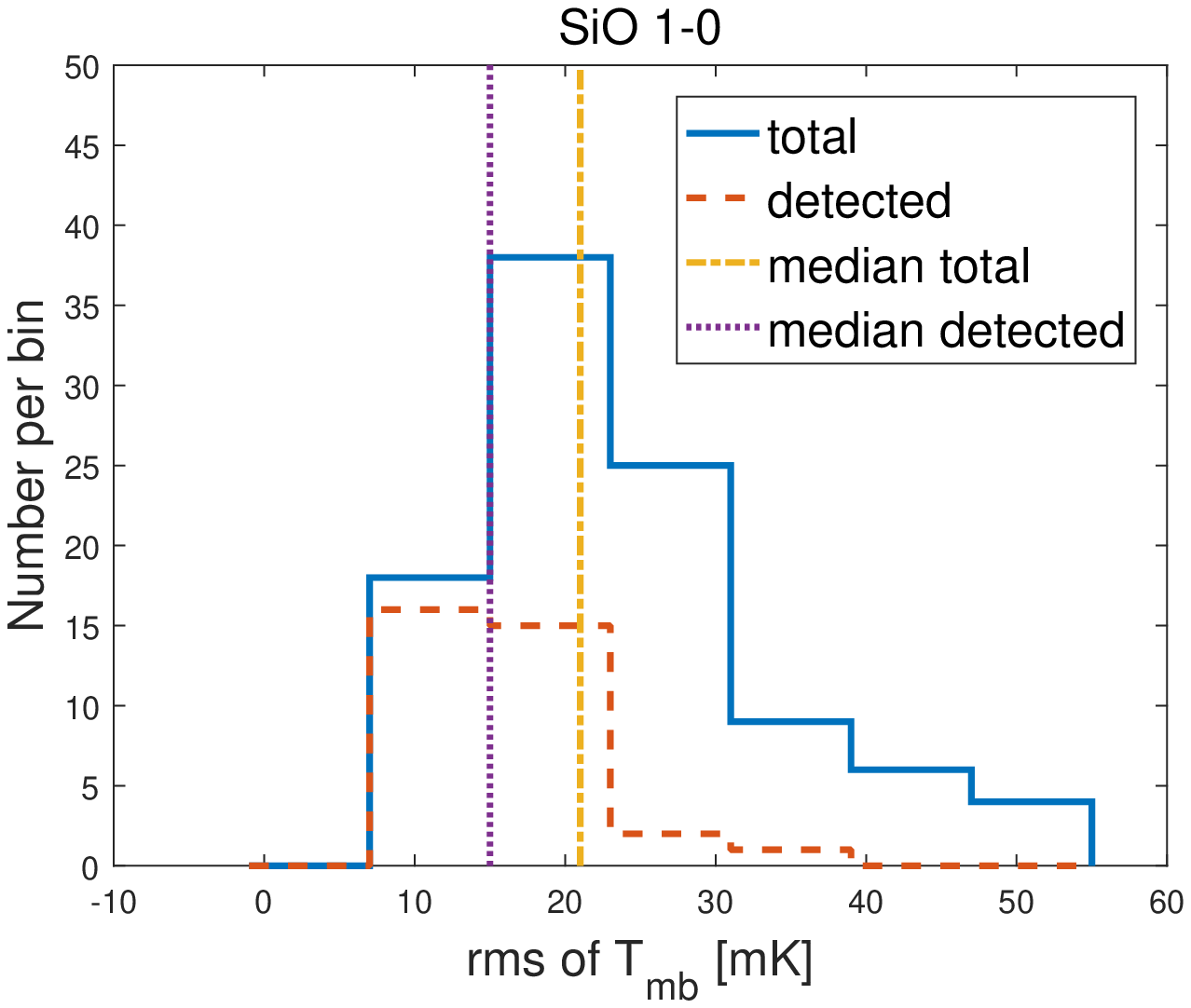}
  \includegraphics[scale=0.5]{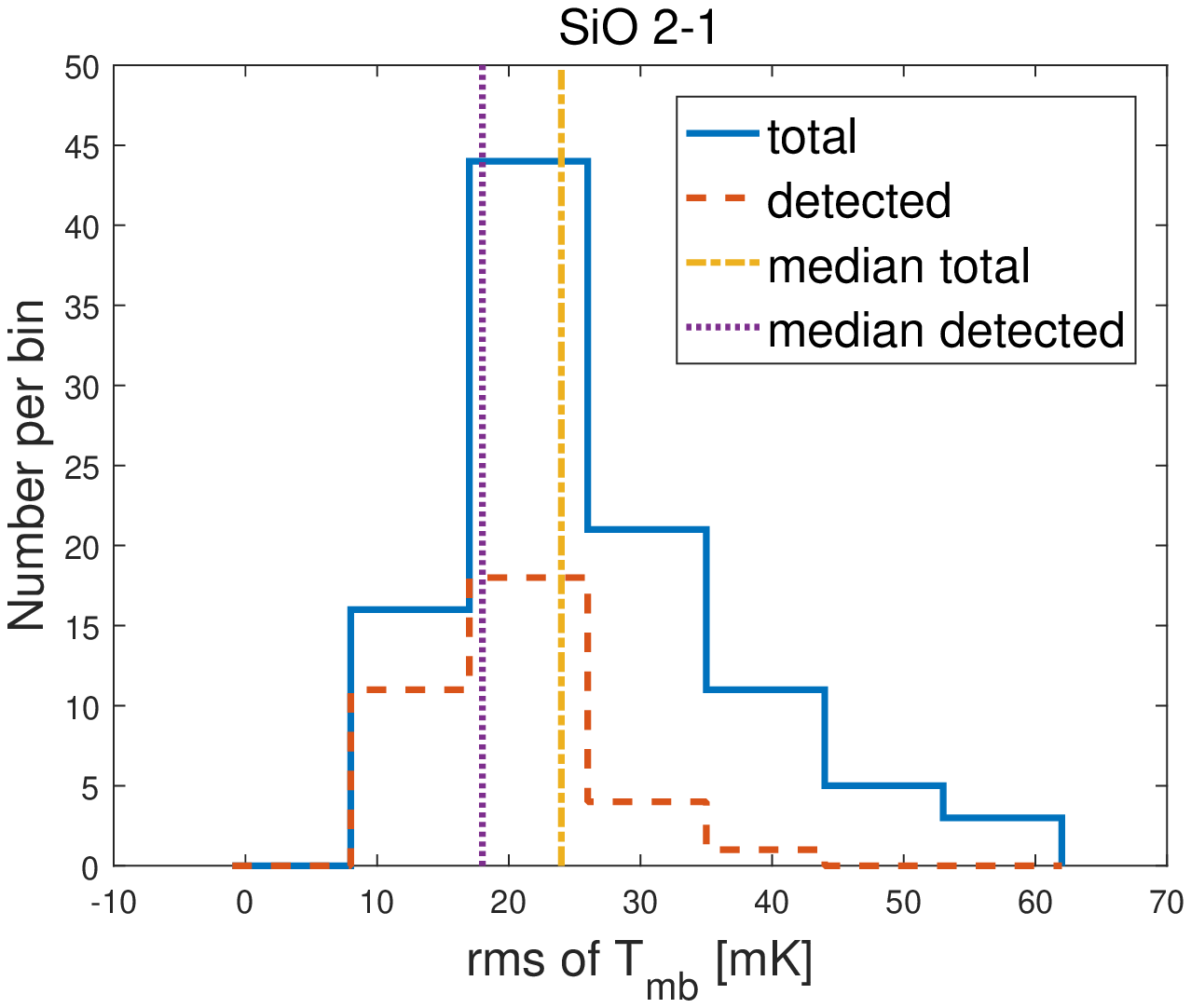}
  \includegraphics[scale=0.5]{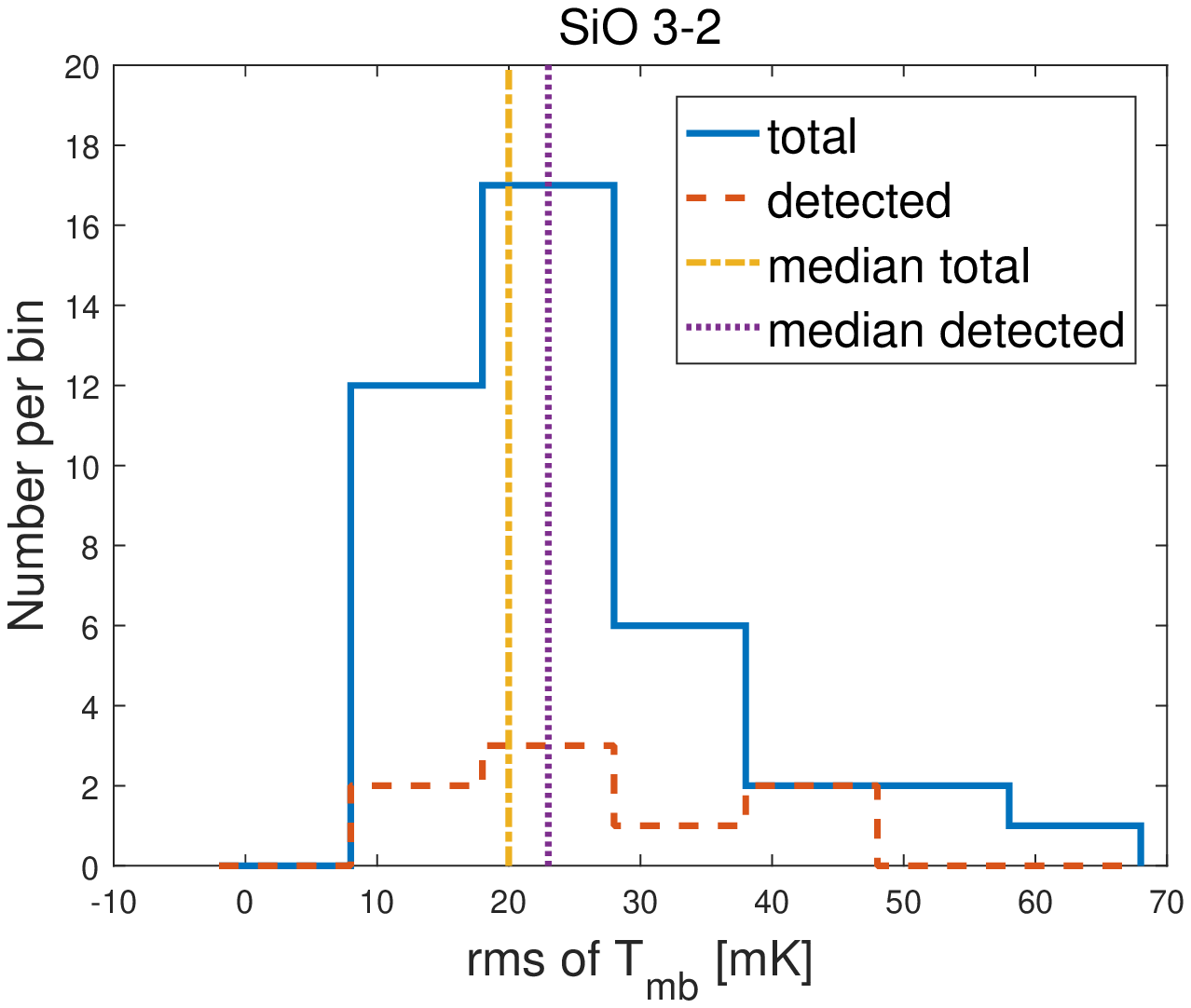}
  \caption{The noise distribution of the observations of the H$_2$CO and SiO lines toward the SCCs in the sample. The vertical dash-dot lines present the median noise levels for the total sample, and the vertical dotted ones show the median values only for the detected sources.}\label{fig:noise}
\end{figure}

In the current work, the integration times and weather conditions of the observations toward individual sources are different. In order to assess this effect on the detection rates of the SiO lines, the rms noise at the velocity resolution of 1 km s$^{-1}$ in the SiO and H$_2$CO lines of the observed sources are calculated and listed in Table \ref{table_rms1}. The noise distributions are presented in Figure \ref{fig:noise}. The median noise levels of the entire sample and the sources with SiO detections are plotted, respectively. The differences between these two kinds of median values are not very significant. So the detection rates are still meaningful. The spectra of the detected SiO lines for individual sources are provided in Figure \ref{fig:siospectrum}.

\begin{table*}\footnotesize
\centering
\caption{The noise levels at the velocity resolution of 1 km s$^{-1}$ in the H$_2$CO and SiO 1-0, 2-1 and 3-2 spectra for individual sources are written. Which instruments used in observations are also presented. YS, US and TN mean Yonsei, Ulsan and Tamna telescopes, respectively.}\label{table_rms1}
\begin{tabular}{|c|c|cccc|}
\hline
\multirow{2}*{Name} & \multirow{2}*{Telescope} & $\delta T_{mb}$(H$_2$CO) & $\delta T_{mb}$(SiO(1-0)) & $\delta T_{mb}$(SiO(2-1)) & $\delta T_{mb}$(SiO(3-2)) \\
 & & [K] & [K] & [K] & [K]  \\
\hline
BGPS 2427 & YS & 0.064 & 0.017 & 0.027 & 0.018 \\
BGPS 2430 & US & 0.053 & 0.026 & 0.030 & 0.031 \\
BGPS 2432 & TN & 0.031 & 0.025 & 0.025 & 0.017 \\
BGPS 2437 & YS & 0.064 & 0.021 & 0.024 & 0.018 \\
BGPS 2533 & US & 0.061 & 0.024 & 0.032 & 0.025 \\
BGPS 2564 & TN & 0.025 & 0.017 & 0.015 & 0.022 \\
BGPS 2693 & US & 0.061 & 0.035 & 0.034 & ... \\
BGPS 2710 & YS & 0.056 & 0.023 & 0.024 & 0.015 \\
BGPS 2724 & US & 0.081 & 0.020 & 0.023 & 0.022 \\
BGPS 2732 & TN & 0.028 & 0.023 & 0.018 & ... \\
BGPS 2742 & US & 0.083 & 0.020 & 0.016 & 0.019 \\
BGPS 2762 & YS+US & 0.064 & 0.027 & 0.023 & 0.018 \\
BGPS 2931 & TN+US & 0.059 & 0.029 & 0.023 & 0.019 \\
BGPS 2940 & YS & 0.026 & 0.021 & 0.032 & ... \\
BGPS 2945 & YS & 0.026 & 0.013 & 0.022 & ... \\
BGPS 2949 & YS & 0.028 & 0.021 & 0.034 & ... \\
BGPS 2970 & YS & 0.033 & 0.015 & 0.024 & 0.056 \\
BGPS 2971 & US & 0.075 & 0.015 & 0.014 & 0.014 \\
BGPS 2976 & TN & 0.022 & 0.013 & 0.014 & 0.019 \\
BGPS 2984 & TN & 0.025 & 0.031 & 0.025 & 0.019 \\
BGPS 2986 & US & 0.064 & 0.039 & 0.027 & 0.031 \\
BGPS 3018 & TN & 0.031 & 0.033 & 0.025 & ... \\
BGPS 3030 & TN & 0.025 & 0.015 & 0.016 & 0.014 \\
BGPS 3110 & US & 0.067 & 0.022 & 0.016 & 0.014 \\
BGPS 3114 & US & 0.017 & 0.013 & 0.018 & 0.044 \\
BGPS 3117 & US & 0.019 & 0.035 & 0.027 & ... \\
BGPS 3118 & YS & 0.033 & 0.013 & 0.020 & 0.044 \\
BGPS 3125 & YS & 0.031 & 0.023 & 0.042 & ... \\
BGPS 3128 & YS & 0.131 & 0.015 & 0.029 & ... \\
BGPS 3129 & TN & 0.047 & 0.027 & 0.021 & 0.014 \\
BGPS 3134 & TN & 0.014 & 0.017 & 0.021 & ... \\
BGPS 3139 & TN & 0.014 & 0.015 & 0.014 & ... \\
BGPS 3151 & TN & 0.017 & 0.035 & 0.034 & ... \\
BGPS 3220 & YS & 0.033 & 0.015 & 0.022 & ... \\
BGPS 3243 & US & 0.042 & 0.041 & 0.046 & ... \\
BGPS 3247 & US & 0.042 & 0.033 & 0.036 & ... \\
BGPS 3276 & US & 0.042 & 0.050 & 0.046 & ... \\
BGPS 3300 & TN & 0.050 & 0.025 & 0.023 & ... \\
BGPS 3302 & TN & 0.036 & 0.042 & 0.052 & ... \\
BGPS 3306 & TN & 0.042 & 0.040 & 0.055 & ... \\
BGPS 3312 & YS & 0.033 & 0.021 & 0.024 & ... \\
BGPS 3315 & US & 0.056 & 0.048 & 0.055 & ... \\
\hline
\end{tabular}
\end{table*}

\begin{table*}\footnotesize
\centering
\caption{continued. }\label{table_rms2}
\begin{tabular}{|c|c|cccc|}
\hline
\multirow{2}*{Name} & \multirow{2}*{Telescope} & $\delta T_{mb}$(H$_2$CO) & $\delta T_{mb}$(SiO(1-0)) & $\delta T_{mb}$(SiO(2-1)) & $\delta T_{mb}$(SiO(3-2)) \\
 & & [K] & [K] & [K] & [K]  \\
\hline
BGPS 3344 & TN & 0.069 & 0.017 & 0.018 & 0.064 \\
BGPS 3442 & YS & 0.028 & 0.010 & 0.017 & ... \\
BGPS 3444 & YS & 0.031 & 0.017 & 0.044 & ... \\
BGPS 3475 & US & 0.042 & 0.048 & 0.046 & ... \\
BGPS 3484 & TN & 0.058 & 0.052 & 0.061 & ... \\
BGPS 3487 & YS & 0.028 & 0.017 & 0.039 & ... \\
BGPS 3534 & YS & 0.036 & 0.017 & 0.042 & ... \\
BGPS 3604 & YS & 0.044 & 0.013 & 0.022 & ... \\
BGPS 3606 & US & 0.042 & 0.044 & 0.032 & ... \\
BGPS 3608 & US & 0.042 & 0.037 & 0.036 & ... \\
BGPS 3627 & US & 0.039 & 0.030 & 0.032 & ... \\
BGPS 3656 & US & 0.050 & 0.024 & 0.032 & ... \\
BGPS 3686 & US & 0.047 & 0.013 & 0.018 & ... \\
BGPS 3705 & TN & 0.036 & 0.029 & 0.039 & ... \\
BGPS 3710 & TN & 0.036 & 0.015 & 0.014 & ... \\
BGPS 3716 & TN & 0.036 & 0.031 & 0.039 & ... \\
BGPS 3736 & TN & 0.033 & 0.023 & 0.043 & ... \\
BGPS 3822 & TN & 0.050 & 0.015 & 0.018 & ... \\
BGPS 3833 & YS & 0.044 & 0.023 & 0.037 & ... \\
BGPS 3892 & YS & 0.028 & 0.019 & 0.032 & ... \\
BGPS 3922 & YS & 0.036 & 0.013 & 0.027 & ... \\
BGPS 3924 & YS & 0.033 & 0.019 & 0.039 & ... \\
BGPS 3982 & YS & 0.036 & 0.008 & 0.015 & ... \\
BGPS 4029 & YS & 0.131 & 0.008 & 0.017 & ... \\
BGPS 4082 & YS & 0.123 & 0.010 & 0.012 & ... \\
BGPS 4095 & YS & 0.105 & 0.010 & 0.017 & ... \\
BGPS 4119 & TN & 0.028 & 0.027 & 0.023 & ... \\
BGPS 4135 & TN & 0.025 & 0.017 & 0.016 & ... \\
BGPS 4140 & US & 0.067 & 0.024 & 0.025 & ... \\
BGPS 4145 & US & 0.058 & 0.015 & 0.018 & ... \\
BGPS 4191 & US & 0.058 & 0.039 & 0.030 & ... \\
BGPS 4230 & TN & 0.025 & 0.021 & 0.023 & ... \\
BGPS 4294 & TN & 0.025 & 0.013 & 0.014 & ... \\
BGPS 4297 & YS & 0.051 & 0.013 & 0.015 & 0.036 \\
BGPS 4346 & TN & 0.025 & 0.027 & 0.025 & ... \\
BGPS 4347 & TN & 0.022 & 0.021 & 0.018 & ... \\
BGPS 4354 & YS & 0.126 & 0.023 & 0.032 & ... \\
BGPS 4356 & YS & 0.095 & 0.010 & 0.020 & ... \\
BGPS 4375 & US & 0.058 & 0.015 & 0.016 & 0.050 \\
BGPS 4396 & US & 0.047 & 0.017 & 0.018 & ... \\
BGPS 4402 & US+YS & 0.031 & 0.015 & 0.021 & ... \\
BGPS 4422 & YS & 0.031 & 0.021 & 0.039 & ... \\
BGPS 4472 & US & 0.017 & 0.017 & 0.027 & 0.028 \\
\hline
\end{tabular}
\end{table*}

\begin{table*}\footnotesize
\centering
\caption{continued. }\label{table_rms3}
\begin{tabular}{|c|c|cccc|}
\hline
\multirow{2}*{Name} & \multirow{2}*{Telescope} & $\delta T_{mb}$(H$_2$CO) & $\delta T_{mb}$(SiO(1-0)) & $\delta T_{mb}$(SiO(2-1)) & $\delta T_{mb}$(SiO(3-2)) \\
 & & [K] & [K] & [K] & [K]  \\
\hline
BGPS 4732 & TN & 0.042 & 0.029 & 0.027 & 0.017 \\
BGPS 4827 & TN & 0.031 & 0.023 & 0.023 & 0.014 \\
BGPS 4841 & TN & 0.036 & 0.023 & 0.027 & 0.014 \\
BGPS 4902 & TN & 0.036 & 0.025 & 0.021 & 0.014 \\
BGPS 4953 & TN & 0.036 & 0.029 & 0.023 & 0.017 \\
BGPS 4962 & US & 0.067 & 0.037 & 0.030 & 0.028 \\
BGPS 4967 & YS & 0.087 & 0.019 & 0.024 & ... \\
BGPS 5021 & YS & 0.036 & 0.013 & 0.015 & ... \\
BGPS 5064 & TN & 0.036 & 0.019 & 0.014 & 0.017 \\
BGPS 5089 & YS & 0.023 & 0.027 & 0.024 & 0.033 \\
BGPS 5090 & YS & 0.056 & 0.025 & 0.020 & 0.023 \\
BGPS 5114 & YS & 0.077 & 0.010 & 0.017 & 0.021 \\
BGPS 5166 & YS & 0.046 & 0.021 & 0.022 & 0.018 \\
BGPS 5183 & YS & 0.051 & 0.017 & 0.024 & 0.023 \\
BGPS 5243 & YS & 0.051 & 0.013 & 0.017 & 0.021 \\
\hline
\end{tabular}
\end{table*}

\begin{figure*}
  \centering

  \includegraphics[scale=0.38]{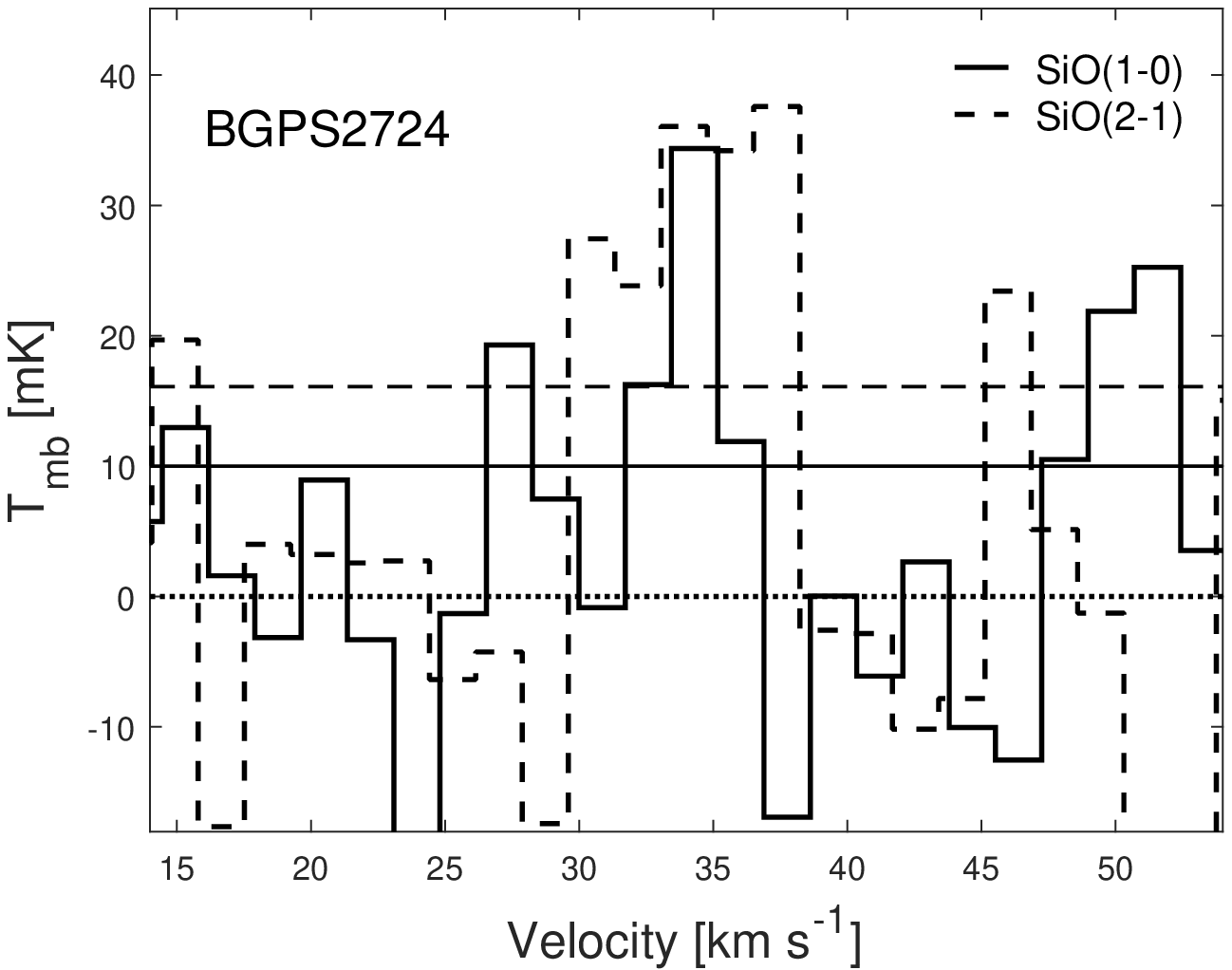}
  \includegraphics[scale=0.38]{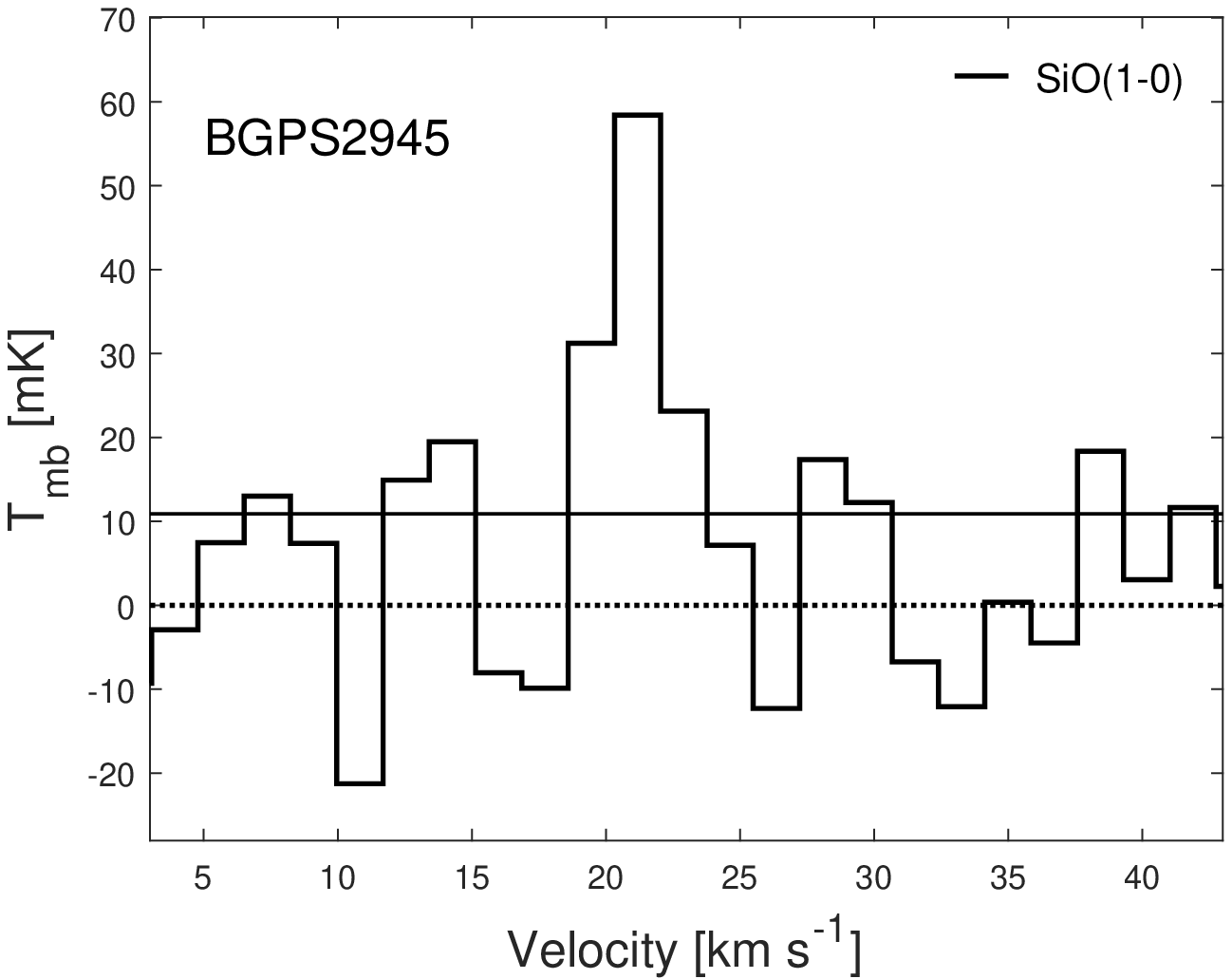}
  \includegraphics[scale=0.38]{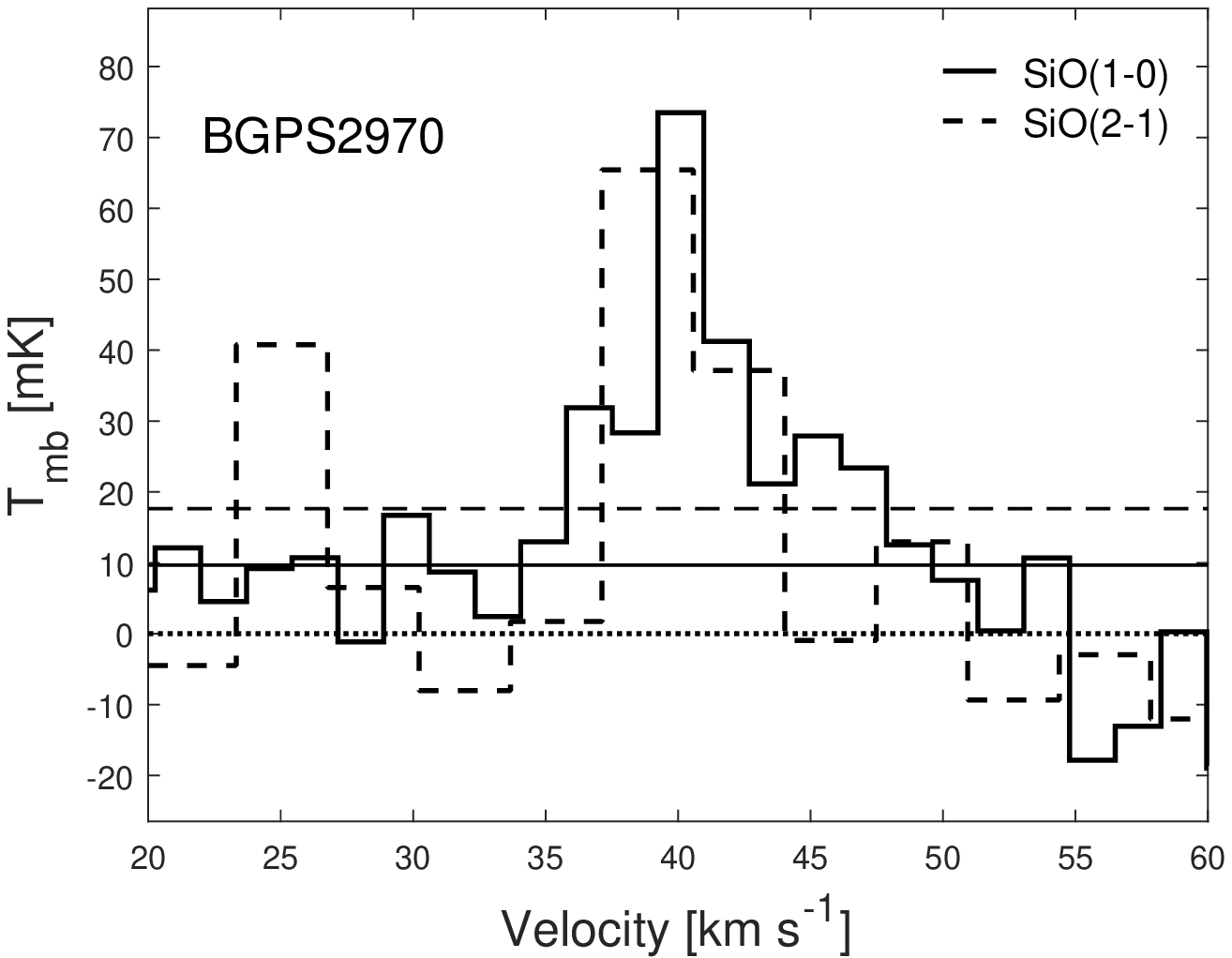}
  \includegraphics[scale=0.38]{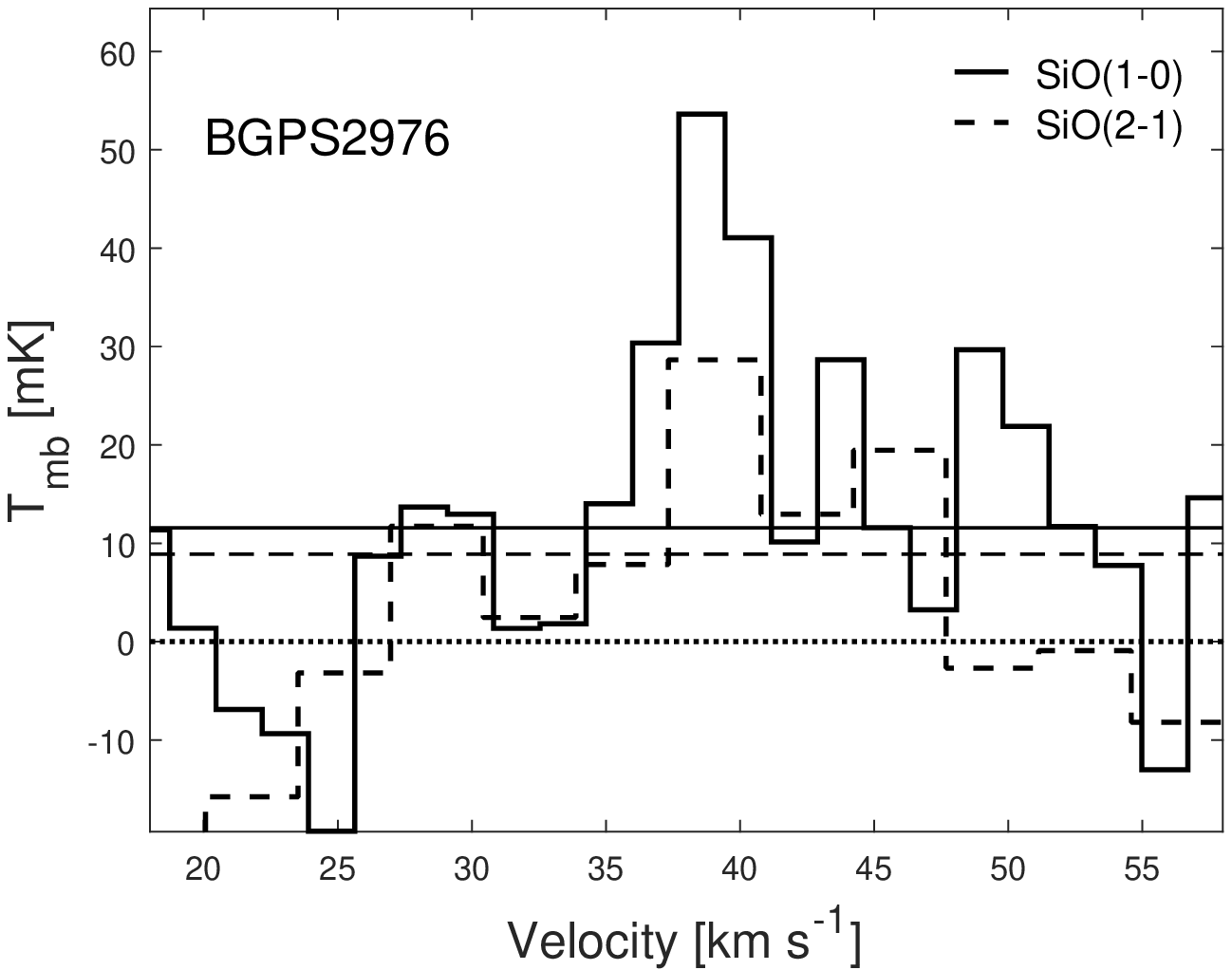}
  \includegraphics[scale=0.38]{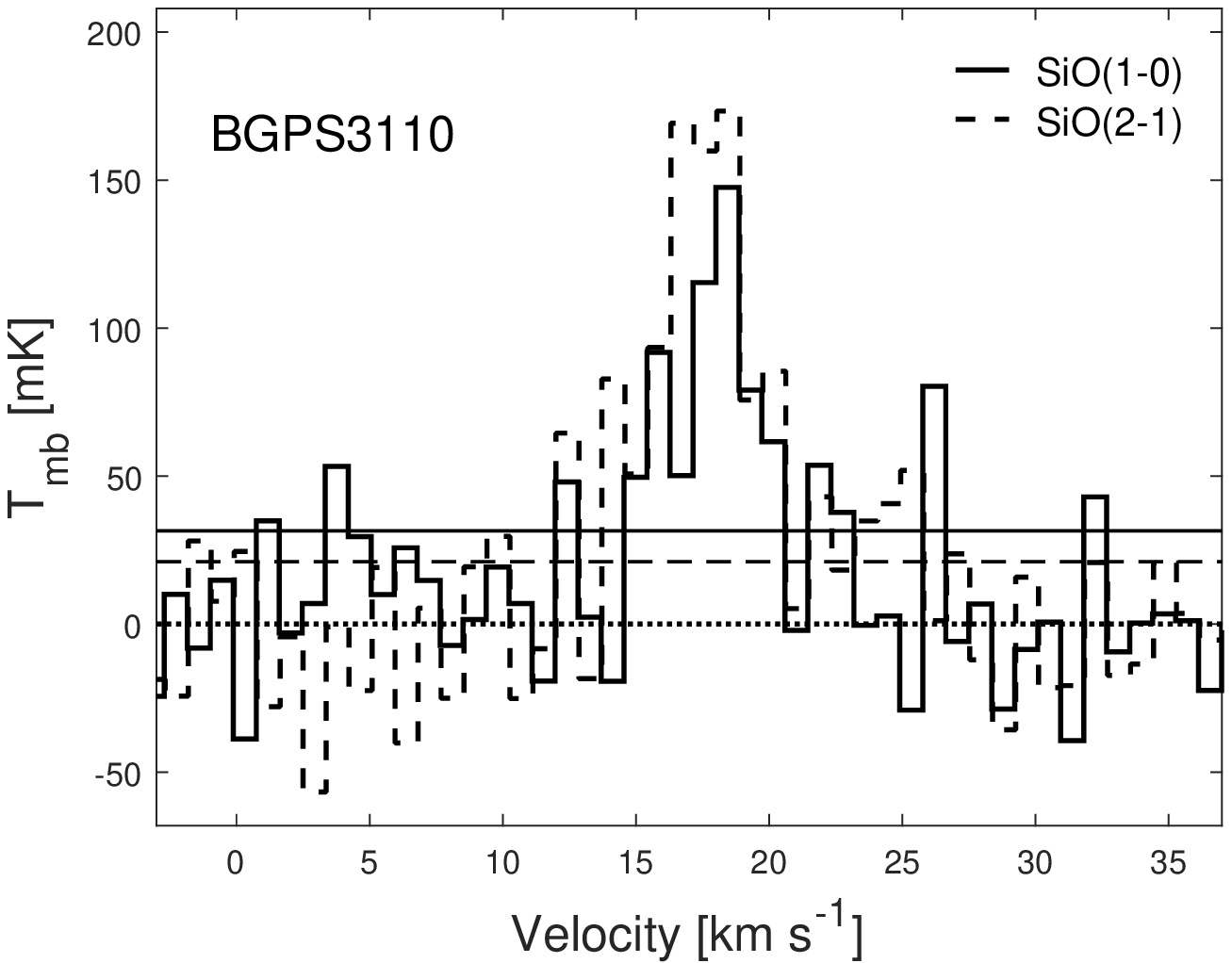}
  \includegraphics[scale=0.38]{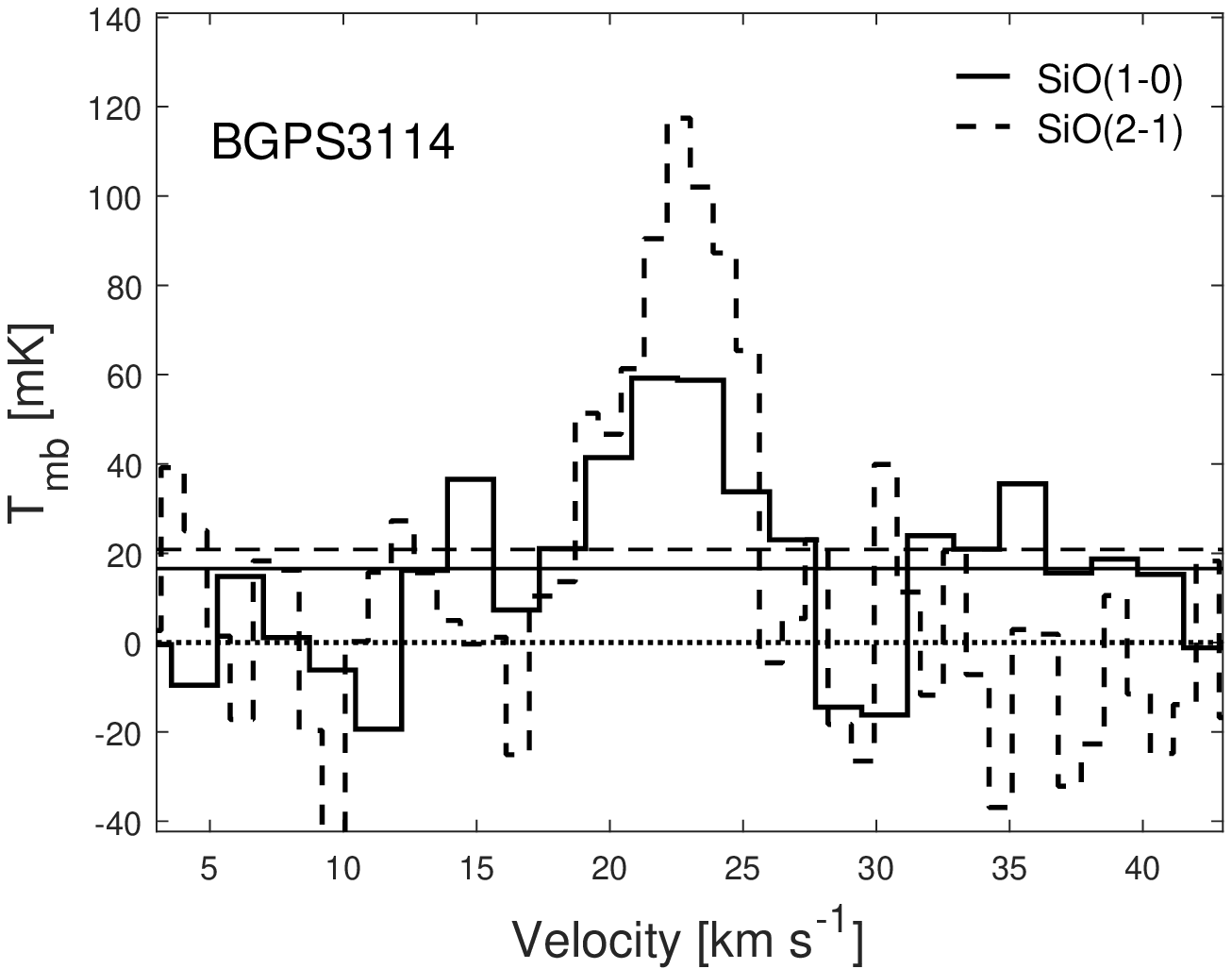}
  \includegraphics[scale=0.38]{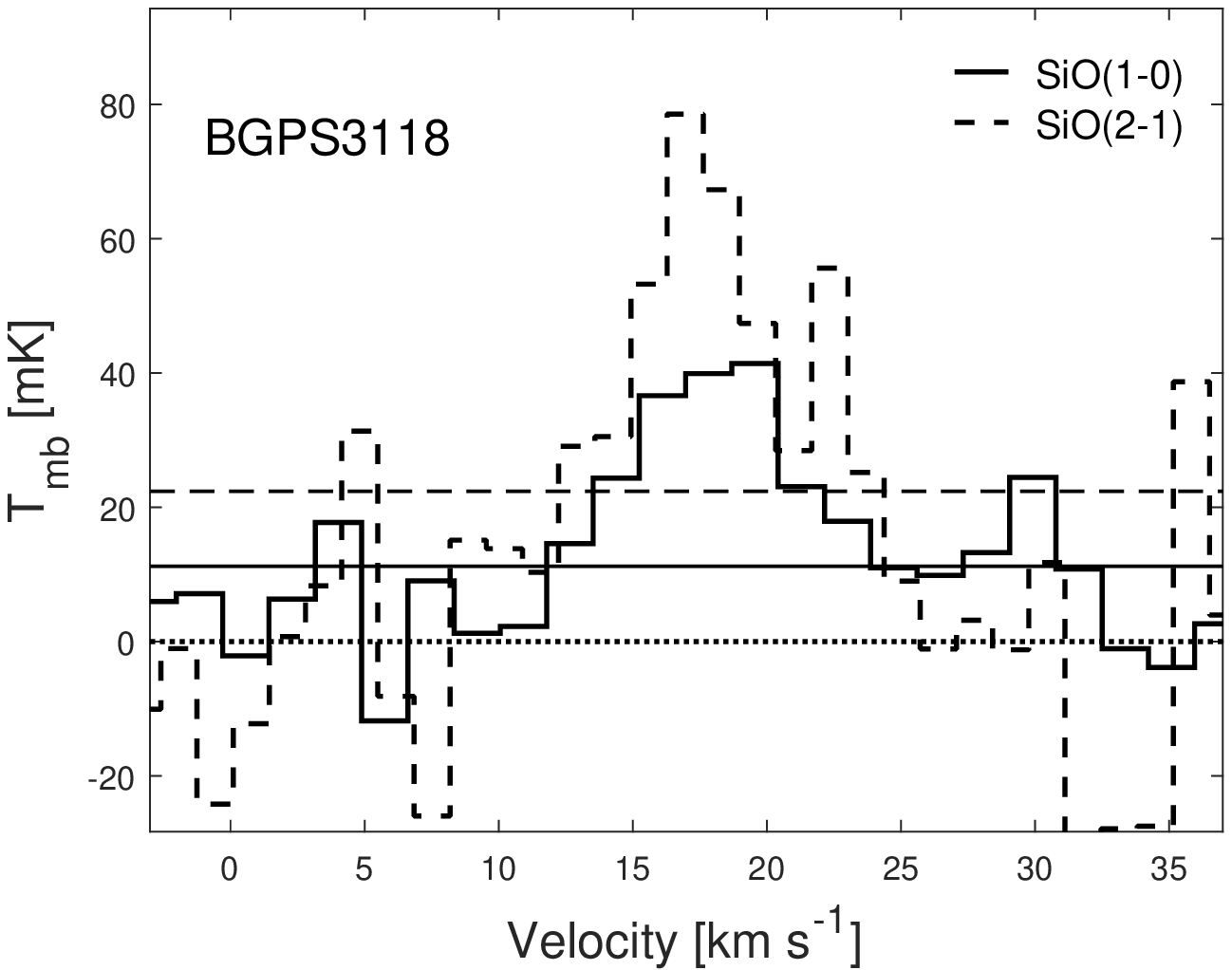}
  \includegraphics[scale=0.38]{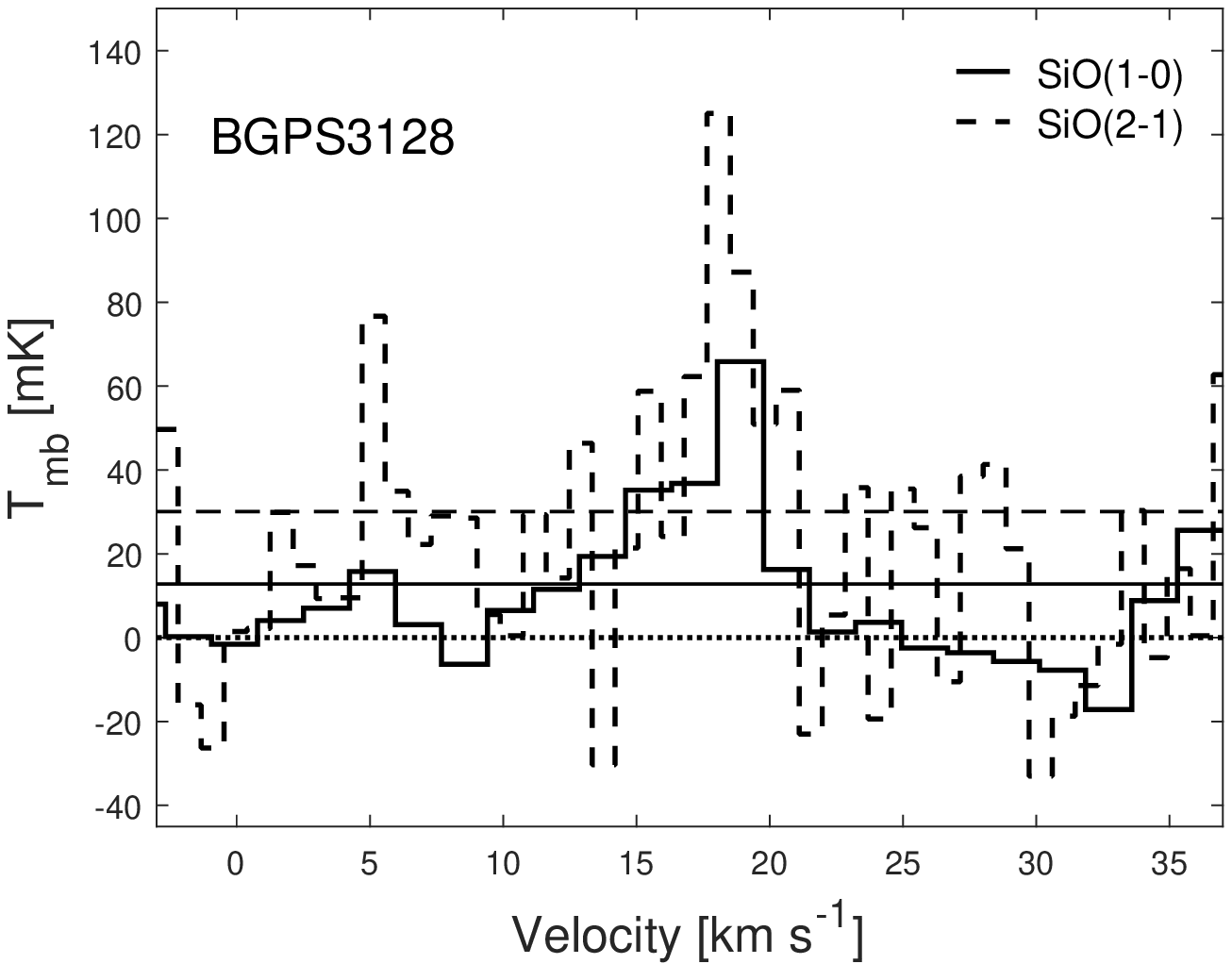}
  \includegraphics[scale=0.38]{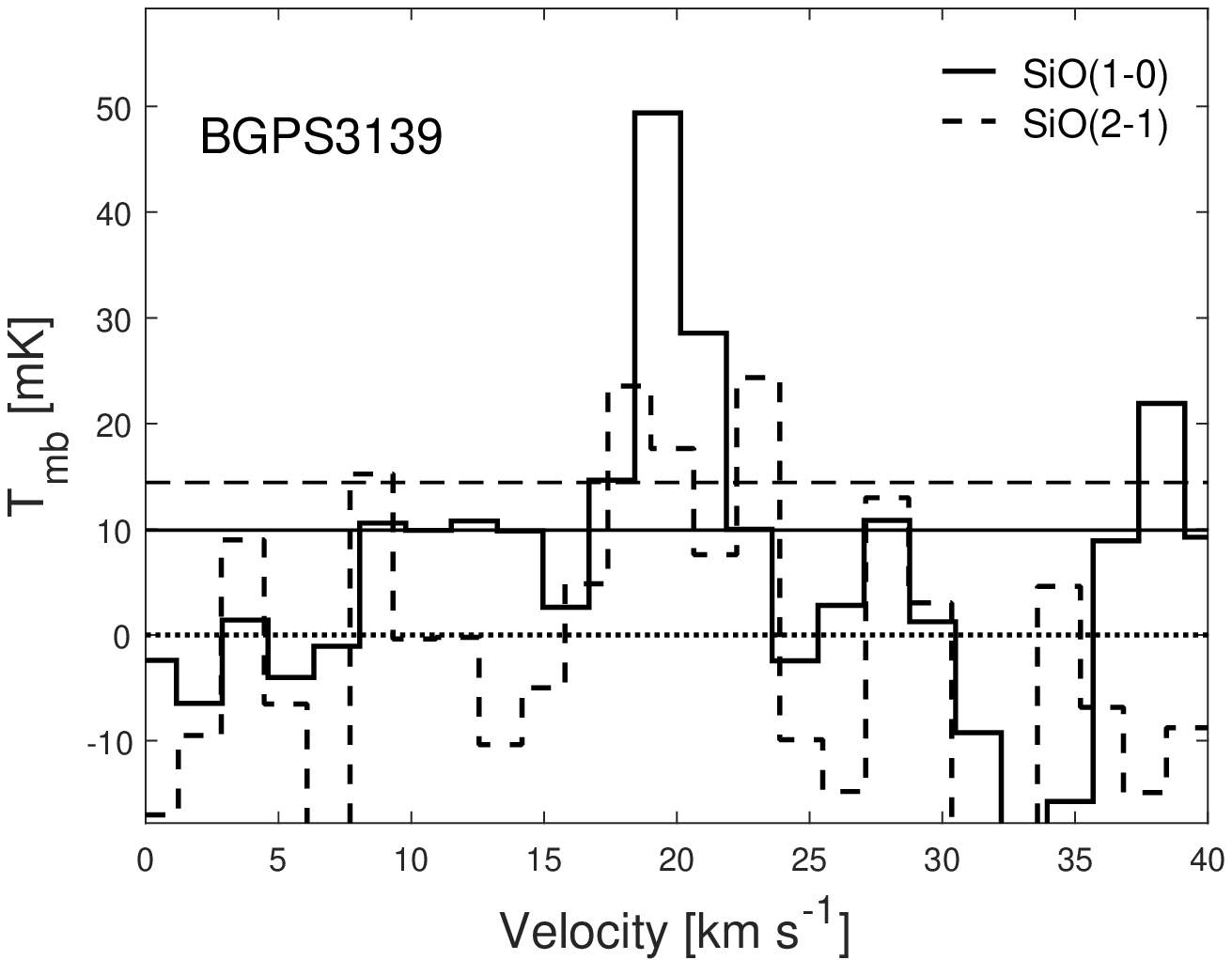}
  \includegraphics[scale=0.38]{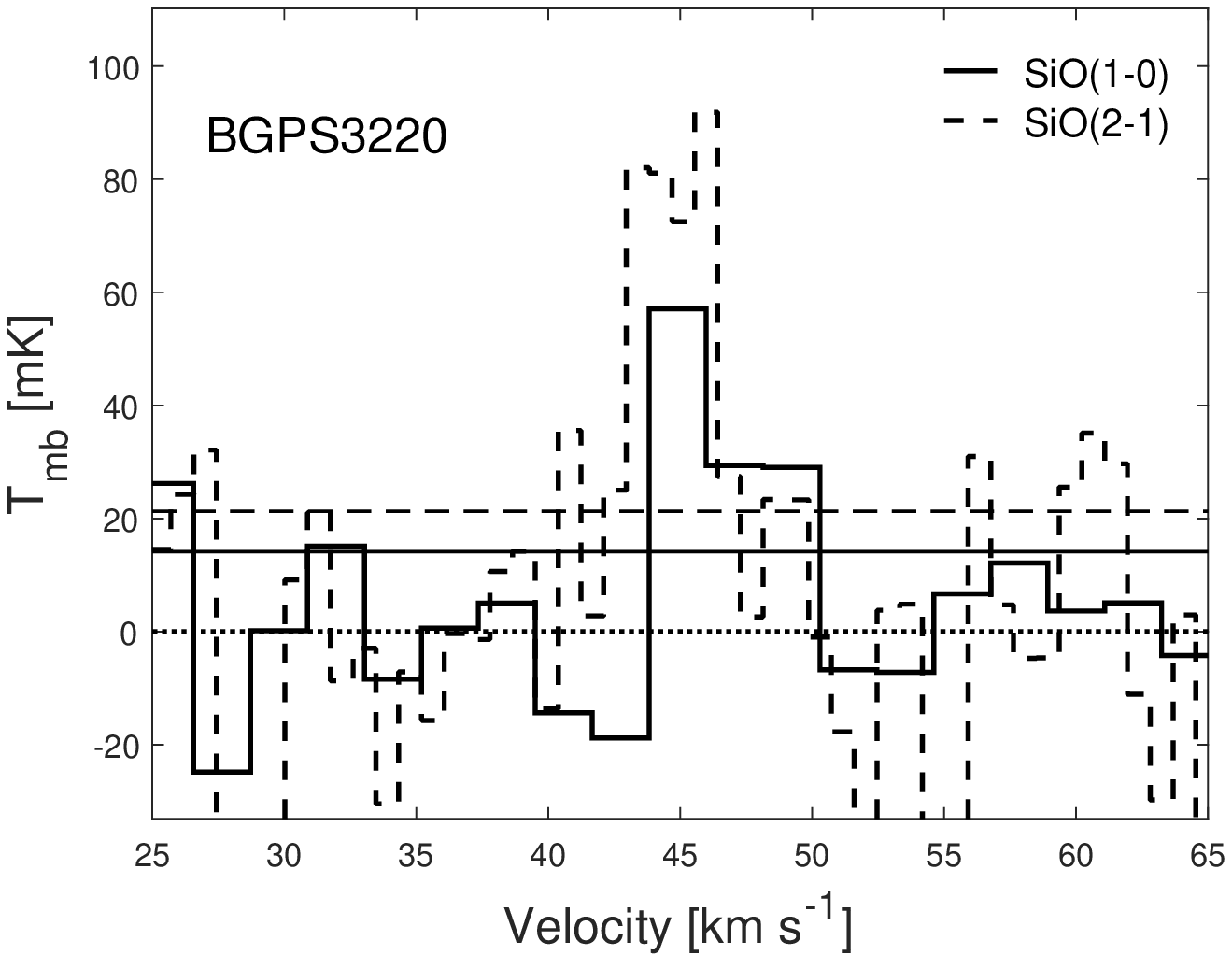}
  \includegraphics[scale=0.38]{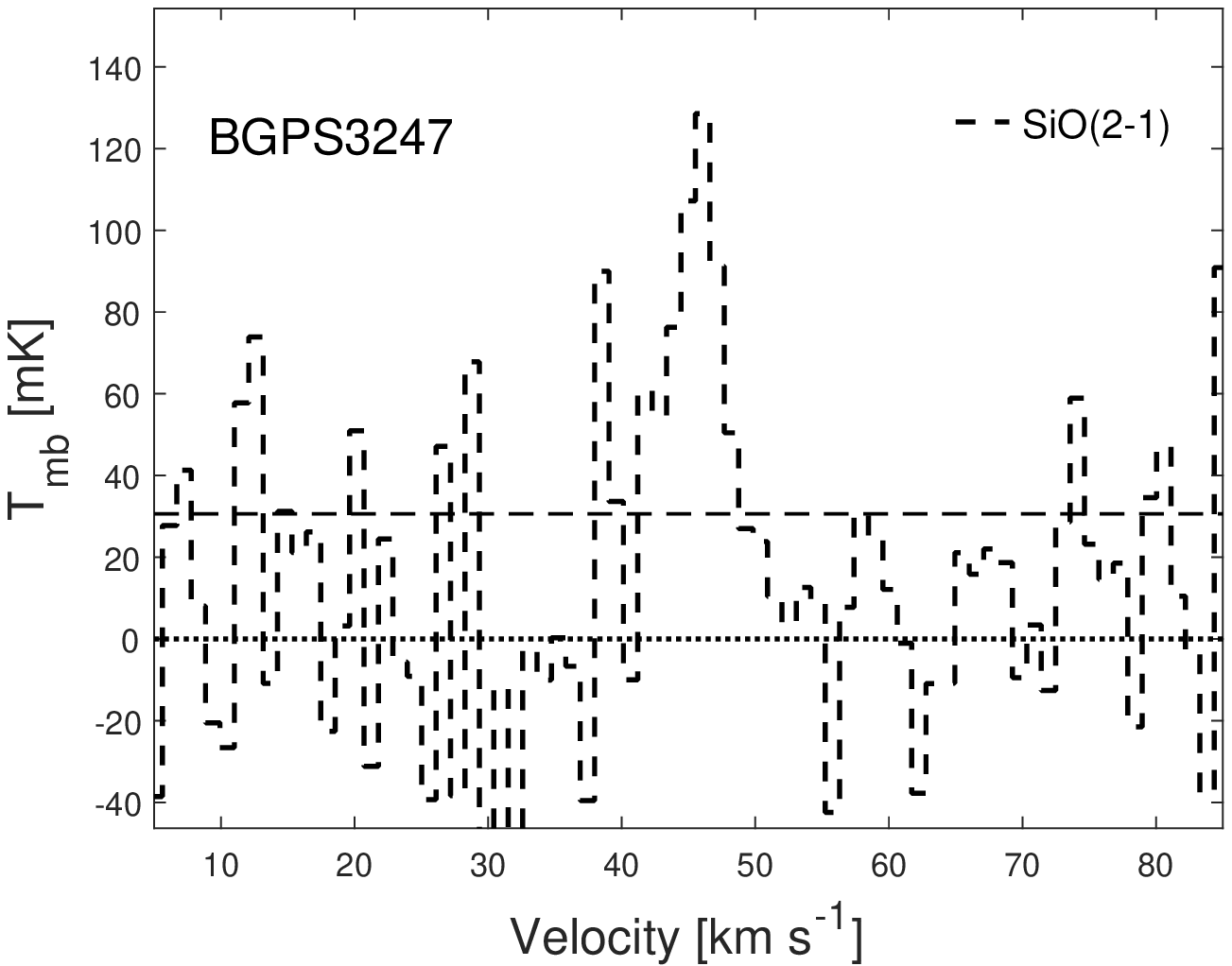}
  \includegraphics[scale=0.38]{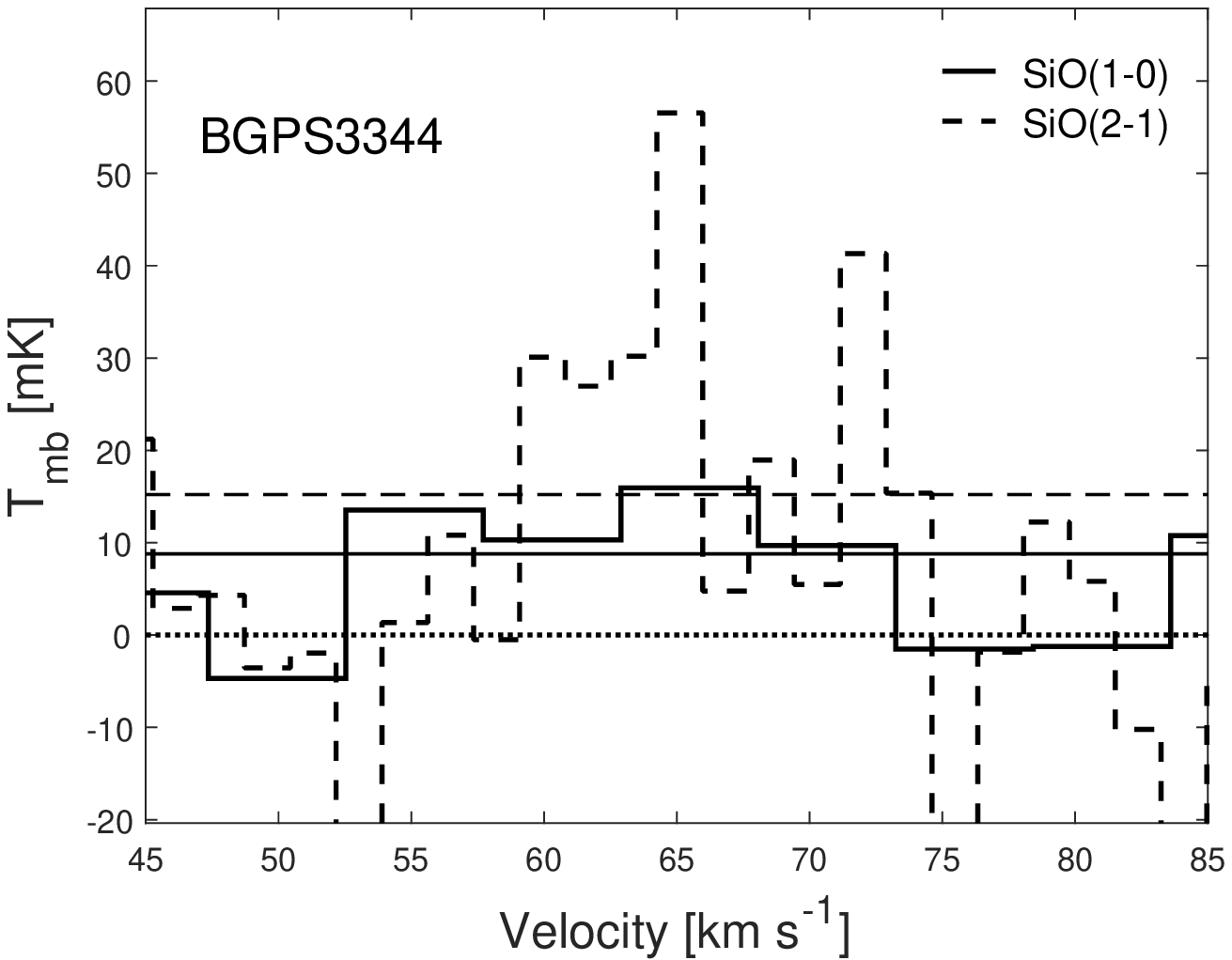}
  \includegraphics[scale=0.38]{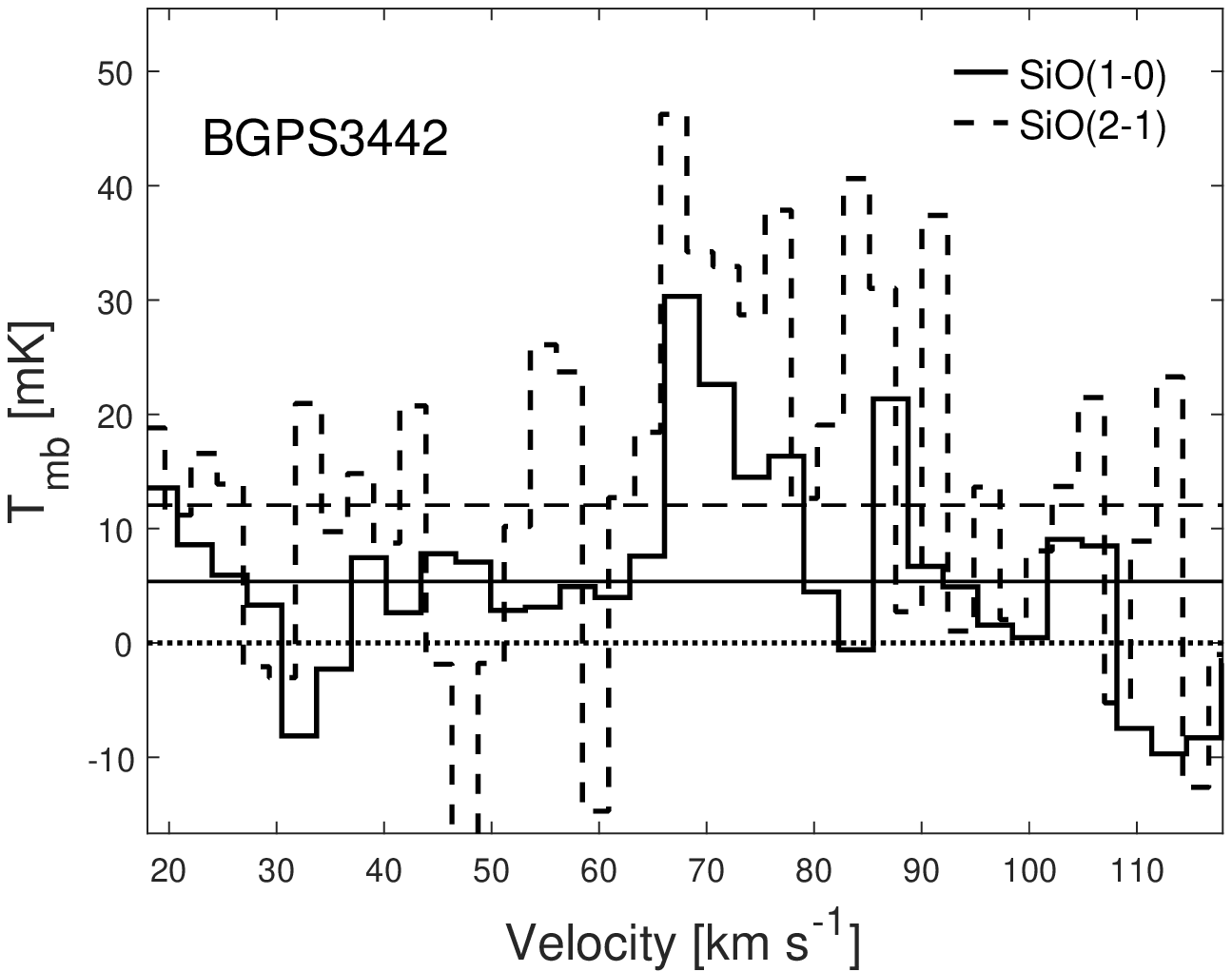}
  \includegraphics[scale=0.38]{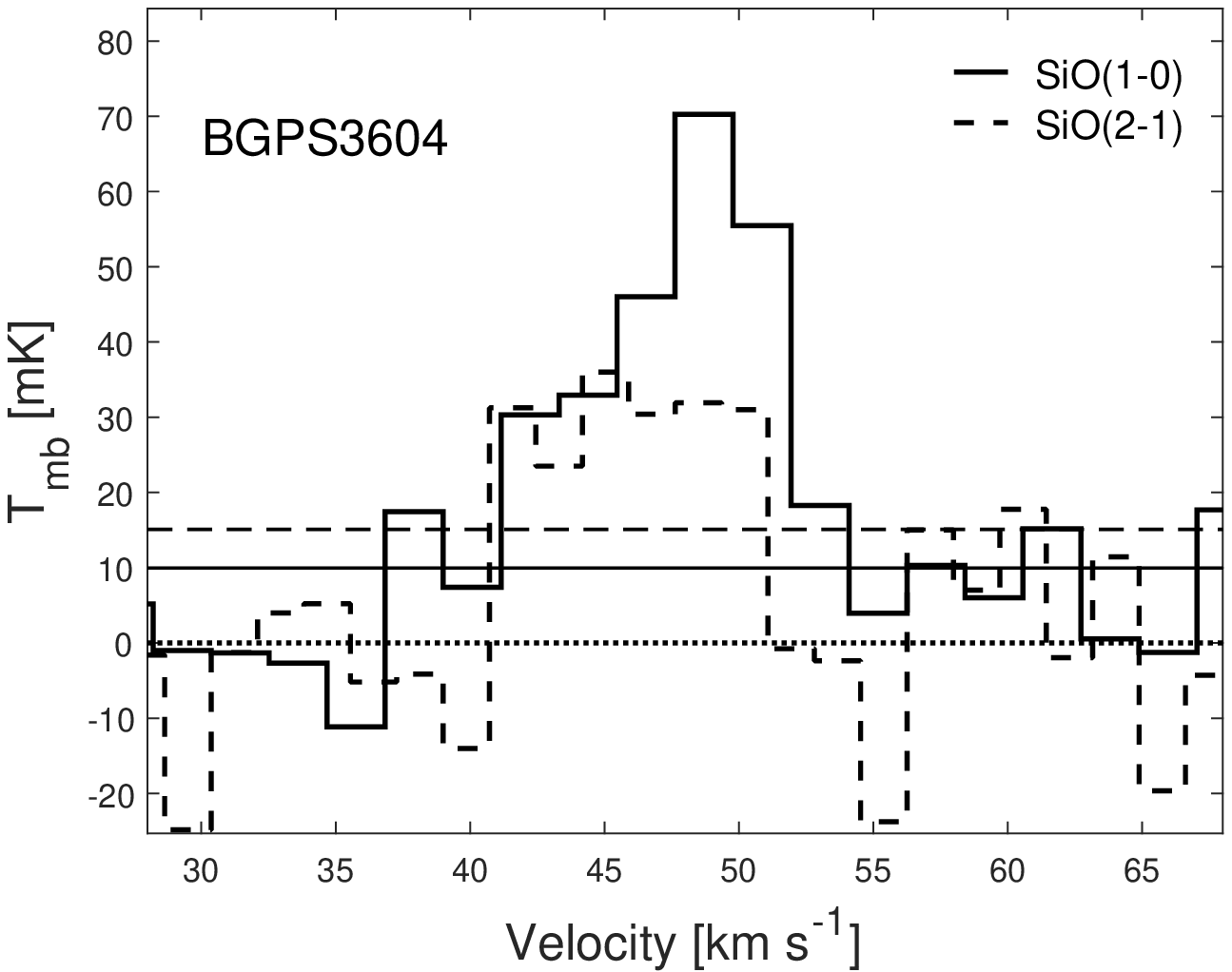}
  \includegraphics[scale=0.38]{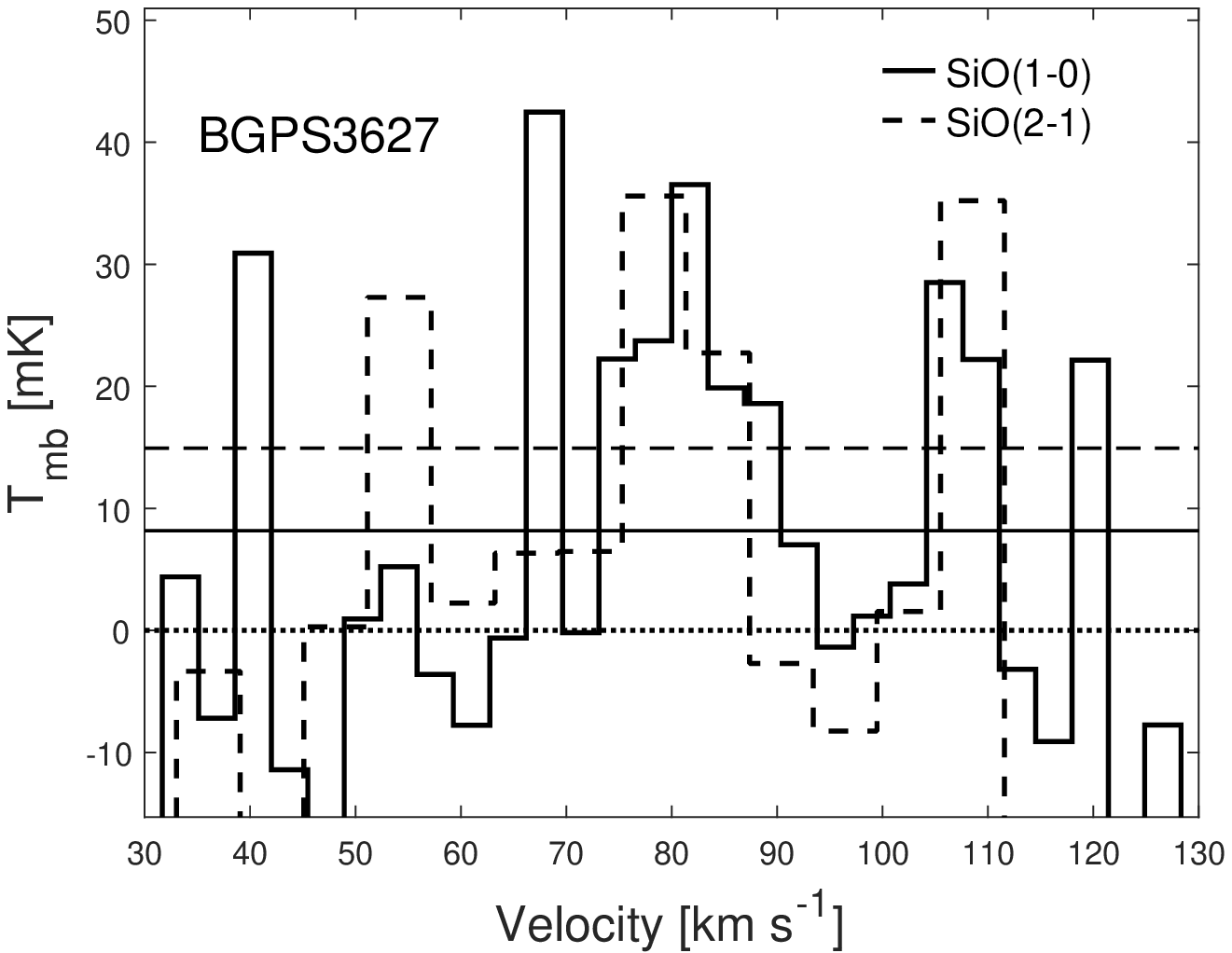}
  \caption{The SiO spectra for individual sources. The solid lines present the SiO 1-0 line profiles, and the dashed ones are the SiO 2-1 line profiles. The horizonal solid and dashed lines represent the rms noise levels of the SiO 1-0 and 2-1 lines at the velocity resolutions of these spectra, respectively. The horizonal dotted lines are baselines.}\label{fig:siospectrum}
\end{figure*}

\begin{figure*}
  \centering
  \includegraphics[scale=0.38]{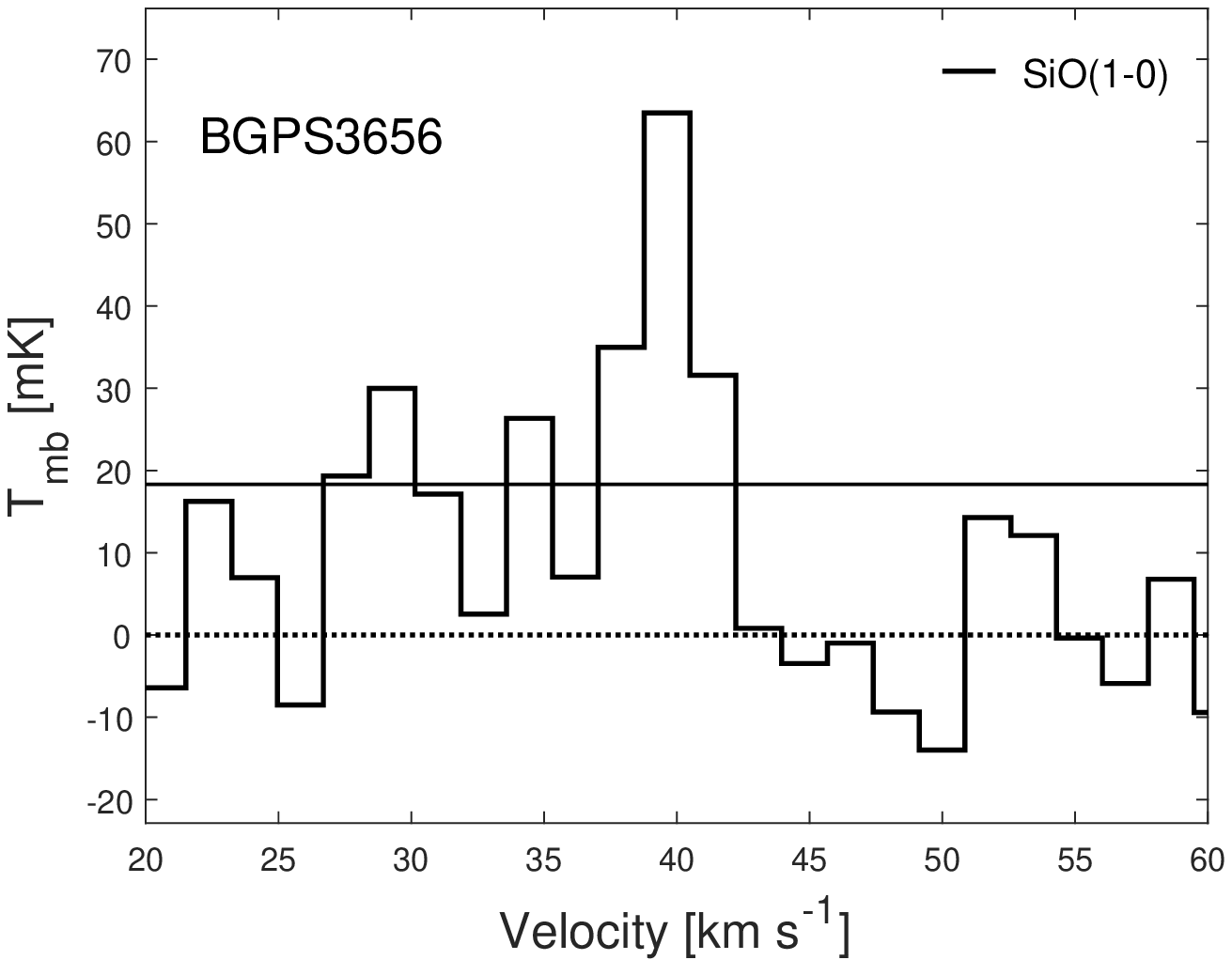}
  \includegraphics[scale=0.38]{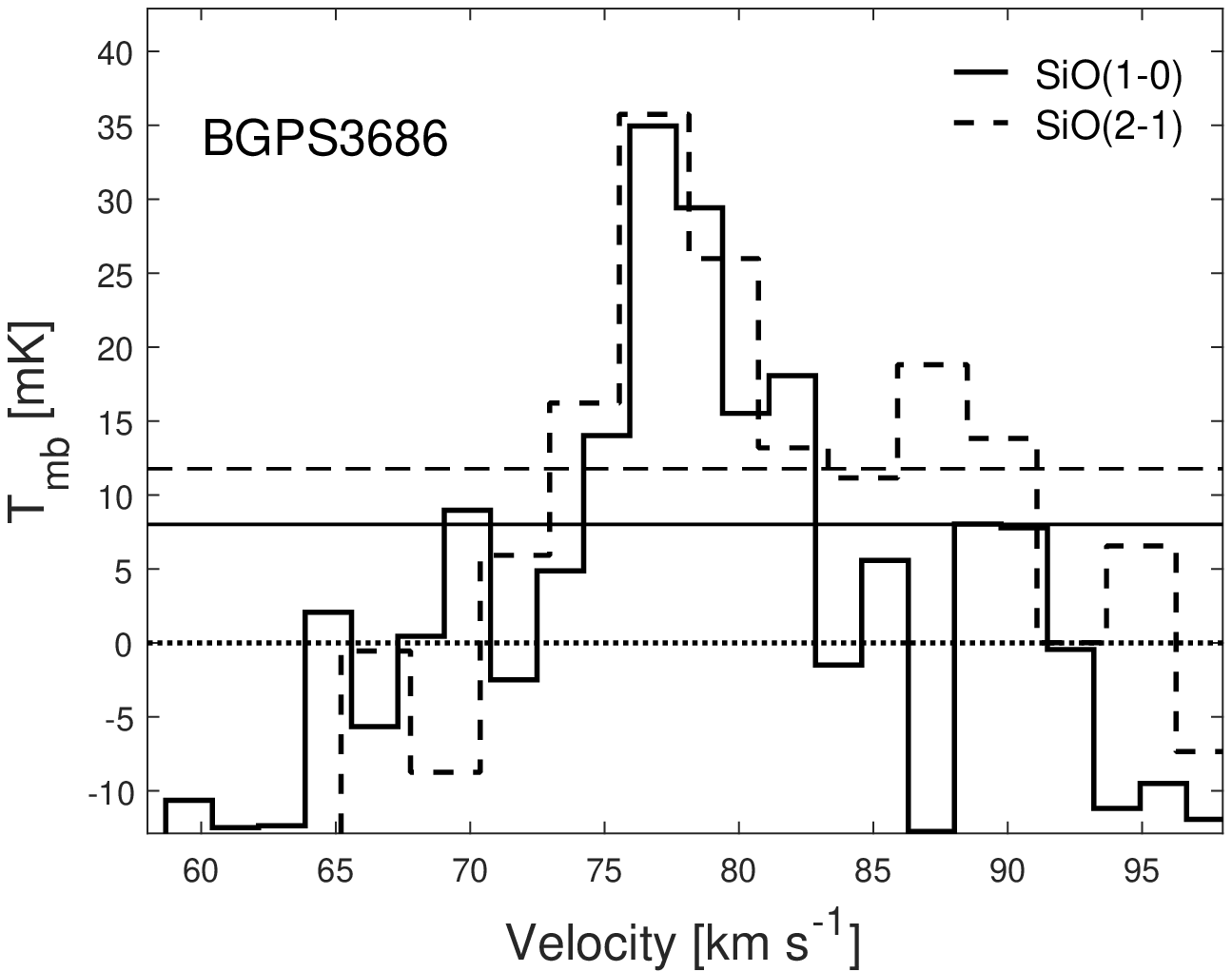}
  \includegraphics[scale=0.38]{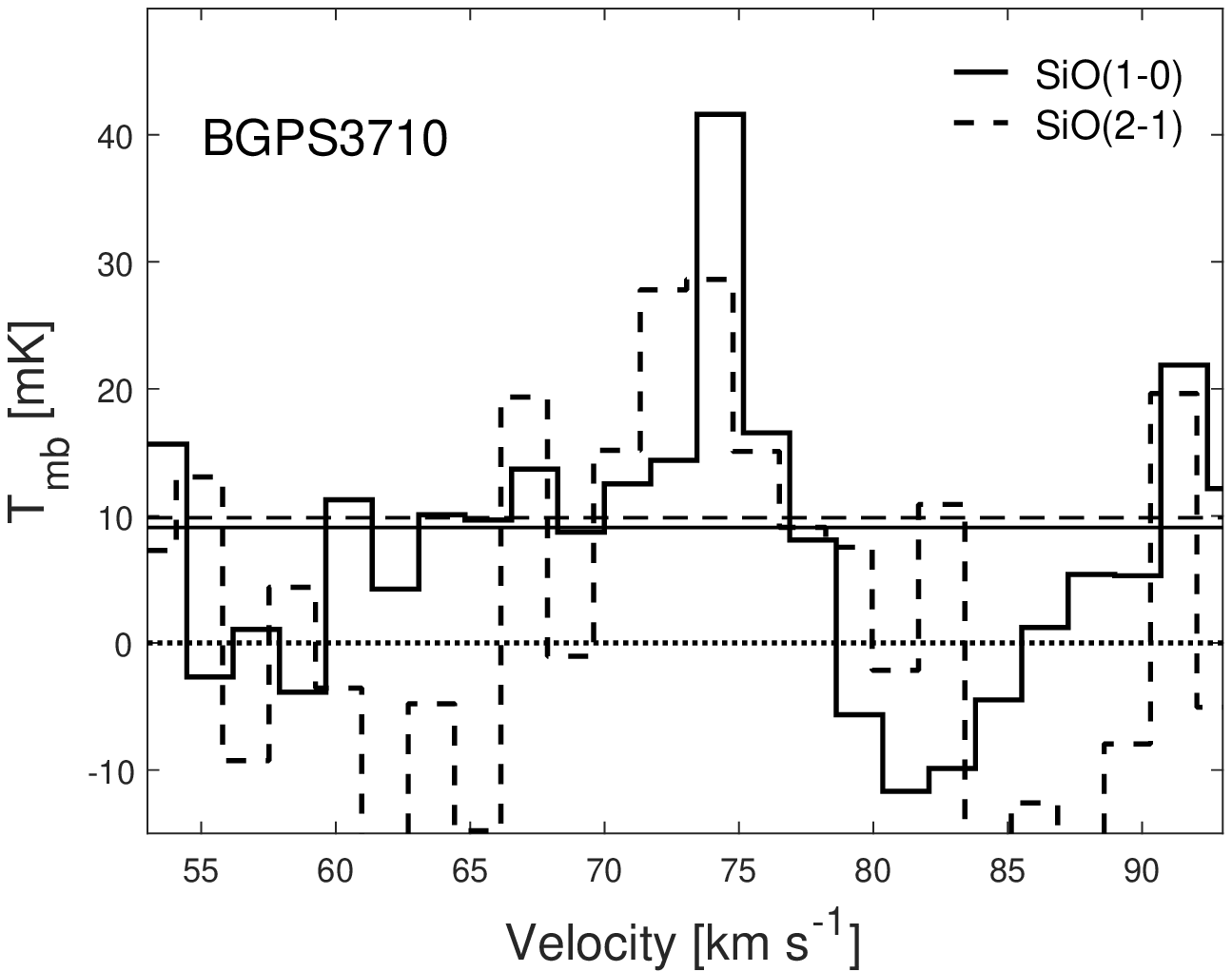}
  \includegraphics[scale=0.38]{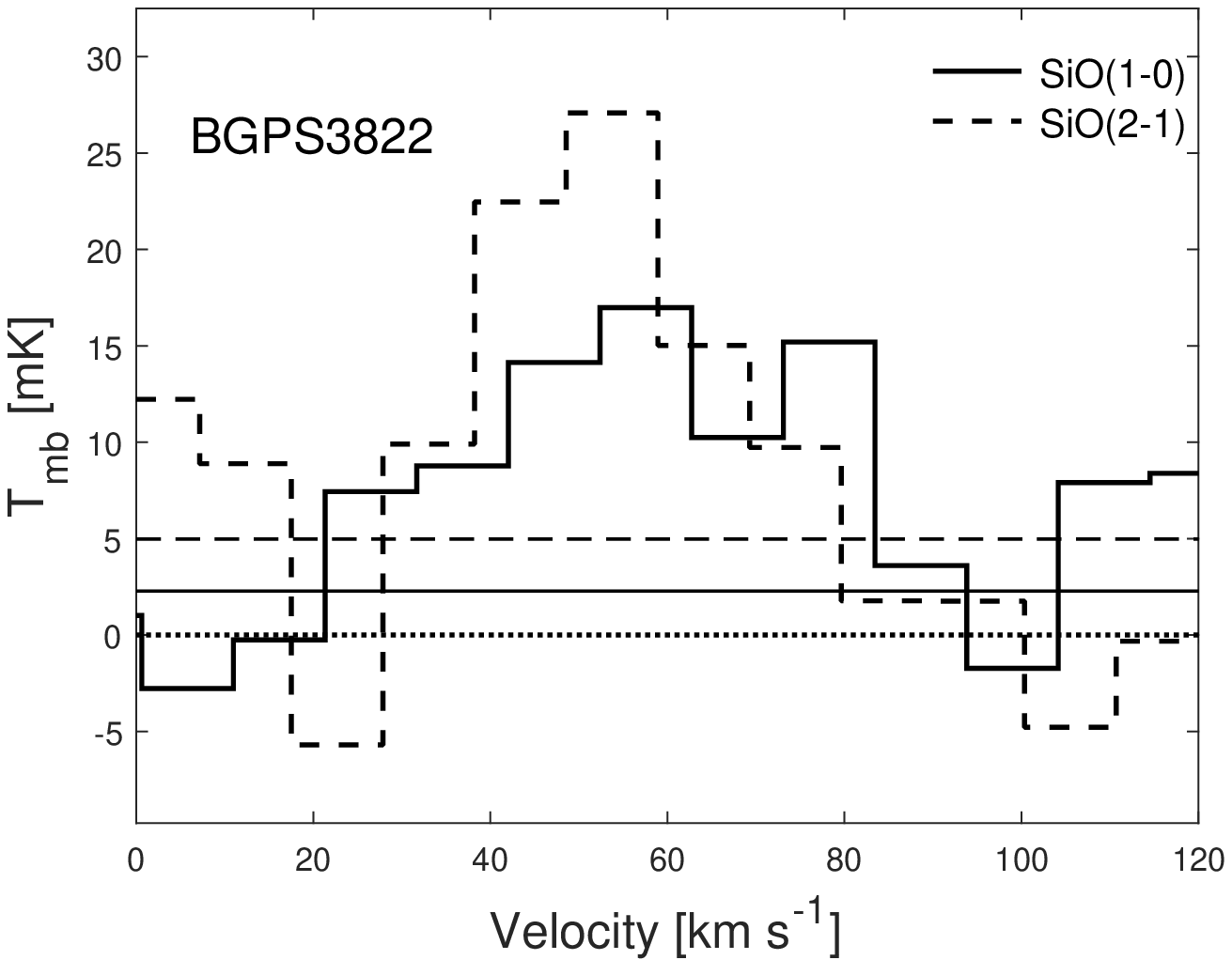}
  \includegraphics[scale=0.38]{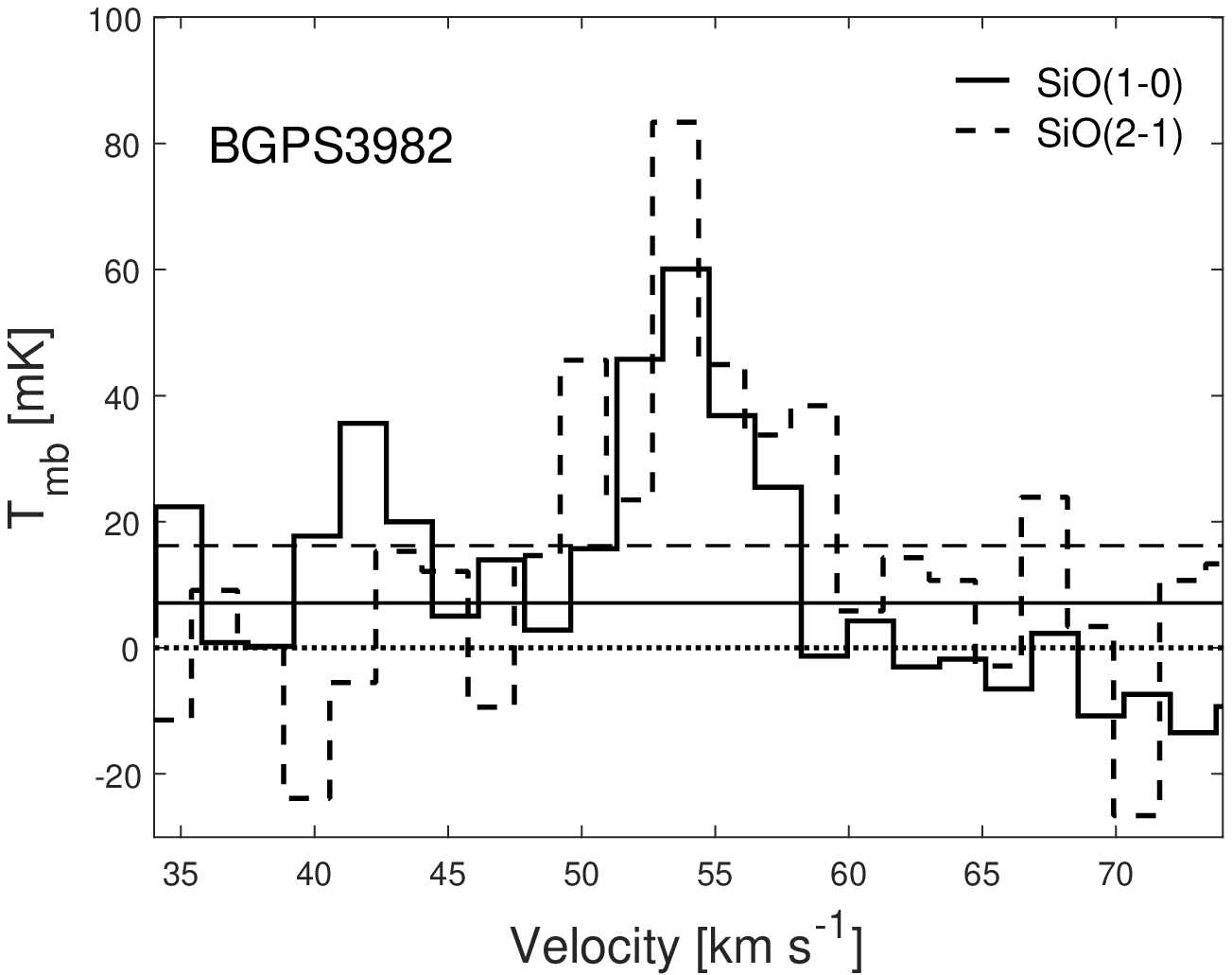}
  \includegraphics[scale=0.38]{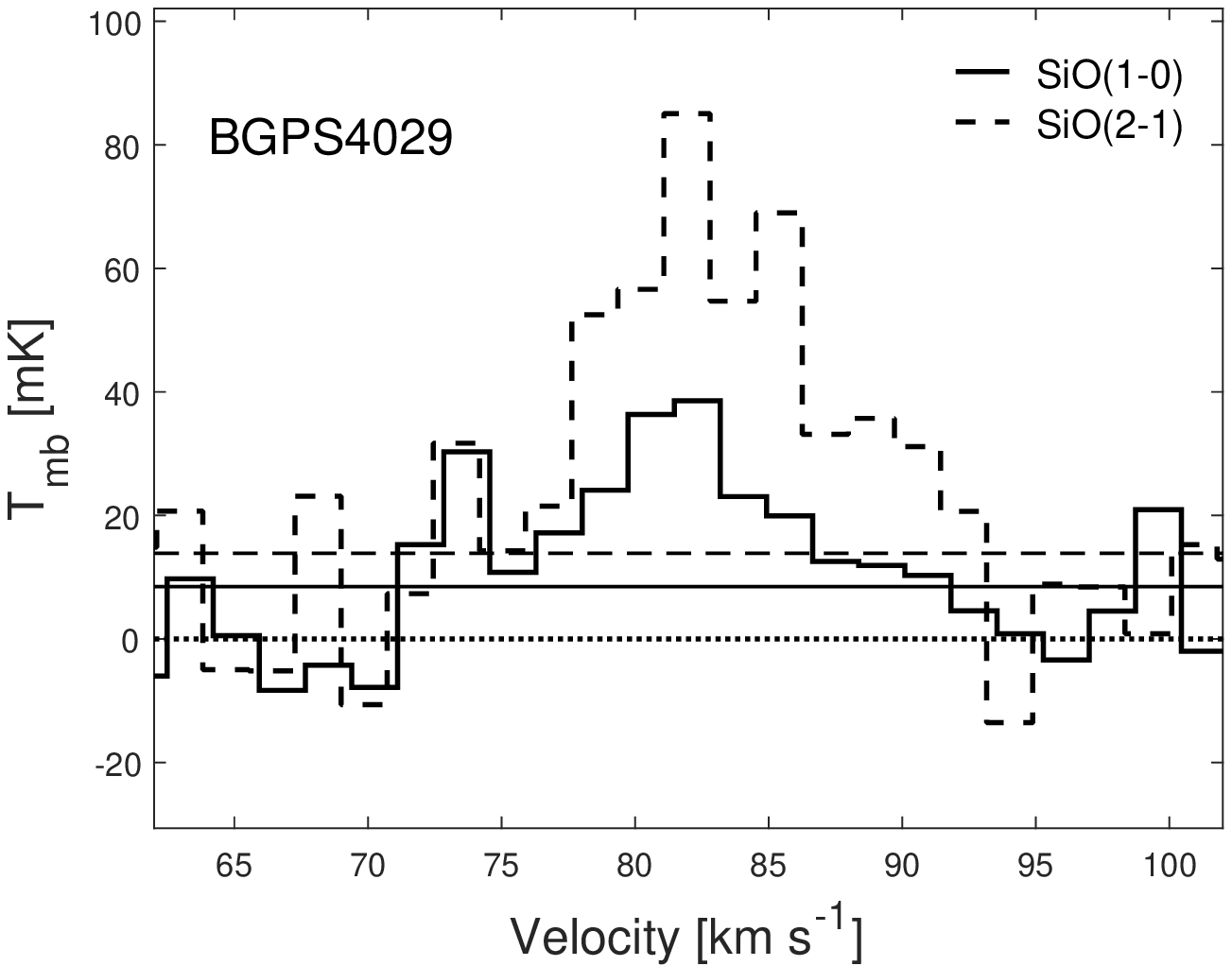}
  \includegraphics[scale=0.38]{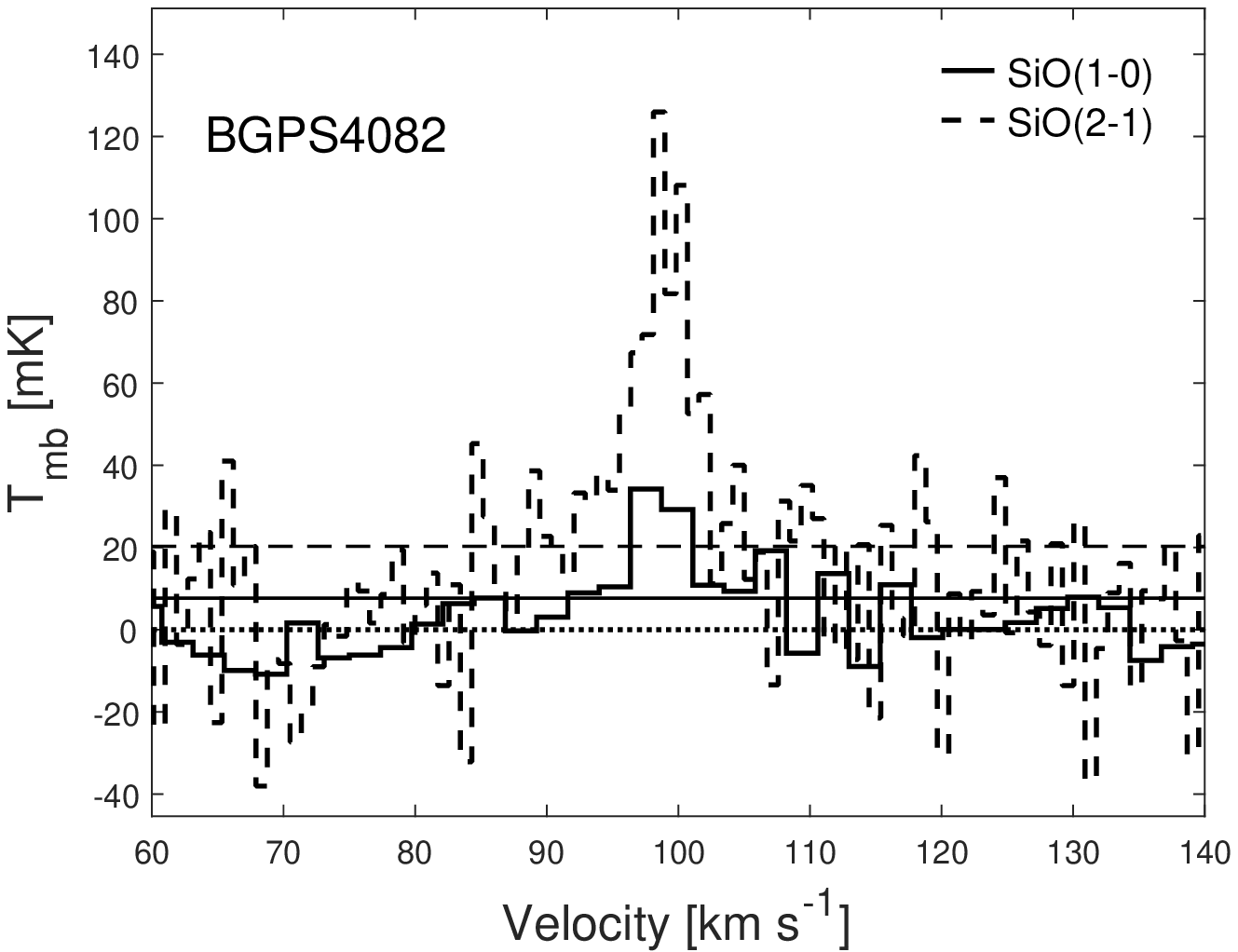}
  \includegraphics[scale=0.38]{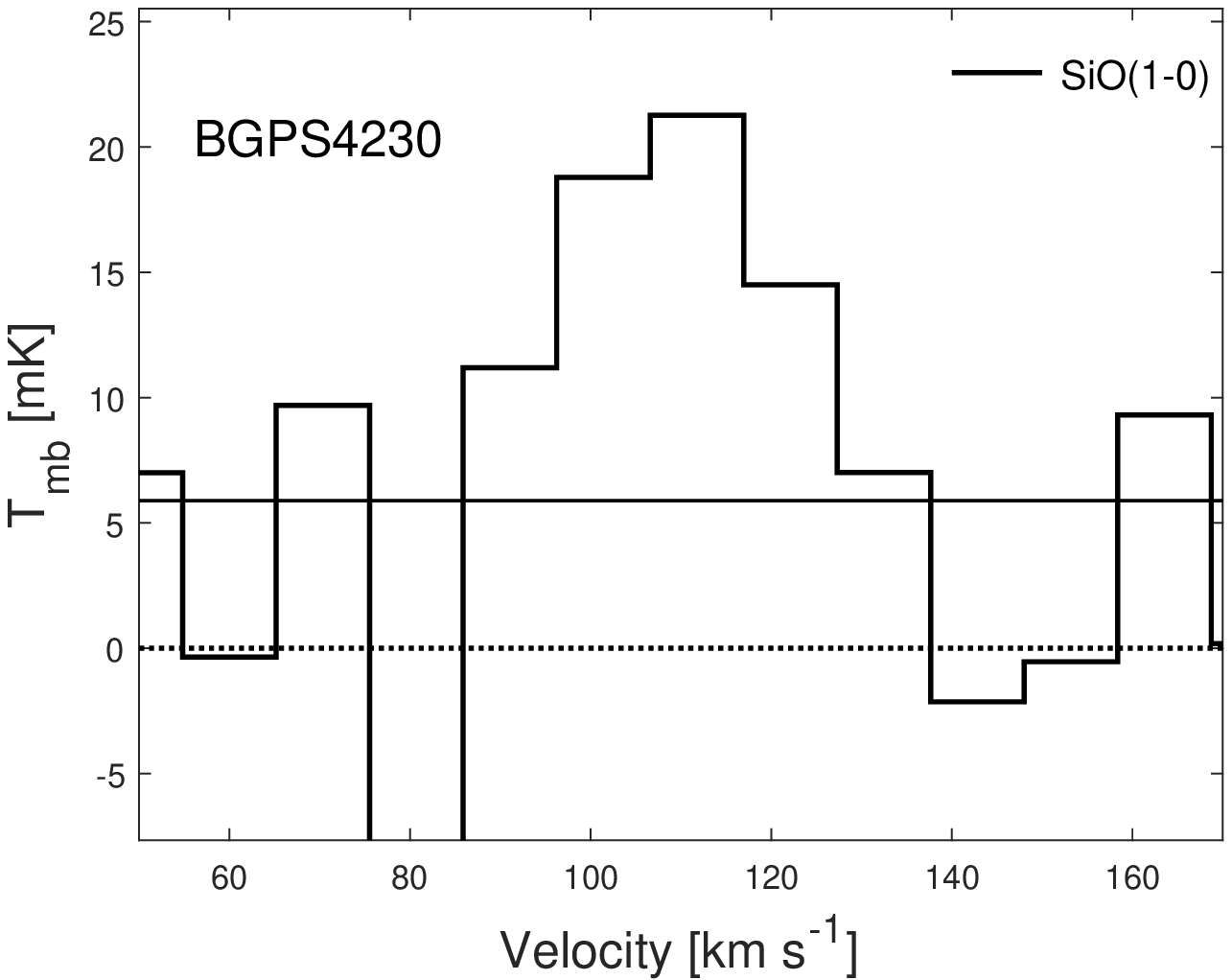}
  \includegraphics[scale=0.38]{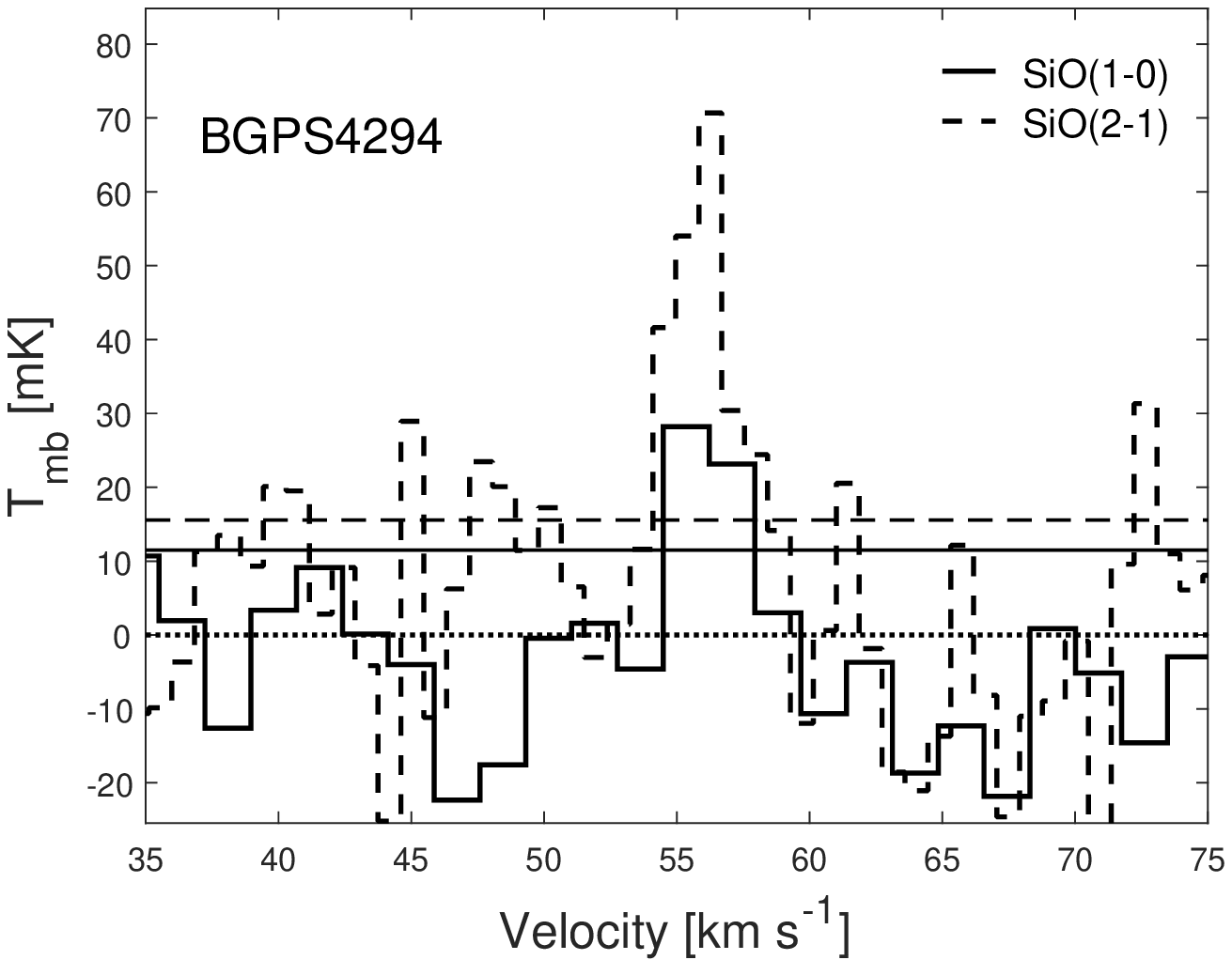}
  \includegraphics[scale=0.38]{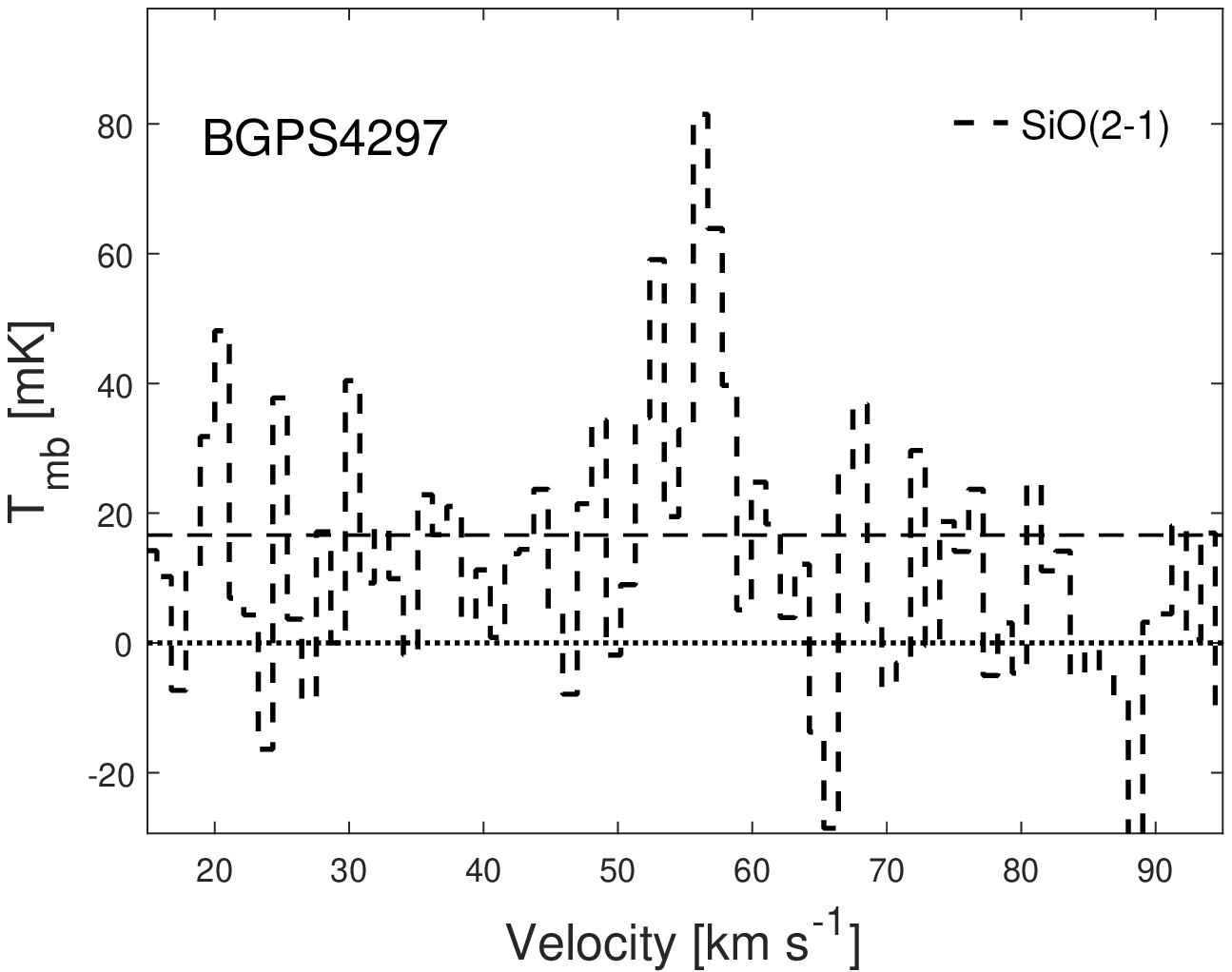}
  \includegraphics[scale=0.38]{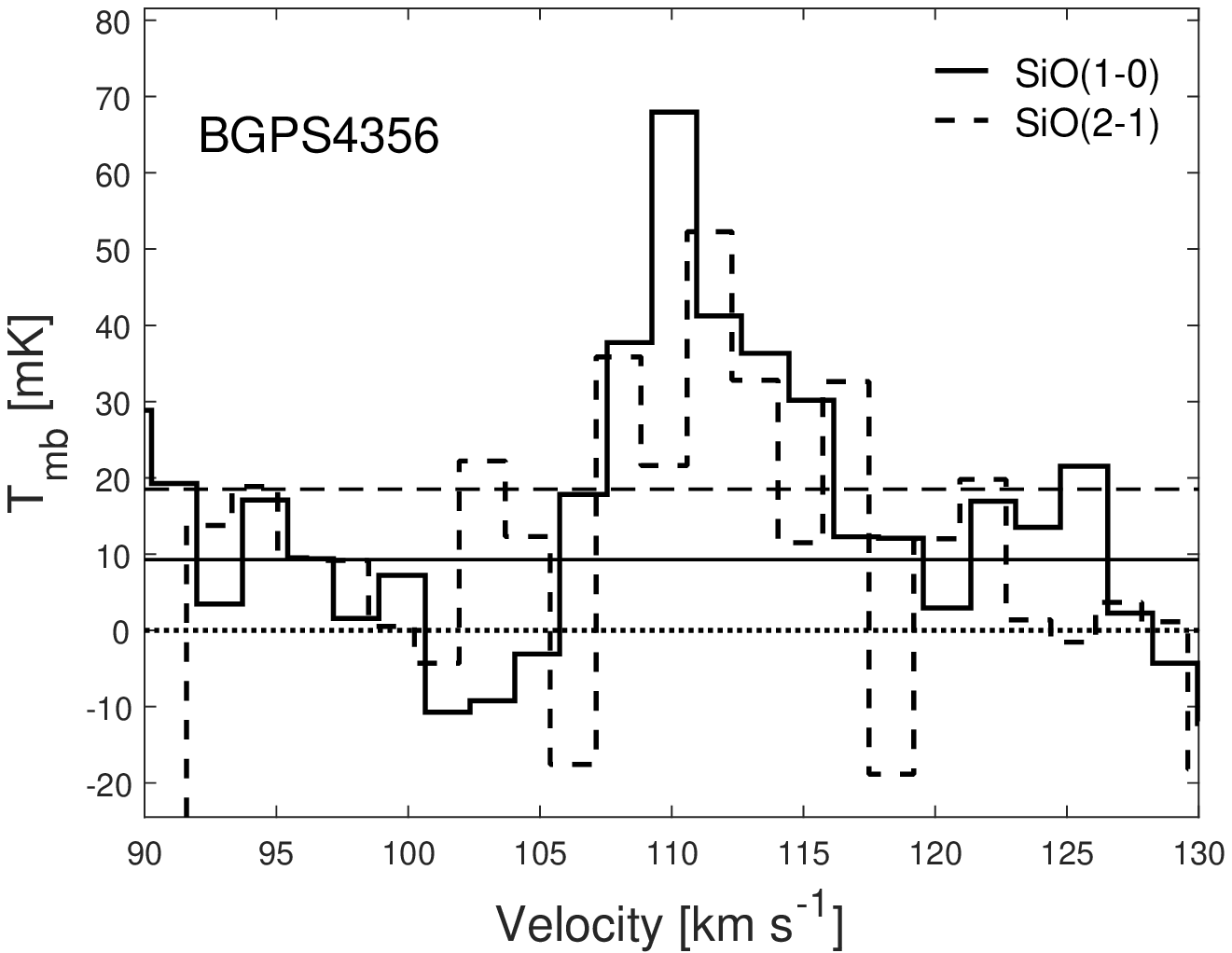}
  \includegraphics[scale=0.38]{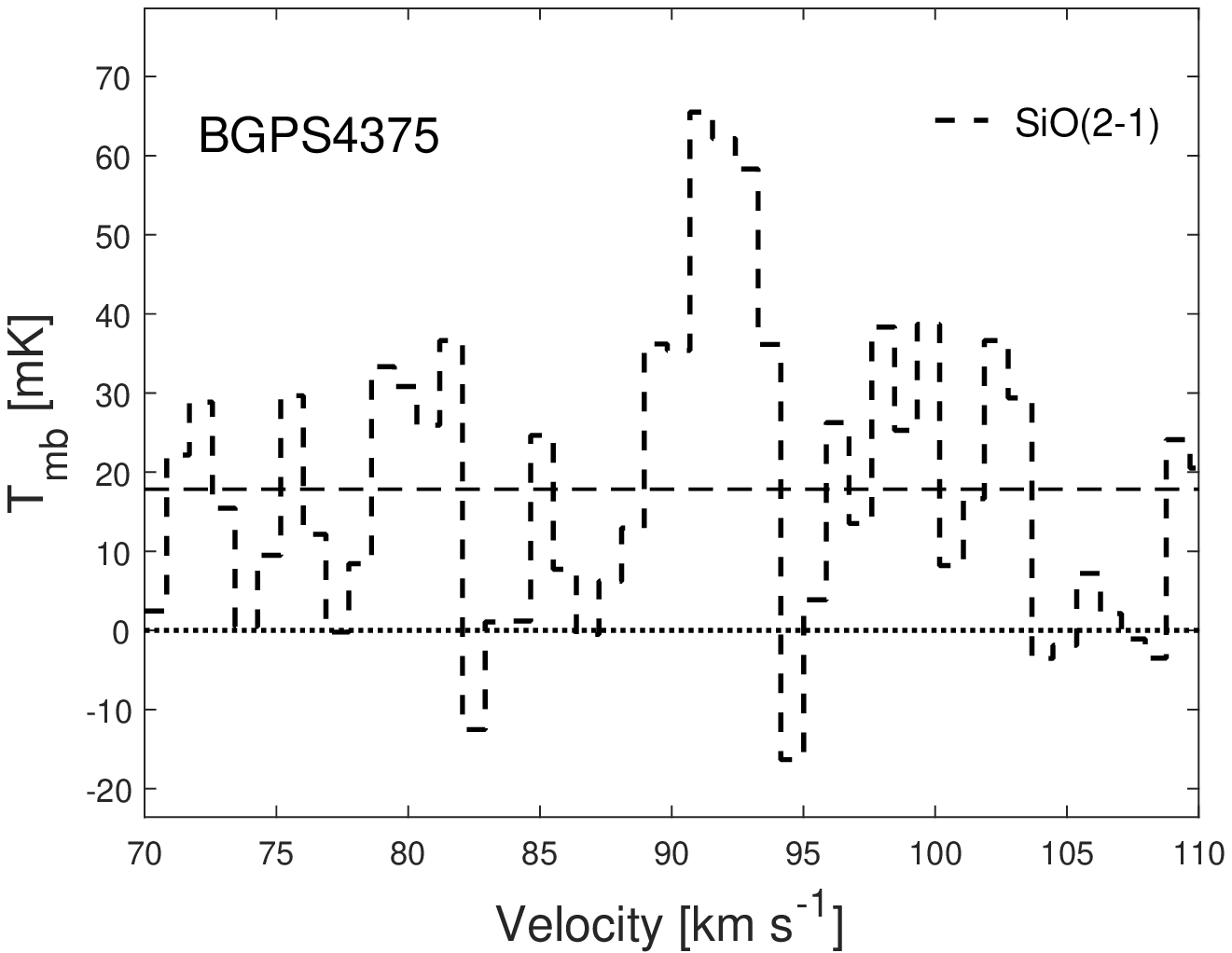}
  \includegraphics[scale=0.38]{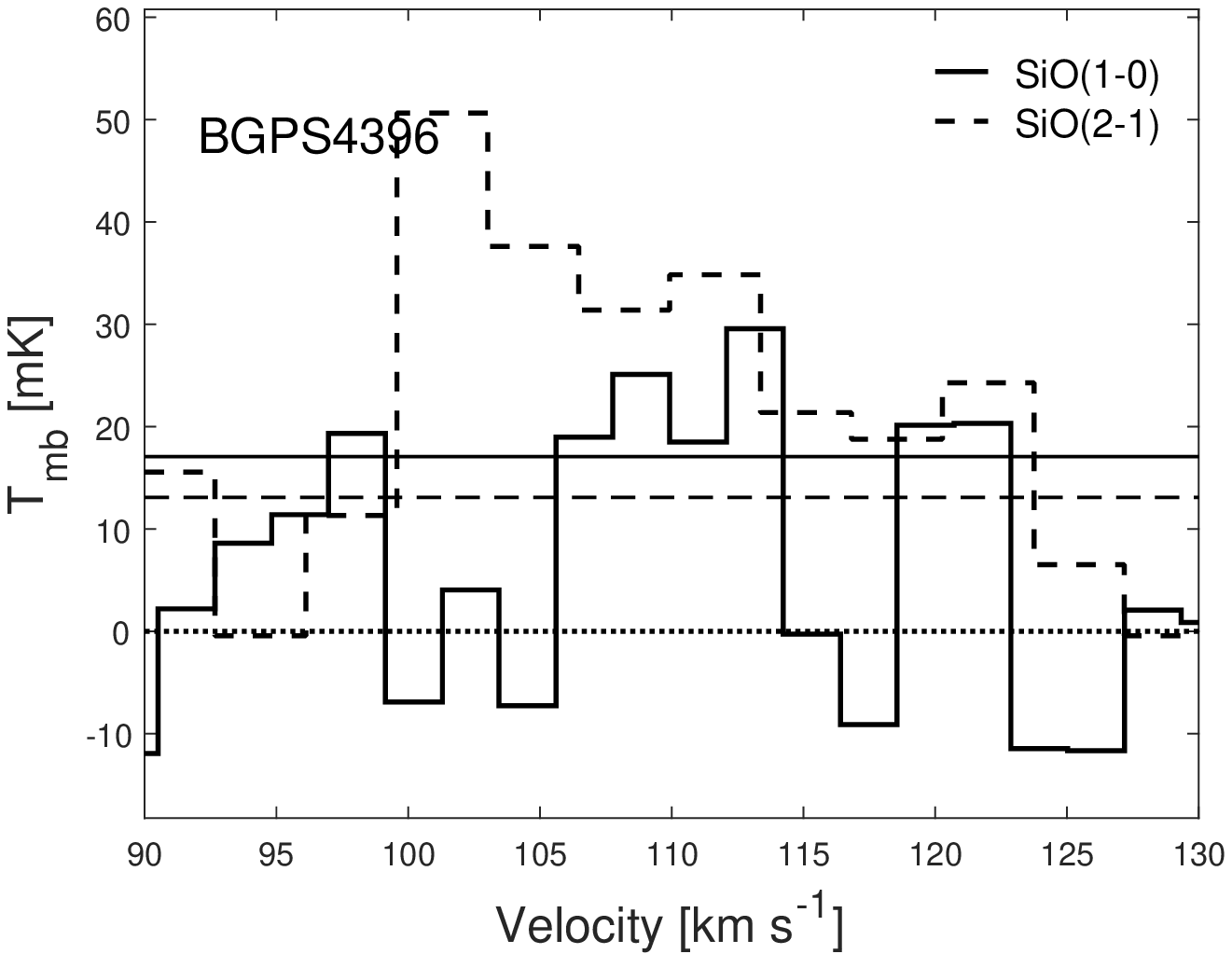}
  \includegraphics[scale=0.38]{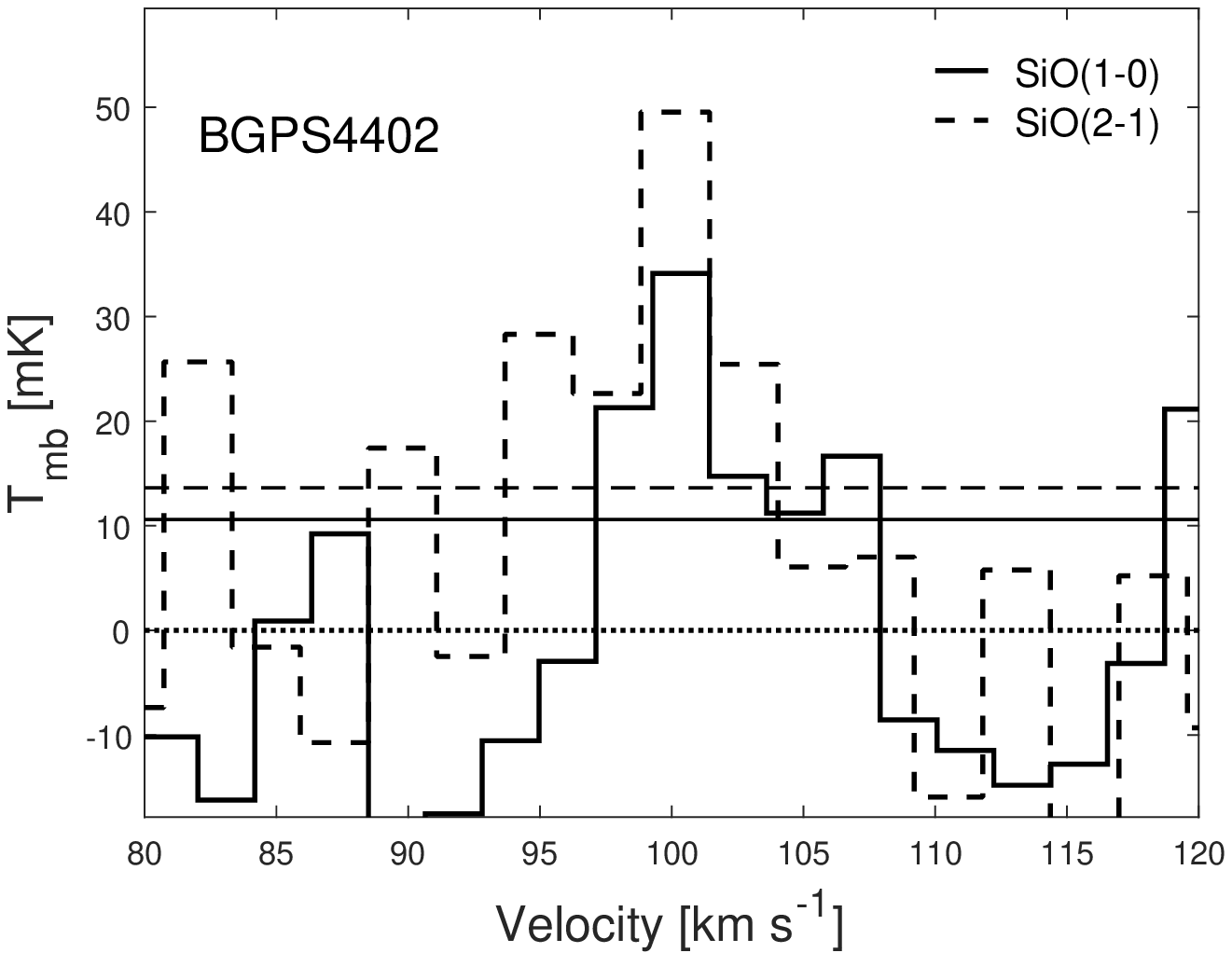}
  \includegraphics[scale=0.38]{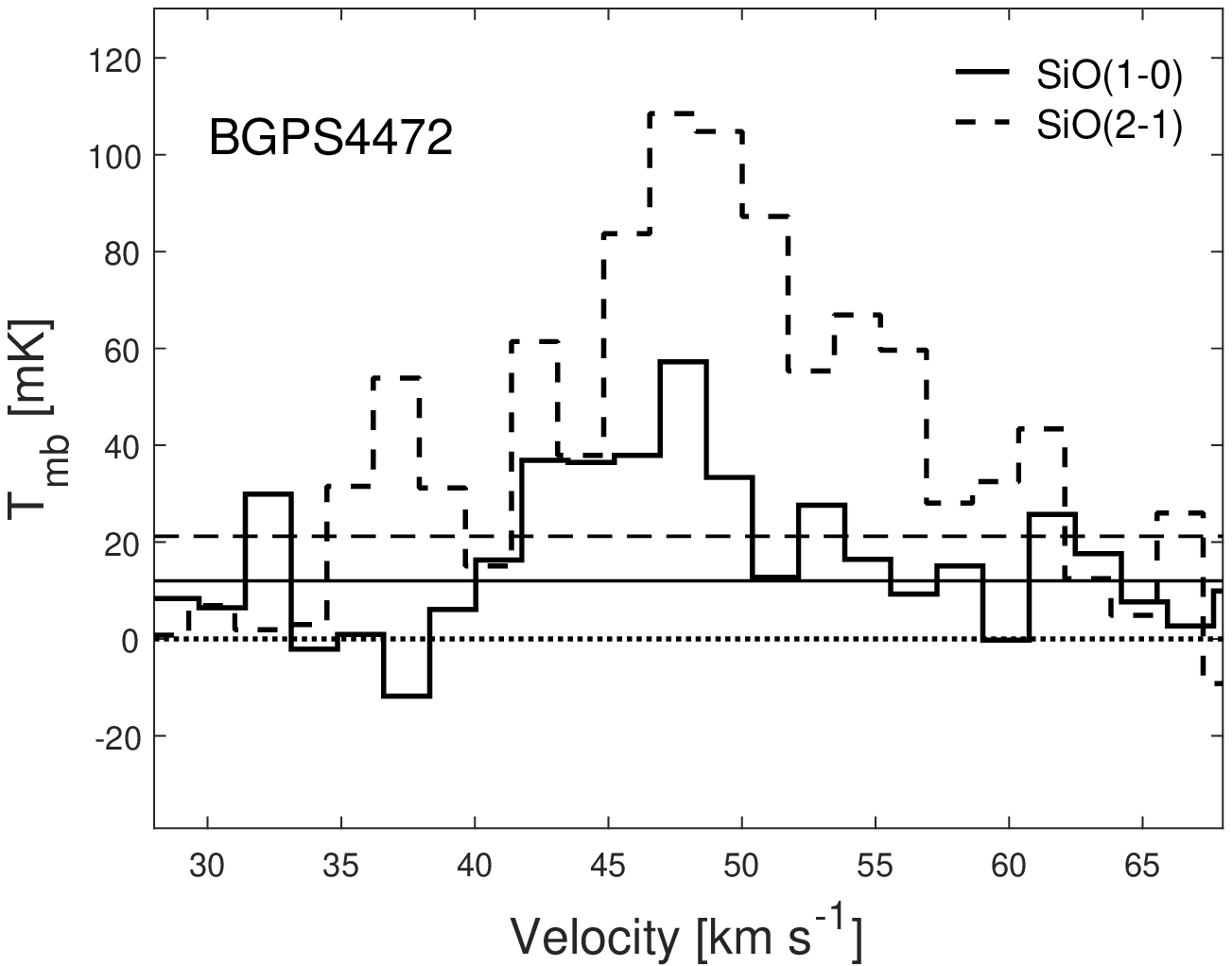}
  \caption{continued.}\label{fig:siospectrum2}
\end{figure*}

\begin{figure*}
  \centering
  \includegraphics[scale=0.38]{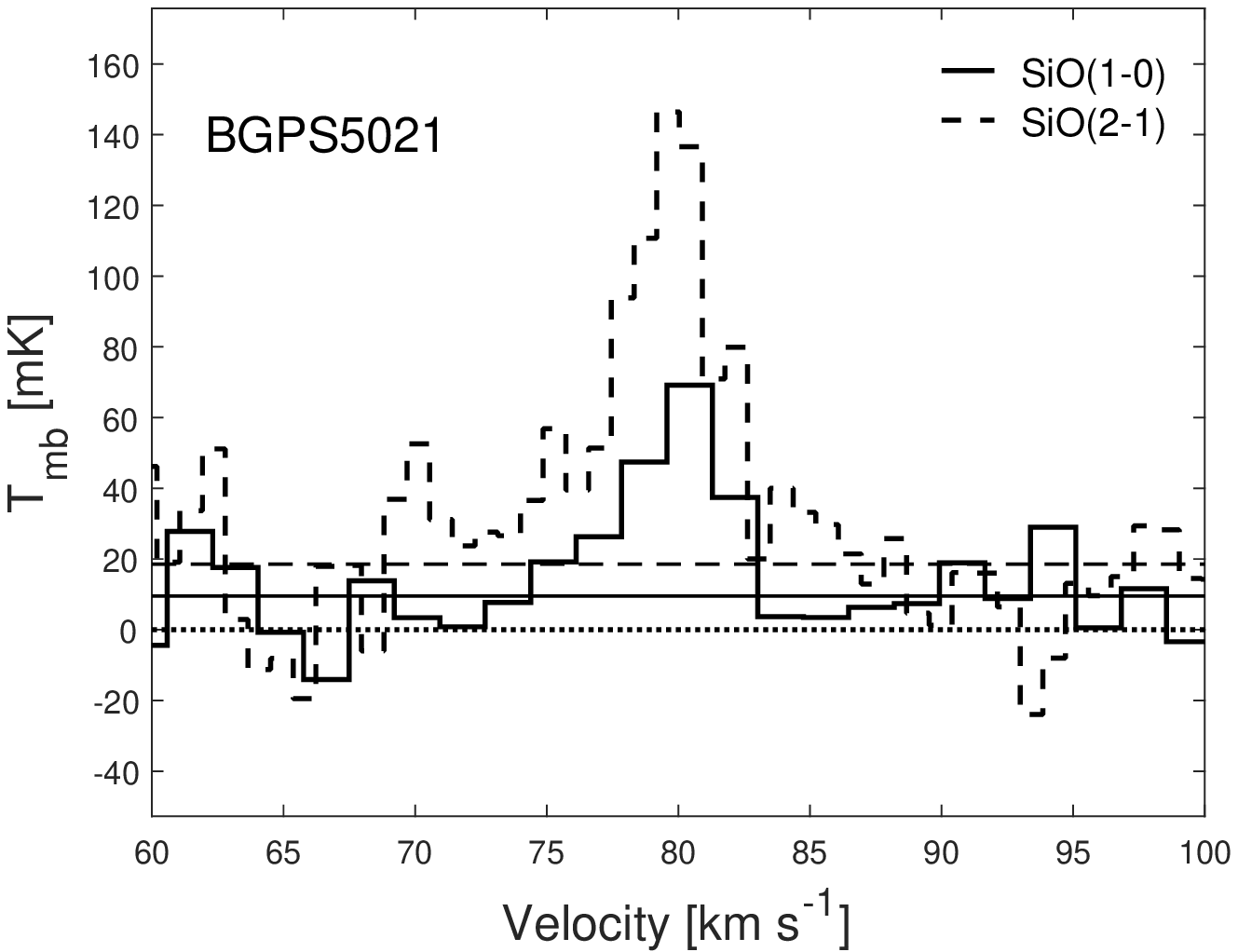}
  \includegraphics[scale=0.38]{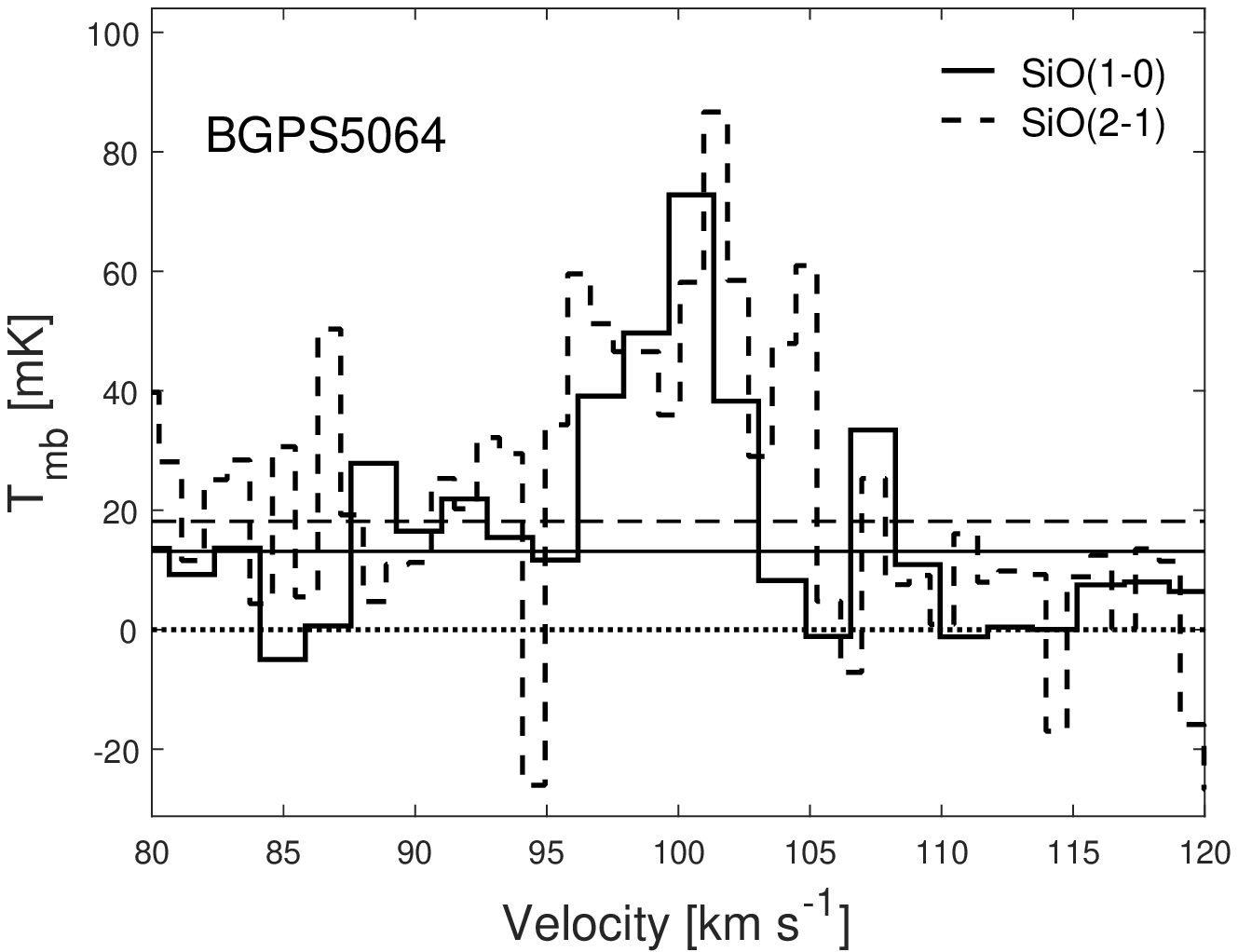}
  \includegraphics[scale=0.38]{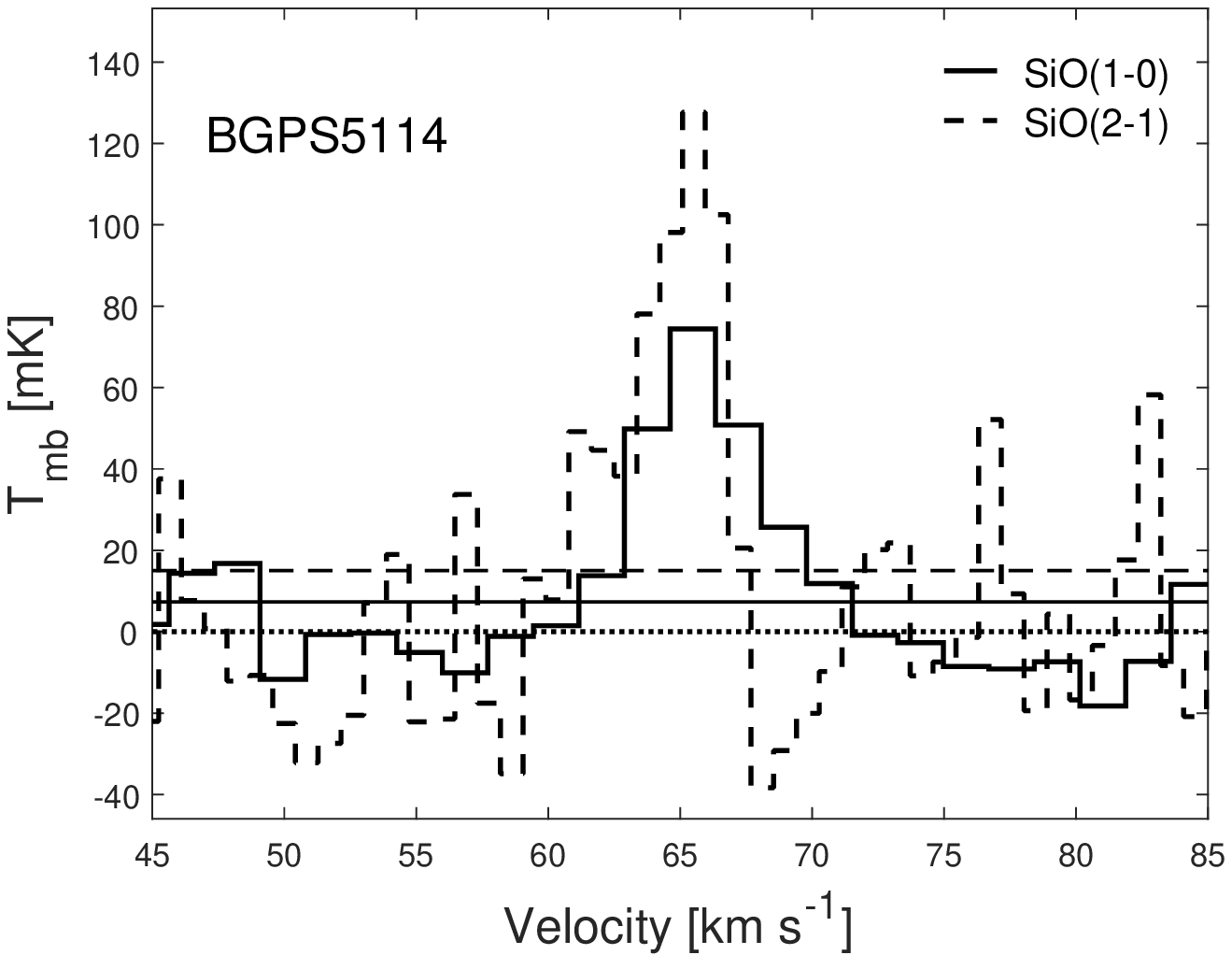}
  \includegraphics[scale=0.38]{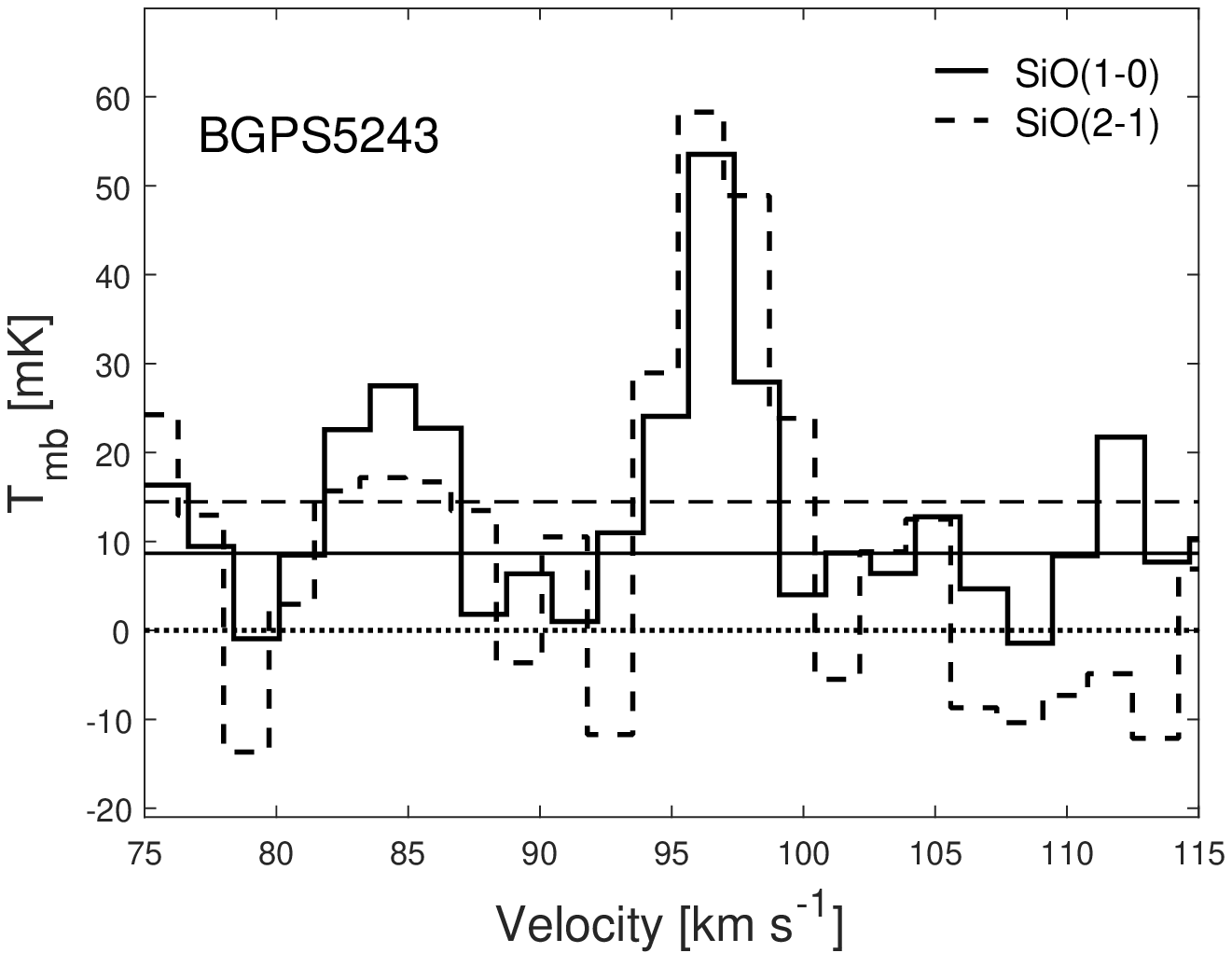}
  \caption{continued.}\label{fig:siospectrum3}
\end{figure*}


\bsp	
\label{lastpage}
\end{document}